\documentclass[11pt,a4paper]{article}

\usepackage{jheppub}
\usepackage{color}
\usepackage[dvipsnames]{xcolor}
\usepackage{graphicx}
\usepackage{wrapfig}
\usepackage{verbatim}
\usepackage{amsmath}
\usepackage{amssymb}
\usepackage{amsfonts}
\usepackage{url}
\usepackage{xspace}
\usepackage{slashed}
\usepackage{multirow}
\usepackage{threeparttable}
\usepackage{paralist}
\usepackage{subcaption}
\usepackage{floatrow}
\usepackage{multirow}

\usepackage{array} 
\usepackage{arydshln}

\usepackage{tikz}
\usetikzlibrary{trees}
\usetikzlibrary{decorations.pathmorphing}
\usetikzlibrary{decorations.markings}
\usetikzlibrary{arrows}

\newcommand{\hc}{\mathrm{h.c.}}

\newcommand{\beq}{\begin{equation}}
\newcommand{\eeq}{\end{equation}}
\newcommand{\bea}{\begin{eqnarray}}
\newcommand{\eea}{\end{eqnarray}}

\newcommand{\Tr}{\mathrm{Tr}}
\newcommand{\Lag}{\mathcal{L}}
\newcommand{\Ham}{\mathcal{H}}

\DeclareRobustCommand{\Sec}[1]{Sec.~\ref{#1}}

\DeclareRobustCommand{\App}[1]{App.~\ref{#1}}
\DeclareRobustCommand{\Tab}[1]{Table~\ref{#1}}

\DeclareRobustCommand{\Fig}[1]{Fig.~\ref{#1}}

\DeclareRobustCommand{\Eq}[1]{\text{Eq.~(\ref{#1})}}
\DeclareRobustCommand{\Eqs}[2]{\text{Eqs.~(\ref{#1}) and (\ref{#2})}}

\newcommand{\cfg}{c_{f\gamma}}

\newcommand{\wtilde}{\widetilde}
\newcommand{\La}{\Lambda}

\newcommand{\add}[1]{\textcolor{Black}{#1}}

\DeclareRobustCommand{\Re}[1]{\mathrm{Re}\left[#1\right]}
\DeclareRobustCommand{\Im}[1]{\mathrm{Im}\left[#1\right]}

\newcommand\Tstrut{\rule{0pt}{2.6ex}}

\notoc

\begin{document}

{\large
\flushright TUM-HEP-1345-21\\
DESY-21-149\\}

\title{Electric dipole moments at one-loop in the dimension-6 SMEFT}

\author[a,b,c]{Jonathan Kley,}
\author[a]{Tobias Theil,}
\author[a]{Elena Venturini,}
\author[a]{Andreas Weiler}

\affiliation[a]{Physik-Department, Technische Universit\"at M\"unchen, James-Franck-Strasse~1, 85748 Garching, Germany}
\affiliation[b]{Deutsches Elektronen-Synchrotron DESY, Notkestr.~85, 22607 Hamburg, Germany}
\affiliation[c]{Institut f\"ur Physik, Humboldt-Universit\"at zu Berlin, Newtonstrasse~15, 12489 Berlin, Germany}


\date{\today}

\abstract{In this paper we present the complete expressions of the lepton and neutron electric dipole moments (EDMs) in the Standard Model Effective Field Theory (SMEFT), up to 1-loop and dimension-6 level and including both renormalization group running contributions and finite corrections. The latter play a fundamental role in the cases of operators that do not renormalize the dipoles, but there are also classes of operators for which they provide an important fraction, $10-20\%$, of the total 1-loop contribution, if the new physics scale is around $\Lambda=5$ TeV.
We present the full set of bounds on each individual Wilson coefficient contributing to the EDMs using both the current experimental constraints, as well as those from future experiments, which are expected to improve by at least an order of magnitude.}

\preprint{}
\maketitle

\newpage
\tableofcontents


\section{Introduction} \label{sec:intro}

Electric dipole moments (EDMs) constitute a set of low energy observables which are extremely sensitive to physics beyond the Standard Model (SM). This is due to the fact that -- as a consequence of their CP violating nature -- EDMs are strongly suppressed within the SM and are far below current experimental sensitivity. Contributions to the EDMs coming from new CP violating physics, however, are typically unsuppressed and expected to be within experimental reach.

\begin{figure}[H]
  \centering
\setkeys{Gin}{width=0.4\textwidth}
  \includegraphics{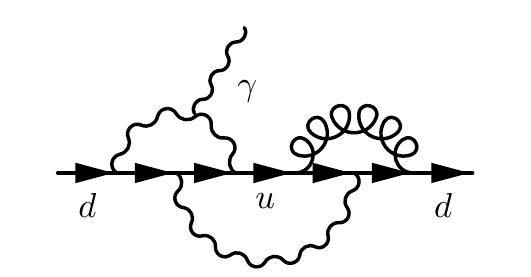}\hspace{0.5cm}%
\setkeys{Gin}{width=0.4\textwidth}
  \includegraphics{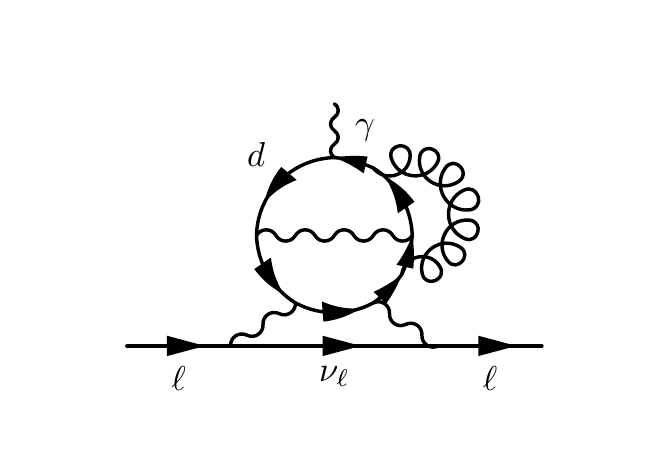}\hspace{0.5cm}%

  \caption{Representative Feynman diagrams for the leading SM contributions to the quark \textit{(left)} and the lepton \textit{(right)} EDMs. For the up-quark EDM the labels $d$ and $u$ have to be exchanged in the left diagram. Unlabeled wiggly lines correspond to $W$ bosons. }
  \label{fig:SM_EDM}
\end{figure}

The experimental sensitivity to EDMs, in particular to those of the electron and neutron, has recently improved by one order of magnitude and is going to further increase in the near future. The current bounds at 90\% C.L. on lepton and neutron EDMs are \cite{Andreev2018,Bennett:2008dy,Grozin:2008nw,PhysRevLett.124.081803}
\begin{equation}
\begin{split}
&|d_e|  < 1.1 \times 10^{-29} ~e\cdot \rm{cm} \, , \\
&|d_\mu| < 1.5 \times 10^{-19} ~e\cdot \rm{cm} \, , \\
&|d_\tau| < 1.6 \times 10^{-18}  ~e\cdot \rm{cm} \, , \\
&|d_n| < 1.8 \times 10^{-26} ~e\cdot \rm{cm} \, ,
\end{split}
\end{equation}
while the prospected bounds on the electron EDM\footnote{Also the bound on the muon EDM might be improved, by three orders of magnitudes, at a future Muon Collider~\cite{Buttazzo:2020eyl}.} at the ACME III experiment and on the neutron EDM at n2EDM are \cite{Doyle:2016,Ayres:2021hoq}
\begin{equation}
\begin{split}
&|d_e|  < 0.3 \times 10^{-30} ~e\cdot \rm{cm} \, ,\\
& |d_n|  <  10^{-27} ~e\cdot \rm{cm} \, .
\end{split}  
\end{equation}
In spite of these incredible sensitivities, the SM values for these observables are many orders of magnitude smaller than the experimental reach. In particular, the electron and neutron EDMs are estimated to be \cite{Pospelov:2013sca,Pospelov:1991zt,Booth:1993af,KHRIPLOVICH1982490,Czarnecki:1997bu} \footnote{\add{Note that the perturbative estimates of the electron EDM could be exceeded by long distance effects by several orders of magnitude \cite{Yamaguchi:2020eub}.}}
\begin{equation}
\begin{split}
&d_e \sim 10^{-48} ~e\cdot \rm{cm} \, ,\\
&d_n \sim 10^{-32} ~e\cdot \rm{cm} \, . 
\end{split}
\end{equation}
How does this surprising suppression arise in the SM? And what effects are expected in a typical Beyond the SM (BSM) scenario? Let us take the electron as an example, whose EDM is \add{estimated} to be
\begin{equation}\label{eldipSM}
d_e \sim e \frac{m_e}{ m_W^2} \frac{g^6 g_s^2}{(16\pi^2)^4} \left( \frac{v}{m_W} \right)^{12}  \, J_{CP}\,.
\end{equation}
\add{This expression makes the small size explicit. Because the EDM violates both chiral symmetry and CP it has to be proportional to the corresponding breaking parameters, which in the SM\footnote{Ignoring non-perturbative effects from the $\theta$ angle of QCD and the PMNS matrix in the neutrino sector.} are the electron mass and the Jarlskog invariant $J_{CP}=\frac{1}{2i}\det\left(\left[y_u y_u^\dagger, y_d y_d^\dagger \right]\right) \sim 10^{-22}$~\cite{Jarlkog85}. The Jarlskog invariant, being antisymmetric in the quark Yukawas, can only be generated at the 4-loop level (c.f. Fig.~\ref{fig:SM_EDM}) in the lepton sector, explaining the additional loop suppression in Eq.~\eqref{eldipSM}}. All in all, this makes a tiny electron EDM in the SM \add{which makes it a promising probes of new physics as the SM background is largely suppressed simplifying the identification of a potential hint of physics beyond the SM}.

Similar considerations apply to the quark EDMs, which feed into the neutron EDM as we show in Sec.~\ref{sec:nEDM}, with the differences that they already arise at \add{the 3-loop level and that} they have a much less severe quark mass suppression~\cite{Smith2017}.

This situation is to be contrasted with what happens in models with new physics, where new sources of CP violation can be present. Taking again the example of the electron EDM, we find in this work contributions like 
\begin{equation}
d_e \simeq -1.1 \times 10^{-29} e \cdot {\rm cm}\, \frac{\Im {C_{\underset{11}{eB}}}}{g'y_e}\left( \frac{1350 \,{\rm TeV}}{\Lambda} \right)^2,
\end{equation}
which does not carry any loop suppression and comes from a tree level Feynman diagram with ${O}_{eB}=\left(\bar{L}_L\sigma^{\mu\nu}e_R\right)HB_{\mu\nu}$ insertion. In the above expression, we divided the Wilson coefficient by its expected size ${g'y_e}$ (more on this in Sec.~\ref{sec:res} and Table~\ref{tab:nat_sizes}), where $g'$ is the $U(1)_Y$ coupling and $y_e$ the electron Yukawa coupling, and $\Lambda$ is the scale of new physics. As the formula shows, if $\Im {C_{\underset{11}{eB}}}\sim g'y_e$ the scale of the CP violating new physics contributing to the EDMs is bounded to be larger than $\sim 10^3$ TeV. One can compare this bound, which is our strongest as we will see, to constraints coming from other CP violating observables, among which some of the most stringent are associated to meson mixings. However, it turns out that the latter~\cite{Bona:2007vi,UTFIT:2016}, are at least one order of magnitude weaker than our bound under similar assumptions (see the last column of Table IV in~\cite{Bona:2007vi}).

The purpose of the present work is to study the lepton and neutron EDMs to 1-loop accuracy in presence of new physics at some scale $\Lambda \gg v$, going to $\mathcal{O}(\Lambda^{-2})$. New physics effects are parametrized in a model independent way within the Standard Model Effective Field Theory (SMEFT), which we expand in the Warsaw basis. We will provide the complete 1-loop expressions of the low energy EDMs observables, for leptons as well as for the neutron, in terms of the Wilson coefficients of the Warsaw basis, including both renormalization group (RG) flow effects and rational terms. In fact, while for extremely large scale separations the logarithmic contributions are expected to be larger than the corresponding finite (rational) terms, for $\Lambda \lesssim 10$ TeV we find them to be comparable. A complete 1-loop result is a step towards a higher accuracy in the theoretical predictions for EDMs observables, which will be measured with increased precision in future experiments. As a matter of fact, having accurate results would turn out crucial in the event of a non-zero measurement of a fermion EDM. 

The constraining power of EDMs has stimulated a lot of different analyses in various UV completions of the SM. There are several studies of the electron and/or neutron EDMs in SUSY models~\cite{Giudice:2005rz,Nakai:2016atk,Cesarotti:2018huy,Aloni:2021wzk}, in Composite Higgs models~\cite{KerenZur:2012fr,Konig:2014iqa,Panico:2015jxa}, in Leptoquark models~\cite{Dorsner:2016wpm,Fuyuto:2018scm,Dekens:2018bci,Altmannshofer:2020ywf}, in complex two-Higgs and three-Higgs doublet models~\cite{Altmannshofer:2020shb,Hou:2021zqq,Logan:2020mdz,Cheung:2020ugr,Chun:2019oix,Davoudiasl:2019lcg,Davoudiasl:2021syn}, in scotogenic models~\cite{Abada:2018zra} and in the context of dark matter~\cite{FileviezPerez:2020oke}. On the model independent side, Ref.~\cite{Panico2018} provides an analysis of the electron EDM including some contributions that arise at 2-loop and at dimension-8 level, while Ref.~\cite{Aebischer:2021uvt} studies the complete 1-loop expression for the lepton EDMs. Ref.~\cite{Haisch2019} studies the neutron EDM in presence of an effective CP violating Higgs-gluon interaction encoded by a dimension-6 SMEFT operator and Ref.~\cite{Haisch:2021hcg} analyzes the contribution to the neutron EDM induced by chromo-dipoles of second and third generation quarks. Other studies of EDMs in presence of dimension-6 interactions involving the Higgs boson and fermion fields -- in particular related to top physics -- are performed in~\cite{Kamenik:2011dk,Brod:2013cka,Brod:2018pli,Fuchs:2020uoc,Fuyuto:2017xup,Cirigliano:2016njn,Cirigliano:2019vfc,Cirigliano:2016nyn,Altmannshofer:2015qra,Chien:2015xha}.

The paper is structured as follows. In Sec.~\ref{sec:hadron} we present the EDM observables and dipole operators, both in the SM and in presence of new physics parametrized by the SMEFT. In Sec.~\ref{highdim} we discuss all the contributions to the dipoles generated by higher dimension-6 SMEFT operators and we furthermore study the neutron EDM in presence of $U(3)^5$ and $U(2)^5$ flavor symmetries for the SMEFT. In Sec.~\ref{sec:calc} we present some important formal and technical aspects of the calculations performed in this work and we finally show the computed bounds in Sec.~\ref{sec:res}.


\section{EDMs}\label{sec:hadron}

\subsection{Electric and magnetic dipole moments of elementary particles}
\label{sec:EDM}

The intrinsic angular momentum of a particle couples to external electric and magnetic fields, with strengths characterized by the electric and magnetic dipole moments respectively. For a spin-1/2 fermion $f$ the non-relativistic Hamiltonian describing these interactions are given by
\begin{equation}
\Ham_{NR}=\frac{a_f\, e Q_f}{2m_f}\,\vec{\sigma}\cdot\vec{B}-d_f\,\vec{\sigma}\cdot\vec{E}\, ,
\label{eq:Dipole_NR}
\end{equation}
where $\vec{\sigma}$ is the vector of Pauli matrices (related to the spin operator $\vec{s}=\vec{\sigma}/2$), $d_f$ and $a_f$ are the electric and magnetic dipole moments of the fermion and $Q_f$ and $m_f$ are its charge and mass. Already from this classical expression we can deduce the transformation properties of the magnetic and electric dipole moments, respectively, under CP: if the theory is invariant under CP, the only term in \Eq{eq:Dipole_NR} which is allowed is the coupling to the magnetic field.
The corresponding relativistic Lagrangian is
\begin{equation}
\Lag=-\frac{a_feQ_f}{4m_f}\bar{\psi}\,\sigma^{\mu\nu}\psi\,F_{\mu\nu}-\frac{i}{2}d_f\bar{\psi}\,\sigma^{\mu\nu}\gamma_5\psi\,F_{\mu\nu}\, .
\label{eq:Dipole_Rel}
\end{equation}
where the second term, barring the factor $i$, changes sign under a CP transformation due to the presence of the $\gamma_5$ matrix.

For the purpose of this work it is more convenient to use chiral fermions and we can rewrite the Lagrangian in \Eq{eq:Dipole_Rel} such that it becomes
\begin{equation}
\Lag=\frac{\cfg}{\La}\,\bar{\psi}_L\,\sigma^{\mu\nu}\psi_R\,F_{\mu\nu}+\hc \, .
\label{eq:LEFT_EDM}
\end{equation}

In the above equation we included explicitly a scale $\La$ for dimensional reasons, such that $\cfg$ is dimensionless. By comparing \Eqs{eq:Dipole_Rel}{eq:LEFT_EDM} we can relate the coefficient $\cfg$ with the dipole moments $a_f$ and $d_f$ and we find
\begin{equation}
a_f = -\frac{4m_f}{e Q_f}\, \frac{\Re{\cfg}}{\La}, \hspace{1cm} d_f = -2\, \frac{\Im{\cfg}}{\La} \, .
\end{equation}

If not further specified, all operators of the form of \Eq{eq:LEFT_EDM}, i.e. also those built from other vector fields, will be called low-energy dipole operators collectively throughout this work. 

\subsection{Dipole operators in the SM and in the SMEFT}
\label{sec:DipoleOp}

Let us now investigate the dipole operators in more detail. \add{As the dipole operator is an irrelevant operator, it is clear that within the SM these operators cannot be generated through RG effects.} Nevertheless, they do acquire finite contributions from loop corrections to the $\bar{\psi}\psi\gamma$ vertex. While the leading contribution to $a_f$ arises already at 1-loop, the EDM $d_f$ receives contributions only starting at three loops, in the quark case, and at four loops, in the electron case, which makes them extremely small as mentioned in \Sec{sec:intro}. 

\add{The situation changes drastically in the presence of heavy new physics where new phases can arise and enter observables through other basis invariants than the Jarlskog invariant. Parameterizing the heavy new physics in the SMEFT, these phases can enter through the Wilson coefficients which allow to build more flavor invariants \cite{Bonnefoy:2021tbt} beyond the Jarlskog invariant. In this paper we study which of the new phases in the SMEFT can enter in the expressions of the lepton and neutron EDMs beyond the log contributions induced by the RG flow which have been studied in the literature.}

Naturally, in the presence of new physics with some heavy new particles the dipole operators can potentially be generated directly together with additional effective operators after integrating out the heavy new physics. In this work we will assume that the scale of new physics lies above the electroweak (EW) scale, which can be quantified using the Higgs vacuum expectation value (VEV) $v$. Then the theory generated upon integrating out the heavy states is the SMEFT\footnote{Here we implicitly assume a linearly realized $SU(3)_c \times SU(2)_L \times U(1)_Y$ gauge symmetry as well as that the physical Higgs is a component of the linearly transforming Higgs doublet. An alternative to this approach is known as the Higgs effective theory (HEFT), where only the $SU(3)_c \times U(1)_{em}$ gauge symmetry is manifest and the physical Higgs is a priori not related to the components of the Goldstone Higgs doublet. For a comparison of the two approaches see for instance \cite{Cohen2020b}.}, which means that the dynamical degrees of freedom are given by the SM field content. \add{In this paper we will work with the \emph{Warsaw} basis \cite{Grzadkowski2010} as the non-redundant operator set, including redundant operators from the so-called \emph{Green's} basis\cite{Jiang:2018pbd,Gherardi:2020det}, whenever necessary at intermediate steps of the computation.}

At low-energies, in particular below the EW scale, the SMEFT can be matched to a low-energy effective theory (LEFT) \cite{Jenkins2017,Dekens2019}, integrating out the top quark, the Higgs boson and the massive EW vectors $Z$ and $W^\pm$; the LEFT, on top of the dipole operators, contains only those built from three gluons or four fermions that are not the top quark. 

Going to higher orders in perturbation theory also other effective operators can contribute to the EDMs through loop effects. So one way to determine the effect on the low-energy observable EDM coming from some new physics, that is matched onto the SMEFT at some scale $\La >  v$, is to calculate the running of the SMEFT dipole operators, to be introduced in the next section, down to the EW scale, to match these operators onto the respective LEFT operators and finally calculate the loop contributions within the LEFT. Partial results of these calculations can be found scattered throughout the literature: the derivation of the renormalization group equation (RGE) within SMEFT has been performed in \cite{Jenkins2013a,Jenkins2013,Alonso2013}, both the tree level matching of the SMEFT to the LEFT as well as the LEFT RGE can be found in \cite{Jenkins2017,Jenkins2017a} and the loop-level matching of the SMEFT to the LEFT has been calculated in \cite{Dekens2019}. Albeit these resources are useful in their own right, they cannot be used to obtain the full 1-loop correction to the EDM.

To perform the full 1-loop calculation we do not choose the multi-stage procedure described above, but instead go directly to the phase of broken EW symmetry, with all the SM fields in the (physical) basis of mass and electric charge eigenstates, and calculate all virtual effects at once, expressing our result in terms of the SMEFT coefficients in the Warsaw basis evaluated at the scale $\La$ above the scale of EW symmetry breaking.

We would like to stress that we are considering, in the SMEFT, tree and 1-loop level contributions to EDMs: in presence of new physics, these CP odd observables can be in general largely enhanced. In fact, allowing for the presence of higher dimensional operators, CP violating effects are no more encoded only by the Jarlskog invariant and a larger variety of complex flavor invariants can be built from Wilson coefficients together with Yukawa matrices \cite{Bonnefoy:2021tbt}. The flavor structure of the effective theory is such that these invariants can indeed be generated at lower loop level with respect to the SM case.

\subsection{Dipole moments of non-elementary particles: neutron EDM}
\label{sec:nEDM}

So far we have considered only fundamental particles within the SMEFT. But, as already mentioned in \Sec{sec:intro}, another prominent observable other than the lepton EDMs is the electric dipole moment of the neutron. Being a composite state built from quarks and gluons we can write the neutron EDM as a function of the constituents' EDMs and chromo-electric dipole moments (cEDMs). The latter are defined as the coefficient of the CP odd operator in \Eq{eq:Dipole_Rel}, but with a gluonic field strength instead of the photonic one. Putting everything together we find \cite{Pospelov:2005pr,Gupta2018,Engel2013,Hisano2012,Haisch2019,deVries:2012ab,Yamanaka:2018uud,Yamanaka:2020kjo,Demir:2002gg,Haisch:2019bml}
\begin{equation}
\label{eq:nEDM}
\begin{split}
d_n=&-(0.204\pm 0.011) ~d_u + (0.784\pm 0.028)~d_d - (0.0027\pm 0.0016) ~d_s +\\
&+ 0.055 (1\pm 0.5) ~\hat{d}_u + 0.111(1\pm 0.5) ~\hat{d}_d-51.2(1\pm 0.5) ~e\cdot \text{MeV} ~\frac{{C}_{\tilde{G}}}{\La^2}+ \\
& -9.22 (1_{-0.67}^{+2.33}) ~e\cdot \text{MeV} ~\frac{\text{Im} [C_{\underset{\, \, \, \, \, \, 11}{Hud}}]}{\La^2}+\\
&-0.615 (1_{-0.75}^{+1}) ~e\cdot \text{GeV} ~\left(\frac{\text{Im} [c_{\underset{1111}{ud}}^{\text{\tiny{(S1,RR)}}}-c_{\underset{1111}{duud}}^{\text{\tiny{(S1,RR)}}}]}{\La^2}+ \frac{\text{Im} [c_{\underset{1111}{ud}}^{\text{\tiny{(S8,RR)}}}-c_{\underset{1111}{duud}}^{\text{\tiny{(S8,RR)}}}]}{\La^2}\right)\, .
\end{split}
\end{equation}
where the "$11$" and "$1111$" subscripts of the Wilson coefficients in the last two lines indicate that the first flavor generation is taken into account.

The first three terms are contributions from the up, down and strange quark EDMs, respectively, the next two terms are the effects of the up and down quark cEDMs and the last term of the second line comes directly from the dimension-6 Weinberg operator \cite{Weinberg1989} built from three gluons,
\begin{equation}
O_{\widetilde{G}}=f^{ABC}G^{A\nu}_\mu G^{B\rho}_\nu \widetilde{G}^{C\mu}_\rho,
\end{equation}
that can be interpreted as the cEDM of the gluon. The contributions in the third and fourth lines are related to the SMEFT operators
\begin{equation}
\begin{split}
O_{Hud} &=   (\bar{u} \gamma_\mu d) (\tilde{H}^\dagger i D^\mu H)\, , \\
O_{quqd}^{\text{\tiny{(1)}}} &=  (\bar{q}^r u) \epsilon_{rs} (\bar{q}^s d) \, , \\
O_{quqd}^{\text{\tiny{(8)}}} &= (\bar{q}^r T^A u) \epsilon_{rs} (\bar{q}^s T^A d) \, .
\end{split}
\end{equation}
Furthermore, $c_{ud}^{\text{\tiny{(S1,RR)}}}$ and $c_{duud}^{\text{\tiny{(S1,RR)}}}$ are the Wilson coefficients of the following operators of the low energy effective field theory \cite{Jenkins2017,Jenkins2017a,Dekens2019}
\begin{equation}
\begin{split}
O_{ud}^{\text{\tiny{(S1,RR)}}} &=  (\bar{u}_L u_R) (\bar{d}_L d_R) \, , \\
O_{duud}^{\text{\tiny{(S1,RR)}}} &= (\bar{d}_L u_R) (\bar{u}_L d_R) \, ,
\end{split}
\end{equation}
which are generated, below the electroweak scale, at tree level by $O_{quqd}^{\text{\tiny{(1)}}}$ and at 1-loop level by $O_{quqd}^{\text{\tiny{(8)}}}$. The tree level matching conditions are the following
\begin{equation}
\begin{split}
O_{\underset{1111}{ud}}^{\text{\tiny{(S1,RR)}}} &=  O_{\underset{1111}{quqd}}^{\text{\tiny{(1)}}} \, , \\
O_{\underset{1111}{duud}}^{\text{\tiny{(S1,RR)}}} &= -O_{\underset{1111}{quqd}}^{\text{\tiny{(1)}}} \, ,
\end{split}
\end{equation}
where the fermion fields are in the mass basis defined in Sec.~\ref{sec:basis}.
Analogously, $c_{ud}^{\text{\tiny{(S8,RR)}}}$ and $c_{duud}^{\text{\tiny{(S8,RR)}}}$ are the Wilson coefficients of
\begin{equation}
\begin{split}
O_{ud}^{\text{\tiny{(S8,RR)}}} &=  (\bar{u}_L T^A u_R) (\bar{d}_L T^A d_R) \, , \\
O_{duud}^{\text{\tiny{(S8,RR)}}} &= (\bar{d}_L T^A u_R) (\bar{u}_L T^A d_R) \, ,
\end{split}
\end{equation}
which are generated at tree level by $O_{quqd}^{\text{\tiny{(8)}}}$ and at 1-loop level by $O_{quqd}^{\text{\tiny{(1)}}}$, with the following tree level matching conditions
\begin{equation}
\begin{split}
O_{\underset{1111}{ud}}^{\text{\tiny{(S8,RR)}}} &=  O_{\underset{1111}{quqd}}^{\text{\tiny{(8)}}} \, , \\
O_{\underset{1111}{duud}}^{\text{\tiny{(S8,RR)}}} &= -O_{\underset{1111}{quqd}}^{\text{\tiny{(8)}}} \, .
\end{split}
\end{equation}

All the terms in the third and fourth lines of \Eq{eq:nEDM}  describe the contributions from CP violating low energy four-fermion interactions. In fact, below the electroweak scale, also $O_{Hud}$ generates four-quark operators through a tree level exchange of a $W$ boson between the right-handed fermion current of the dimension-6 operator and a left-handed current which has a SM coupling with the $W$.
All the coefficients appearing in the above expression should be evaluated at the hadronic scale \add{$\mu_H$} that characterizes the neutron EDM. To be more rigorous, in the case of $C_{Hud}$ what is evaluated at \add{energy scales below the EW scale} are the coefficients of the four-quark operators generated after \add{integrating out the heavy SM particles from the SMEFT}, which is to say $(\bar{u}_L\gamma_\mu d_L) (\bar{d}_R \gamma^\mu u_R)$ at tree level and $(\bar{u}_L\gamma_\mu T^A d_L) (\bar{d}_R \gamma^\mu T^A u_R)$ at 1-loop level. Note that while $C_{\widetilde{G}}$, $C_{Hud}$, $c_{ud}^{\text{\tiny{(S1(8),RR)}}}$ and $c_{duud}^{\text{\tiny{(S1(8),RR)}}}$ are dimensionless, the fermionic dipole coefficients have the dimension of an inverse energy ($d_i,\hat{d}_i\sim v/\Lambda^2$). 

As we have just discussed, the neutron EDM does not only receive contributions from the EDMs and cEDMs of the quarks. This allows operators to be probed, that would otherwise only be available at higher loop orders if at all. One example would be the Yukawa type operators $\psi^2H^3$. At the 1-loop level, they cannot be accessed by EDMs of elementary particles, as they contribute only starting at the 2-loop order. However, as they give 1-loop contributions to $O_{Hud}$, which enters the neutron EDM also at tree level, one can probe them at a lower order as naively expected.

In the expression of the neutron EDM we implicitly assumed a Peccei-Quinn mechanism \cite{PhysRevD.16.1791} to remove the contribution from the well-known QCD $\theta-$term 
\begin{equation}
\Lag_\theta \sim \bar{\theta} ~ \Tr\left[G^{\mu\nu}\widetilde{G}_{\mu\nu}\right],
\label{eq:QCD_theta}
\end{equation}
which otherwise would give the dominant effect on the neutron EDM. Here $\bar{\theta}$ is a linear combination of a bare $\theta$ parameter and the argument of the determinant of the quark Yukawa couplings \cite{Hook:2018dlk}. On top of introducing the usual term that removes the contribution of the QCD $\theta$-term, the Peccei-Quinn mechanism induces a shift on the axion potential due to the presence of the chromo dipole operators. In return, this shift modifies the coefficients for the light quark cEDMs and completely cancels the effect of the strange quark cEDM~\cite{Pospelov1999,Pospelov2000,Engel2013}.

At this point we want to stress that the results presented in this paper can in principle be used for any function of the neutron EDM in terms of quark (c)EDMs, which might differ from \eqref{eq:nEDM}.


\section{Higher dimensional operators}
\label{highdim}

\subsection{1-loop effects}

Within the SMEFT framework, 
in the phase of unbroken EW symmetry, 
the relevant dipole operators are the ones containing the hypercharge and weak gauge bosons $B$ and $W^I$, respectively. To ensure gauge invariance, these operators have to contain an additional Higgs doublet compared to the expression in \Eq{eq:Dipole_Rel} to compensate for the transformation of the left-handed fermion doublet. They have the form

\begin{equation}
O_{fB} = \left(\bar{\psi}_L\sigma^{\mu\nu}\psi_R\right)\overset{(\sim)}{H}B_{\mu\nu} \hspace{0.4cm}\text{and}\hspace{0.4cm} O_{fW} = \left(\bar{\psi}_L\sigma^{\mu\nu}\sigma^{I}\psi_R\right)\overset{(\sim)}{H}W^I_{\mu\nu}.
\end{equation}
In these equations, $\psi_{L(R)}$ describes a left(right)-handed $SU(2)_L$ doublet (singlet) and the (conjugate) Higgs doublet has to be used if the fermion species in question sits in the (upper) lower component of the doublet. After EW symmetry breaking and the transition from the gauge basis to the mass basis, we see that the Wilson coefficients of the EW SMEFT dipoles are related to the photonic one defined in \Eq{eq:Dipole_Rel} via the relation 

\begin{equation}
\cfg=\frac{v}{\Lambda}\left(c_w ~C_{fB} + 2T^3_f s_w ~C_{fW}\right),
\end{equation}
where we defined the trigonometric function of the weak mixing angle $c_w\equiv\cos\theta_w$ and $s_w\equiv\sin\theta_w$ and $T^3_f$ is the third component of the weak isospin for the respective fermion and is non-zero only for left-handed chiralities. 

Of course, for the neutron EDM also the gluonic dipole operators are relevant, as is obvious from \Eq{eq:nEDM}. But since these are not affected by the EW symmetry breaking, except for effectively setting $H \to v$ and picking the relevant component from the $SU(2)_L$ quark doublet, we can see immediately that 
\begin{equation}
c_{qg}=\frac{v}{\Lambda}C_{qG} \quad \text{and} \quad \hat{d_q}=-2\frac{v}{\Lambda^2}\Im{C_{qG}}.
\end{equation}

\begin{figure}[t]
\centering
\includegraphics{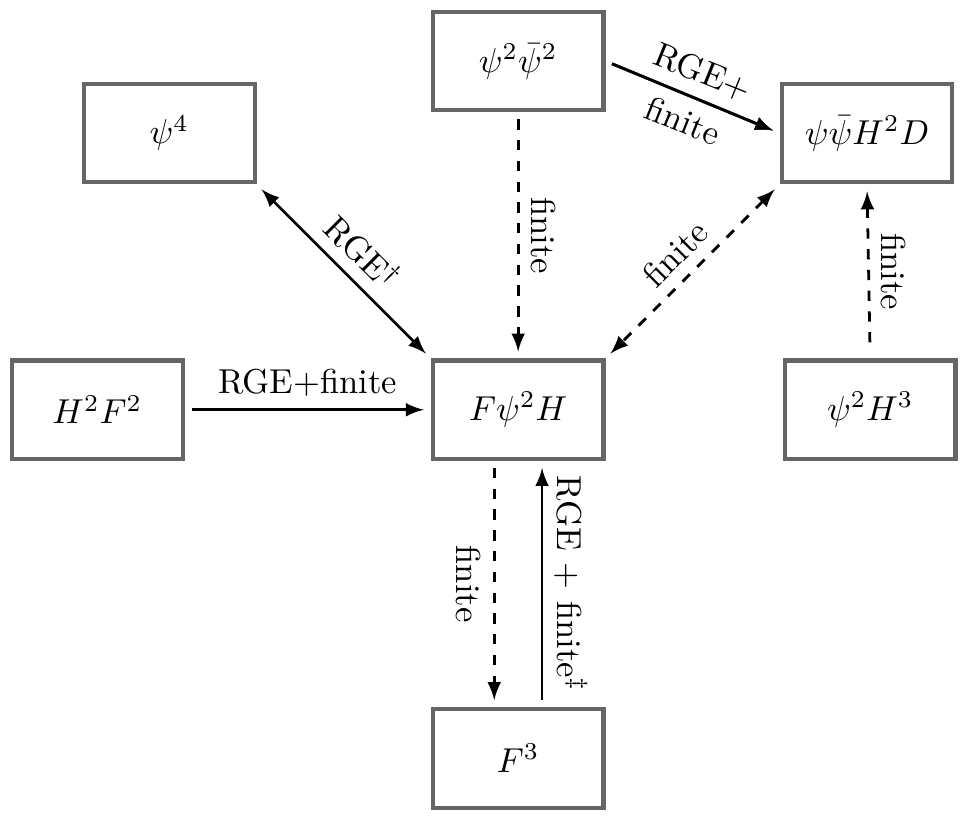}
\caption{Contributions to the operators entering the dipoles. Operators connected with solid arrows enter the RGEs, while dashed arrows describe purely finite effects. The $\dagger$ indicates that the operator $O_{lequ}^{\text{\tiny{(1)}}}$ is not included in the $\psi^4$ class here. Interestingly, we find that the others operators in this class only enter via the RGE, generating no rational terms. The $\ddagger$ shows that the operator $O_{\widetilde{W}}$ gives only rational terms.}
\label{fig:ContributionPattern}
\end{figure}

\add{At the loop level various other higher-dimensional operators can enter the EDMs, either through RG mixing or through finite effects. Even though it is easy to see that many operators cannot contribute to the EDMs there exist a few powerful criteria the operators have to satisfy to be able to enter the expressions of the EDM.}
We summarize all the relevant effective operators in Table~\ref{tab:operatorsSMEFT}, show all the different kinds of contributions in Fig.~\ref{fig:ContributionPattern} and explain the selection rules in the following\add{, using contributions to the dipole operators as an example. Since there are no conceptual differences regarding the selection rules for the other operators, we do not go through them here explicitly.} The full analysis is performed in the SMEFT at the dimension-6 level. We do not consider $O(1/\Lambda^4)$ corrections to the EDMs in this work and refer to Ref.~\cite{Panico2018} for a related discussion, where a partial study of the contributions to the electron EDM from dimension-8 operators is presented.

\renewcommand{\arraystretch}{1.35}
\begin{table}[t]
  \begin{center}{}
  \begin{tabular}{|c|}
\hline
$O_{\underset{ab}{eB}} = \left(\bar{L}^a_L\sigma^{\mu\nu}e_R^b\right)HB_{\mu\nu}$\\
$O_{\underset{ab}{eW}} = \left(\bar{L}^a_L\sigma^{\mu\nu}\sigma^Ie_R^b\right)HW^I_{\mu\nu},$ \\
$O_{\underset{ab}{uB}} = \left(\bar{Q}^a_L\sigma^{\mu\nu}u_R^b\right)\widetilde{H}B_{\mu\nu}$\\
$O_{\underset{ab}{uW}} = \left(\bar{Q}^a_L\sigma^{\mu\nu}\sigma^Iu_R^b\right)\widetilde{H}W^I_{\mu\nu},$ \\
$O_{\underset{ab}{dB}} = \left(\bar{Q}^a_L\sigma^{\mu\nu}d_R^b\right)HB_{\mu\nu}$\\
$O_{\underset{ab}{dW}} = \left(\bar{Q}^a_L\sigma^{\mu\nu}\sigma^Id_R^b\right)HW^I_{\mu\nu}$\\

$O_{\underset{ab}{uG}} = \left(\bar{Q}^a_L\sigma^{\mu\nu}T^Au_R^b\right)\widetilde{H}G^A_{\mu\nu}$\\
$O_{\underset{ab}{dG}} = \left(\bar{Q}^a_L\sigma^{\mu\nu}T^Ad_R^b\right)HG^A_{\mu\nu},$ \\

\hline
  \end{tabular}
  \hspace{5mm}
  \begin{tabular}{|c|}
\hline
$O_{\underset{abcd}{lequ}}^{\text{\tiny (3)}} = \big(\bar L^{ja}_L\, \sigma_{\mu\nu} e_R^b \big)\epsilon_{jk}\big(\bar Q^{kc}_L\, \sigma_{\mu\nu} u_R^d \big)$ \\
$O_{\underset{abcd}{quqd}}^{\text{\tiny (1)}} = \big(\bar Q^{ja}_L\, u_R^b \big)\epsilon_{jk}\big(\bar Q^{kc}_L\, d_R^d \big)$ \\
$O_{\underset{abcd}{quqd}}^{\text{\tiny (8)}} = \big(\bar Q^{ja}_L\, T^A u_R^b \big)\epsilon_{jk}\big(\bar Q^{kc}_L\, T^A d_R^d \big)$ \\[2ex]
\hdashline
$O_{\underset{abcd}{le}} = \big(\bar L^a_L \gamma_\mu L^b_L \big)\big(\bar e^c_R \gamma_\mu e^d_R \big)$ \\
$O_{\underset{abcd}{qu}}^{\text{\tiny (1)}} = \big(\bar Q^a_L \gamma_\mu Q^b_L \big)\big(\bar u^c_R \gamma_\mu u^d_R \big)$ \\
$O_{\underset{abcd}{qu}}^{\text{\tiny (8)}} = \big(\bar Q^a_L \gamma_\mu T^A Q^b_L \big)\big(\bar u^c_R \gamma_\mu T^A u^d_R \big)$ \\
$O_{\underset{abcd}{qd}}^{\text{\tiny (1)}} = \big(\bar Q^a_L \gamma_\mu Q^b_L \big)\big(\bar d^c_R \gamma_\mu d^d_R \big)$ \\
$O_{\underset{abcd}{qd}}^{\text{\tiny (8)}} = \big(\bar Q^a_L \gamma_\mu T^A Q^b_L \big)\big(\bar d^c_R \gamma_\mu T^A d^d_R \big)$ \\
$O_{\underset{abcd}{ud}}^{\text{\tiny (1)}} = \big(\bar u^a_R \gamma_\mu u^b_R \big)\big(\bar d^c_R \gamma_\mu d^d_R \big)$ \\
$O_{\underset{abcd}{ud}}^{\text{\tiny (8)}} = \big(\bar u^a_R \gamma_\mu T^A u^b_R \big)\big(\bar d^c_R \gamma_\mu T^A d^d_R \big)$ \\[2ex]

\hline
  \end{tabular}
  \end{center}
  \vspace{2mm}
  \begin{center}
  \begin{tabular}{|c|}
\hline
$O_{\widetilde{W}} = \epsilon^{IJK}\widetilde{W}^{I\nu}_\mu W^{J\rho}_\nu W^{K\mu}_\rho$ \\
$O_{\widetilde{G}} = f^{ABC}\widetilde{G}^{A\nu}_\mu G^{B\rho}_\nu G^{C\mu}_\rho$ \\
\hline  
  \end{tabular} 
    \hspace{5mm}
  \begin{tabular}{|c|}
\hline
$O_{H\widetilde{B}} = H^\dagger H B^{\mu \nu}\widetilde B_{\mu \nu}$ \\
$O_{\widetilde{W}} = H^\dagger H W^{I \mu \nu}\widetilde W^I_{\mu \nu}$ \\
$O_{HW\widetilde{B}} = (H^\dagger \sigma^I H)W^{I \mu \nu}\widetilde B_{\mu \nu}$ \\

$O_{H\widetilde{G}} = H^\dagger H G^{A\mu \nu}\widetilde G^A_{\mu \nu}$ \\
\hline
  \end{tabular}
  \end{center}
  \vspace{2mm}
  \begin{center} 
  \hspace{5mm}
  \begin{tabular}{|c|}
\hline
$O_{\underset{ab}{Hud}} = i \left(\tilde{H}^\dagger D_\mu\, H\right)\left(\bar{u}^a_R\gamma^\mu d^b_R\right)$ \\
\hline  
  \end{tabular} 
  \hspace{5mm}
  \begin{tabular}{|c|}
\hline
$O_{\underset{ab}{dH}} = H^\dagger H\,\left(\bar{Q}^a_L\,d_R^b \,H\right)$ \\
$O_{\underset{ab}{uH}} = H^\dagger H\,\left(\bar{Q}^a_L\,u_R^b \,\tilde{H}\right)$ \\
\hline  
  \end{tabular}
  \end{center}
  
  \caption{Set of dimension-6 SMEFT operators relevant in this paper, grouped in six different boxes corresponding to the different classes discussed in the main text. {The operators $O_{ud}^{(1,8)}$ as well as the $\psi^2 H^3$ type operators can only be probed at the 1-loop level through the neutron EDM.} The dashed line separates the 4-fermion operators of the form $\psi^4$ and those of the form $\psi^2\bar{\psi}^2$. We use the usual definitions $\widetilde{H}=i\sigma^2H^\ast$ and $\widetilde{F}_{\mu\nu}=\frac{1}{2}\epsilon_{\mu\nu\alpha\beta}F^{\alpha\beta}$ for $F$ any of the gauge bosons. For the operators $O_{lequ}^{\text{\tiny (3)}}$ and $O_{quqd}^{\text{\tiny (1,8)}}$ we show SU(2) indices $j,k$ explictly. For the vector operators in the 4-fermion class the only CP violation can arise if flavors of the fermions in each current are not identical, hence we explicitly give the generation indices $a,b,c,d$ explicitly.} 
  \label{tab:operatorsSMEFT}
\end{table}
\renewcommand{\arraystretch}{1}

\paragraph{1-loop contributions to the dipole operators: CP violation\\}

The first criterion might also be the most trivial. Since we are only interested in the EDM, we need to consider only those operators that can give contributions to the imaginary part of the dipole coefficient. This significantly reduces the number of  relevant operators.

\paragraph{1-loop contributions to the dipole operators: helicity selection rules\\}

In general we can use helicity \cite{Cheung2015,Azatov2016,Craig2019} and angular momentum \cite{Jiang2020} arguments to derive selections rules, essentially making it possible to pinpoint only the relevant operators. Using the former, one finds that an operator generating a contact interaction with $n$ external legs and total helicity $\sum h$ can only be renormalized by another operator with $(n',\sum h')$ if the relations
\begin{equation}
n'\leq n \hspace{0.4cm}\text{and}\hspace{0.4cm} \left|\sum h - \sum h'\right| \leq n - n'
\label{eq:HelSelRules}
\end{equation}
hold \cite{Cheung2015,Craig2019}. The dipole operators are of the form $F\psi^2H$, where $F$ and $\psi$ are positive helicity field strength tensor and fermion respectively, so we can characterize them by $(n, \sum h)=(4,2)$. Using the above relations we see that the only operators able to renormalize the dipoles within the Warsaw basis of the SMEFT are:

\begin{itemize}
\item Operators with $(n, \sum h)=(3,3)$, i.e. operators of the class $F^3$;
\item Operators with $(n, \sum h)=(4,2)$. This includes operators of the form $F^2H^2$, $\psi^4$ and of course the dipole operators $F\psi^2H$ themselves.
\end{itemize}

Although there is an exception to \Eq{eq:HelSelRules}, we can show that it does not change the set of renormalizing operators given above. It is related to the existence of the so-called exceptional, four-dimensional $\psi^4$ amplitude with $(n, \sum h)=(4,2)$.\footnote{This amplitude is proportional to the product of up and down quark Yukawa couplings and is the only SM 4-point amplitude having total helicity different from zero} \cite{Cheung2015,Azatov2016,Craig2019}. It can be shown that an insertion of this exceptional amplitude could potentially lead to the renormalization of the dipoles from higher-dimensional operators with $(n,\sum h) = (4,0)$. Operators with this number of legs and total helicity in the Warsaw basis are of the form $\psi^2\bar{\psi}^2$, $\psi\bar{\psi}H^2D$ and $H^4D^2$. Hence we see that we can not build a loop amplitude with the particle content of the dipole in the external states by combining these higher-dimensional contact amplitudes with the four-dimensional exceptional amplitude.

While these helicity selection rules provide a helpful tool when aiming to calculate the RGEs of various operators there, they have one major shortcoming if one is interested in a full 1-loop calculation. This is related to the fact that helicity arguments deal only with the renormalization of operators and are not able to tell if there are operators that contribute only through rational terms\footnote{While there are no helicity selection rules for rational terms, they can still be calculated using helicity amplitudes. However, this would require to perform all possible multi-particle cuts either in D dimensions, see e.g. \cite{Anastasiou2006}, or using massive loop propagators \cite{Badger2008,ArkaniHamed2017}}. 

Interestingly, although the operator $O_{\widetilde{W}}$ belongs to the $F^3$ class, hence could renormalize the dipole operators, it instead gives only a finite, rational contribution. This was computed using both Feynman diagrams \cite{Boudjema1991,Gripaios2013,Panico2018} as well as on-shell methods \cite{Baratella2020}. Its gluonic counterpart on the other hand does also enter the dipole operator RGE.

\paragraph{1-loop contributions to the dipole operators: angular momentum selection rules\\}

We can alleviate the problem of rational terms by augmenting the helicity selection rules with angular momentum considerations \cite{Jiang2020}. So far these were used as an alternate way to derive the pattern of renormalization among operators using the conservation of angular momentum of external states, instead of employing the cut-based factorization of loop amplitudes that is used to arrive at the above helicity selection rules. \add{As the name suggests, at the core of this approach lies the conservation of angular momentum during a scattering process. For every scatterig amplitude we find (at least) one scattering channel where the total angular momentum $j$ of the scattering particles is conserved. Given this, operators can renormalize one another if they share at least one such channel.} Note that this approach is complementary to the helicity selection rules in the sense that operators allowed by the former can be forbidden if we additionally use the latter and vice versa. \add{To highlight a few features of this method, we again use the dipole operators as an example, for which we find the following scattering channels}
\begin{itemize}
\item the $VH\to\psi\psi$ channel, with $j=1$;
\item and the $V\psi\to H\psi$ channel, with $j=1/2$.
\end{itemize}
\add{Considering now, e.g. the renormalization of the leptonic from the $\psi^4$ class operators, helicity selection rules tell us that the only viable ones are \begin{equation}
O_{lequ}^{\text{\tiny (1)}} = \big(\bar L^j_L e_R \big)\epsilon_{jk}\big(\bar Q^k_L u_R \big)
\end{equation}
and its tensorial cousin $O_{lequ}^{\text{\tiny (3)}}$ defined in \Tab{tab:operatorsSMEFT}. It turns out that the former has $j=0$ in the fermion channel, while the latter has $j=1$, telling us that the scalar operator can in fact not renormalize the dipole, something oblivious to the helicity rules.}

Because the angular momentum argument does not rely on performing cuts in the loop integral but only on the angular momentum of external states it should be possible to extend the procedure to rational terms. While we are not aware of a rigorous proof for this and leave any detailed investigation for later work, we checked a few cases and the procedure worked for all of them. \add{One example would be the $\psi\bar{\psi}H^2D^2$ class of operators, whose renormalization to the dipoles is forbidden by helicity selection rules. Notice, that the only operator in this class that can give CP odd contributions is $O_{Hud}$, which contributes only to the chromo-dipoles. Looking at the angular momentum structure we see that $O_{Hud}$ shares the $\psi H$ channel with the dipoles\cite{Jiang2020}. So even though helicity selection rules forbid renormalization angular momentum conservation allows for rational contributions, which we indeed find.}

There is, however, a caveat that might be worth investigating in a future work and is related to the existence of so-called evanescent operators. These are operators that generate non-vanishing amplitudes in $d\neq4$ space-time dimensions that then vanish in the limit $d\to4$ and often arise in the context of 4-fermion operators and Fierz identities that change for $d\neq4$. In particular, let us look at the operator $O_{le}$ in the Warsaw basis and its counterparts with quarks defined in \Tab{tab:operatorsSMEFT}; the connection of evanescent operators to the dipole through this particular operator was already mentioned in \cite{Panico2018}. It lives in the operator class $\psi^2\bar{\psi}^2$, having $j=0$ in the $\psi\psi$ channel, so by angular momentum conservation it neither can renormalize the dipole nor give only rational contributions. On the other hand, it is straightforward to compute the loop diagram with a single insertion of this operator and see that it, against all odds, does in fact give a rational contribution. This apparent contradiction with angular momentum conservation can be resolved by realizing that we can apply a Fierz identity to rewrite this operator as
\begin{equation}
O_{le}=\big(\bar L_L \gamma_\mu L_L \big)\big(\bar e_R \gamma_\mu e_R \big)\propto\big(\bar{L}_Le_R\big)\big(\bar{e}_RL_L\big).
\end{equation}
Again, we can calculate the corresponding diagram with an insertion of this operator after the Fierzing and we indeed find a vanishing result, in accordance with angular momentum conservation. At this point, we have to stress that the above Fierzing does only hold in $d=4$, in a general number of space-time dimensions the identity reads \cite{Dekens2019}
\begin{equation}
O_{le}=\big(\bar L_L \gamma_\mu L_L \big)\big(\bar e_R \gamma_\mu e_R \big) = 2\big(\bar{L}_Le_R\big)\big(\bar{e}_RL_L\big) + E_{LR}^{\text{\tiny{(2)}}},
\end{equation}
where $E_{LR}^{\text{\tiny{(2)}}}$ is an evanescent operator that vanishes in 4 dimensions. This additional operator then gives a rational term when inserted into the loop integral.

In this paper we use the Warsaw basis without any Fierzing, so the contribution from this kind of operators appears explicitly in the final result, however, keeping in mind that it is related to the presence of an evanescent operator.

\subsection{2-loop effects}
Although the main focus of this paper lies on the full 1-loop calculation we want to briefly discuss possible higher order effects. Formally, these are suppressed by additional loop factors and more powers of some coupling, so naively these higher order processes are always suppressed compared to the leading order term. In a realistic theory like the SM and extensions thereof with many different couplings and likely a large hierarchy among them, this does not hold in general. A prominent example is the Barr-Zee diagram for fermion EDMs in the presence of additional Higgs doublets \cite{Barr1990}. Here, the leading order diagram is suppressed by two powers of the fermion Yukawa, $y_f^2$, due to two couplings of a Higgs to the fermion line. However, by going to the next order in perturbation theory one of these Yukawas can be traded for a factor of $\frac{g^2}{16\pi^2}$, by adding e.g. an additional top quark loop. 
In this case, for light external fermions, this factor is still larger than a potentially tiny Yukawa, hence the formally subleading 2-loop effects can actually dominate over the leading order contribution. 


\begin{figure}[H]
  \centering
\setkeys{Gin}{width=0.35\linewidth}
  \includegraphics{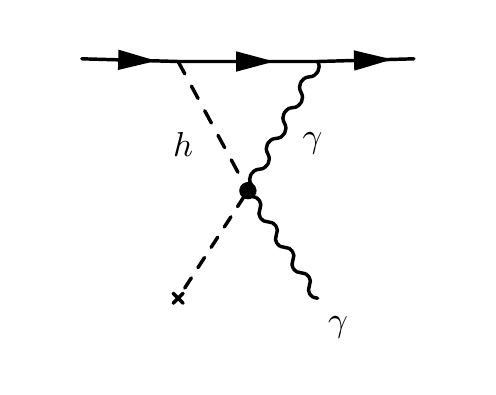}\hspace{0.5cm}%
  \caption{Representative diagram giving a contribution of the $H^2F^2$ class of operators to the dipole operators.}
  \label{fig:H2F2}
\end{figure}

\add{In a similar spirit, we might expect higher loop contributions to dominate over the formally leading ones by the virtue of larger couplings. In particular we consider the $H^2F^2$ operators, specifically their contribution to the dipole operators. A representative diagram for these is shown in \Fig{fig:H2F2} and all other diagrams generated by this class of operators can be found in \App{app:diags}. We see that the mass dependence is generated directly by the coupling of the Higgs to the fermion line. Contrary to the Barr-Zee diagrams, it turns out that there is no 2-loop diagram generated by this class of operators where this Higgs-fermion coupling is not present and still a contribution to the EDM is generated; so the 1-loop contribution is indeed the leading one and in comparison all 2-loop effects remain suppressed. Note that the same happens for the $F^3$ class operators, where the proportionality to the fermion mass or equivalently the Yukawa is more evident in the phase of unbroken symmetry. Further note that this dependence on the external fermion mass could have already been anticipated using flavor symmetries.}

\add{Having said this, in the case of the nEDMs the $H^2G^2$ operator can indeed enter through 2-loop diagrams without a suppression from small Yukawas\footnote{We kindly thank the referee for pointing this out.}. It has been shown that the $H^2G^2$ mixes with the Weinber operator at 2-loops and $\mathcal{O}(\alpha^2)$ \cite{Haisch2019}. This mixing can than feed into the nEDM through the tree-level contribution of the Weinberg operator, see equation \eqref{eq:nEDM}. Since the lepton EDMs are given purely by the dipole operator coefficients, we do not expect similar effects to appear in those cases.}

\subsection{Transition from the gauge to the mass basis}
\label{sec:basis}

As already explained, all calculations in this paper were performed directly in the phase of spontaneously broken electroweak symmetry and it is convenient to go from the gauge to the mass basis for fermion\add{, which is necessary in order to deal with the propagating degrees of freedom. We want to briefly discuss how this basis change affects the Wilson coefficients.}

We diagonalize the fermion mass matrices by rotating each chiral fermion, which are triplets in generation space, using unitary transformations in flavor space,
\begin{equation} 
\psi'\,^i_{L/R} = U^{i}_{L/R}\, \psi^i_{L/R}
\end{equation}
with (un)primed fields in the (mass) gauge basis and $i$ denotes any of the fermion flavors. \add{Note, at the order we are working at, it is sufficient to take the SM part from the matrices $U^{i}_{L/R}$, ignoring their dimension-6 pieces.} \add{These rotations can be absorbed by redefining the Wilson coefficients, as shown in Table~\ref{tab:Redefs}.} \add{Because the components of the electroweak quark doublets need to be transformed differently in order to diagonalize all Yukawa matrices, there is no way to redefine the Wilson coefficients such that all rotation matrices are absorbed. This is because the Wilson coefficients are defined in the unbroken phase, where a $U(3)_Q$ flavor transformation acts on the full $SU(2)_L$ doublet.} Possible choices, for the gauge basis in which the SMEFT Wilson coefficients are defined in the unbroken phase, are the absorption either of the up-type or of the down-type rotation: we denote them in the following as up- and down-quark bases, in which the up- and down-quark Yukawa matrices are diagonal respectively. Then, after EW spontaneous symmetry breaking and full rotation to the mass basis, in the quark sector this generates the CKM matrix, defined as
\begin{equation} 
V\equiv V_{CKM}=\left(U^{u}_L\right)^\dagger U^d_L,
\end{equation}
and the precise terms where it appears are given by the choice of the definition of the Wilson coefficients in the gauge basis\footnote{If we would relax our assumption of massless neutrinos the PMNS matrix would be generated accordingly in the lepton sector.}. 
We choose, in our work, to present the final expressions for the EDMs in terms of the Wilson coefficients in the mass basis, defined in Table~\ref{tab:Redefs}, in such a way that the least amount of CKM matrices appear explicitly. Furthermore, also the bounds are set here on the mass basis Wilson coefficients, even if in presenting these constraints in Table~\ref{tab:nEDMBounds4FWC} and~\ref{tab:nEDMBounds4FLam}, the $C'$ coefficients in the gauge basis are shown explicitly, choosing the up-quark basis and consequently $U^{d}_L = V$ and $U^{u}_{L/R}=\mathbf{1}$.

\renewcommand{\arraystretch}{1.35}
\begin{table}[t]
	\begin{center}
	\begin{tabular}{|c|}
\hline
$C_{dW}=\left(U^{d}_L\right)^\dagger C'_{dW} U^{d}_R$   \\[2ex]
$C_{dB}=\left(U^{d}_L\right)^\dagger C'_{dW} U^{d}_R$   \\[2ex]
$C_{dG}=\left(U^{d}_L\right)^\dagger C'_{dW} U^{d}_R$   \\[2ex]
$C_{dH}=\left(U^{d}_L\right)^\dagger C'_{dH} U^{d}_R$   \\[2ex]
$C_{uW}=\left(U^{u}_L\right)^\dagger C'_{uW} U^{u}_R$   \\[2ex]
$C_{uB}=\left(U^{u}_L\right)^\dagger C'_{uW} U^{u}_R$   \\[2ex]
$C_{uG}=\left(U^{u}_L\right)^\dagger C'_{uW} U^{u}_R$   \\[2ex]
$C_{uH}=\left(U^{u}_L\right)^\dagger C'_{uH} U^{u}_R$   \\[2ex]
$C_{Hud}=\left(U^{u}_R\right)^\dagger C'_{Hud} U^{d}_R$   \\

\hline
	\end{tabular}
	\hspace{10mm}
	\begin{tabular}{|c|}
\hline
$C_{\underset{abcd}{lequ}}^{\text{\tiny (3)}} = \delta_{ia}\delta_{jb}\left(U^{u}_L\right)^\dagger_{ck} \left(U^{u}_R\right)_{ld}C{'}_{\underset{ijkl}{lequ}}^{\text{\tiny (3)}}$ \\[2ex]
$C_{\underset{abcd}{quqd}}^{\text{\tiny (1,8)}} = \left(U^{d}_L\right)^\dagger_{ai} \left(U^{u}_R\right)_{jb}\left(U^{u}_L\right)^\dagger_{ck}\left(U^{d}_R\right)_{ld}C{'}_{\underset{ijkl}{quqd}}^{\text{\tiny (1,8)}}$ \\[2ex]
$C_{\underset{abcd}{qu}}^{\text{\tiny (1,8)}} = \left(U^{u}_L\right)^\dagger_{ai} \left(U^{u}_L\right)_{jb}\left(U^{u}_R\right)^\dagger_{ck}\left(U^{u}_R\right)_{ld}C{'}_{\underset{ijkl}{qu}}^{\text{\tiny (1,8)}}$ \\[2ex]
$C_{\underset{abcd}{qd}}^{\text{\tiny (1,8)}} = \left(U^{d}_L\right)^\dagger_{ai} \left(U^{d}_L\right)_{jb}\left(U^{d}_R\right)^\dagger_{ck}\left(U^{d}_R\right)_{ld}C{'}_{\underset{ijkl}{qd}}^{\text{\tiny (1,8)}}$ \\[2ex]
$C_{\underset{abcd}{ud}}^{\text{\tiny (1,8)}} = \left(U^{u}_R\right)^\dagger_{ai} \left(U^{u}_R\right)_{jb}\left(U^{d}_R\right)^\dagger_{ck}\left(U^{d}_R\right)_{ld}C{'}_{\underset{ijkl}{ud}}^{\text{\tiny (1,8)}}$ \\

\hline
	\end{tabular}
	\end{center}
	
	\caption{Definitions of Wilson coefficients of fermionic operators used in this work. We suppress flavor indices \add{whenever} their contraction is non-ambiguous. (Un)primed coefficients denote the ones in the (mass) gauge basis. The specific form of the $U$ unitary matrices, needed to the transformation to the mass basis, depends on the specific choice for the gauge basis in which the $C'$ coefficients are defined: for example, $U^{u}_{L/R}=1, \, U^{d}_L = V$ ($U^{d}_{L/R}=1, \, U^{u}_L = V^\dagger$) in the up (down) - quark basis. Here we already assumed a diagonal lepton Yukawa, hence $U^{e}_{L/R}=1$ and $C_{eV}=C{'}_{eV}$.} 
	\label{tab:Redefs}
\end{table}
\renewcommand{\arraystretch}{1}


\subsection{Neutron EDM Bounds under the Light of Flavor Symmetries}\label{sec:neutronflavor}

Since a lot of different flavor components of the fermionic Wilson coefficients enter through the various quark EDMs in the expression of the neutron EDM in Eq.~\eqref{eq:nEDM}, in particular in the gauge basis, due to the misalignement with the mass basis in the quark sector, as shown in the previous section, it is interesting to study the effects of flavor symmetries that relate the components of the flavor tensors.
\add{In the absence of Yukawa interactions, the SM, with massless neutrinos, is invariant under the global flavor symmetry \cite{SEKHARCHIVUKULA198799}}
\begin{equation}
G_F=U(3)^5=U(3)_Q\times U(3)_u\times U(3)_d\times U(3)_L\times U(3)_e,
\label{eq:GF}
\end{equation}
\add{where $i=Q,u,d,L,e$ stands for the left-handed quarks, right-handed up and down quarks and left- and right-handed leptons. The only breaking of this symmetry is due to the SM Yukawa couplings. One simple assumption on the flavour structure of the UV is to assume that the breaking of flavour universality in the BSM sector is also only due to the SM Yukawas. This goes under the name of Minimal Flavour Violation (MFV) \cite{DAmbrosio2002,Isidori2012,Smith2017}.}

By enforcing this symmetry on the SMEFT and promoting the Wilson coefficients to spurions, we can completely determine the flavor tensors of the fermionic Wilson coefficients in terms of Standard Model parameters and a reduced number of flavor indipendent coefficients, one for each term in the spurionic Yukawa expansion of the SMEFT operators \cite{SMEFTFlavorSym}. Notice, that this expansion is performed for Wilson coefficients in the gauge basis. By only including terms up to $\mathcal{O}(y_{u,d,e}^2)$, we can reduce the Wilson coefficient of most fermionic operators to one complex coefficient while the operators $O_{qu}^{\tiny{(1,8)}},O_{qd}^{\tiny{(1,8)}}$, combining two chirality-conserving currents, are forbidden at the considered order. E.g. for the dipole operator we find
\begin{equation}
C^\prime_{\substack{uB\\pr}}O^\prime_{\substack{uB\\pr}} = C^\prime_{\substack{uB\\pr}} \left( \bar{Q}^\prime_p \sigma^{\mu\nu} u^\prime_r \right) \widetilde{H} B_{\mu\nu} \longrightarrow F_{uB} \left( \bar{Q}^\prime_p \sigma^{\mu\nu} u^\prime_r \right) \widetilde{H} B_{\mu\nu} \left( \left( y_u \right)_{pr} + \mathcal{O}( y_{u,d,e}^3) \right) \, ,
\end{equation}
\add{where we use the notation introduced in the last section to denote the choice of basis.}
\add{Since the top Yukawa is $\mathcal{O}(1)$ it is technically not a soft breaking of $G_F$, so it would be more accurate to not consider it as part of the symmetry. Further, there are many BSM models where the third generation plays a special role motivating the complete removal of the third generation from the flavor group \cite{SMEFTFlavorSym2,FlavorU2}}
The simplest smaller allowed symmetry group is then 
\begin{equation}
H_F = U(2)^5 = U(2)_Q \times U(2)_L \times U(2)_u \times U(2)_d \times U(2)_e \, .
\end{equation}
\add{In this case we need to introduce more spurions than in the MFV scenario to make all interactions formally invariant under $H_F$.}
We can parametrize the Yukawa matrices in terms of these spurions as follows \cite{SMEFTFlavorSym2} 
\begin{equation}
\label{eq:YukawaU2}
y_u =  \lambda_{t} \begin{pmatrix}
\Delta_u & x_{t} V_q \\
0 & 1 
\end{pmatrix}\qquad
y_d =  \lambda_{b} \begin{pmatrix}
\Delta_d & x_{b} V_q \\
0 & 1 
\end{pmatrix}\qquad
y_e =  \lambda_{\tau} \begin{pmatrix}
\Delta_e & x_{\tau} V_l \\
0 & 1 
\end{pmatrix} \, .
\end{equation}
where the $\Delta_i$ are in general $2\times2$ complex matrices, the $V_i$ are complex 2-vectors and the $\lambda_i$ \add{and $x_i$ are free} complex parameters expected to be of $\mathcal{O}(1)$. The spurions have the following transformation properties under $H_F$
\begin{equation}
\begin{split}
\Delta_u \sim (2,1,\bar{2}&,1,1) \qquad \Delta_d \sim (2,1,1,\bar{2},1) \qquad \Delta_e \sim (1,2,1,1,\bar{2}) \\ 
& V_q \sim (2,1,1,1,1) \qquad V_l \sim (1,2,1,1,1) \, .
\end{split}
\end{equation}
\add{As before, one can relate the spurions to parameters in the SM where this time some parameters remain unconstrained (see App.~\ref{app:flavorsym} for details). }

\add{Again, the assumption is that the can Wilson coefficients can be written in terms of the above spurions, finding e.g. for the dipole operator \cite{SMEFTFlavorSym}}
\begin{equation}
\label{eq:UpDipSpurion}
C^\prime_{\substack{uB\\pr}} O^\prime_{\substack{uB\\pr}} \supset  f_{uB} \left[ \alpha_1 \bar{q}^\prime_3 \sigma^{\mu\nu} u^\prime_3 \widetilde{H} B_{\mu\nu} + \beta_1 \bar{Q}^{\prime p} V_q^p \sigma^{\mu\nu} u^\prime_3  \widetilde{H} B_{\mu\nu} + \rho_1 \bar{Q}_p \sigma^{\mu\nu} (\Delta_u)_{pr} U_r \widetilde{H} B_{\mu\nu} \right]
\end{equation}
where the fields in capitals denote fields from the first two generations and the field with the subscript 3 a field from the third generation. For convenience we introduced the notation $C_X(Y) = f_X Y$ in analogy to Ref.~\cite{SMEFTFlavorSym}. We will work at an accuracy of $\mathcal{O}(10^{-2})$ which means that we have to keep terms up to $\mathcal{O}(\Delta_{u,d,e},V_{q,l}^2)$. We have listed an exemplary set of expansions of all relevant Wilson coefficients in App.~\ref{app:flavorsym}. 

Since more fermionic operators appear in the expression of the up-type quark dipoles we choose the up-quark basis for the SMEFT operators in the unbroken phase. Then, e.g. the $\beta_1$ component in Eq.~\eqref{eq:UpDipSpurion} can be ignored for the dipole Wilson coefficients in the mass basis but is generated for the down-type dipole after relating the Wilson coefficient in the mass basis to the one in the gauge basis.



\section{Loop calculation} \label{sec:calc}
\subsection{Scheme definitions}

By performing a simple power counting we can easily verify the expectation that most of the diagrams we encounter, see \App{app:diags}, are UV divergent. We regularize these by using dimensional regularization and evaluate all loop diagrams in $D=4-2\epsilon$ space-time dimensions performing the limit $\epsilon\to0$ at the end of the calculation. In this regularization scheme 1-loop UV divergences manifest themselves as simple poles in the expansion for small $\epsilon$ and we subtract these poles with appropriate counterterms in the $\overline{\text{MS}}$ scheme. The only exception to this procedure are the scalar tadpoles, loop contributions to the Higgs one-point function, that renormalize the Higgs VEV and are present only in the broken phase of the SMEFT. To deal with this type of diagrams, we chose the tadpole counterterm such that it cancels the tadpole diagrams completely, analogously to what was done in~\cite{Hartmann:2015oia}. The result is that no such diagrams have to be calculated and the loop contributions to the Higgs VEV are given by the loops in the physical Higgs 2-point function. In addition, due to the photon and gluon being massless, we encounter a few IR divergent diagrams. We regularize these by assigning both these bosons an infinitesimally small mass $m$ and keeping only terms that are regular in the limit $m\to0$. \add{Note, the IR divergencies can, of course, be regulated using dimensional regularization, analogous to the UV divergencies. Nevertheless, we chose the finite mass regulator to make distinguishing between UV and IR divergencies and logarithms straightforward. However, we checked explicitly that the rational terms presented in this work are independent of the chosen regulator.}

\add{The SMEFT, being a chiral theory, requires special care when dealing with $\gamma_5$ in the context of dimensional regularization.} \add{In this paper we use the \emph{naive dimensional regularization} (NDR) scheme \cite{Hooft1972,Breitenlohner1977,Bonneau1980}. It is well-know that this introduces a scheme dependence of certain rational terms, in particular those arising from diagrams in which traces with an odd number of $\gamma_5$ matrices appear. In this work these appear in diagrams with insertions of four-fermion operators as well as the insertions of the dipole operators into the gluon 3-point function. The usage of NDR} also fixes the treatment of the Levi-Civita symbol in an arbitrary number of space-time dimensions, by treating its properties the usual way, pretending as we would be working in four dimensions. This can lead to possible issues for diagrams containing the CP odd operators from the $H^2F^2$ and $F^3$ classes. These would arise mainly from contractions of two or more Levi-Civita symbols, but they can be avoided by performing the loop integral before contracting any of the indices of the Levi-Civita symbol, leaving only four four-dimensional indices to be contracted and hence no source of any inconsistencies remain \cite{Dekens2019}. In fact, we explicitly checked that for the $H^2F^2$ operators the result does not change if the indices are contracted from the beginning. 

Additional care has to be taken when calculating the contributions of the CP odd $F^3$ operators to the dipoles, independently of the gauge bosons they are built from. By investigating the respective diagram and performing a power counting we note that its most singular piece is linearly divergent and from the treatment of axial anomalies it is known that such diagrams are not necessarily independent of the choice of momentum routing in the loop. Together with the NDR scheme this leads to the result for e.g. the $W^3$ operator, 
\begin{equation}
\frac{d_\psi}{e}\times \Lambda^2 \supset \frac{3-A}{32\pi^2}\frac{e\, m_\psi}{s_w}~  C_{\widetilde{W}}.
\end{equation}
Here $A$ is a constant, arbitrary shift of the loop momentum in the convention where the fermion in the loop carries the momentum $q+A~p_1$, where $q$ is the loop momentum and $p_1$ the incoming fermion momentum. In this calculation the choice $A=0$ corresponds to the known result found in the literature \cite{Boudjema1991,Gripaios2013,Panico2018}. The same dependence on $A$ appears in the rational part of the gluonic diagram if it is calculated in this naive way, while the divergent structure is independent of the loop momentum routing. To circumvent this issue, we proceed as mentioned above and explained in \cite{Dekens2019} and keep the Levi-Civita symbol external to the loop integral and contract its indices only after evaluating said integral. However, contrary to \cite{Dekens2019}, we extract the $W^+ W^-\gamma$ vertex by treating all the legs of the operator $O_{\widetilde{W}}$ to be on-shell and in $D=4$, such that we can use properties of the the Levi-Civita symbol to simplify the vertex rule. This procedure reproduces the results in \cite{Boudjema1991,Gripaios2013,Panico2018}, where the authors start from a $W^+ W^-\gamma$ operator, but does not capture the the contribution of an evanescent operator, see \cite{Dekens2019}.

\subsection{Gauge invariance and redundant operators}

\subsubsection{Gauge invariance and BFM}

Being built upon the SM, the SMEFT is imbued with the same gauge symmetry, hence our results respect this gauge invariance as well.

\add{In order to not have to deal with a large number of gauge-variant counterterms, instead of using the usual $R_\xi$ gauges, we make use of the background field method (BFM) \cite{Abbott1981,Abbott1983,Denner1994,Denner1996} by splitting all the fields into a classical background as well as quantum fluctuations around that background. Because the gauge for these two pieces can be chosen independently, we} choose a linear $R_\xi$ gauge and in particular the Feynman gauge ($\xi=1$) for quantum and unitary gauge ($\xi\to\infty)$ for classical fields. We will not go into further detail about the BFM and refer the reader to \cite{Abbott1981,Abbott1983,Denner1994,Denner1996}. See also \cite{Helset2018,Corbett2020,Corbett2020a} for the BFM in the context of gauge fixing the SMEFT. 

In practice, the classical fields correspond to external fields while the quantum fields describe fields running in loops and differences to the conventional gauge fixing procedure can arise only in Feynman rules containing both classical and quantum fields. In fact, because we are dealing only with CP odd dimension-6 operators that are not directly affected by gauge fixing, the only modifications we encounter involve only the gauge boson self-interactions, Goldstone-gauge and ghost-gauge vertices within the SM.

\subsubsection{Redundant operators and choice of basis}
\add{It is well-known that in order to take care of all operator structures appearing in loop calculations within EFTs, additional operators have to be included, even if we started with a non-redundant set of operators.}

\add{This is also the case for the loop contributions to the EDMs. In particular, we consider as redundant set the so-called \textit{Green's} basis \cite{Jiang:2018pbd,Gherardi:2020det}, which is given by all the operators, independent under integration by parts, which are directly generated by 1PI Green's functions. Of course, at the end of the calculations, the redundant operators have to be removed using the field redefinitions, shifting the coefficients of the operators in the non-redundant basis.}

\add{We demonstrate the procedure using the example of the electron EDM. Consider the loop contributions to the fermion 2-point function, especially the middle and right diagram in Fig.~\ref{fig:Fermion_2pt}, ignoring the left diagram which is purely an SM effect. The diagrams containing the electron dipole operator, on the other hand, give rise to not only the usual SM structures but also one that is proportional to the fermion momentum squared, $p^2$. Clearly, there is no corresponding operator in both the SM or the Warsaw basis, however, there is one in the Green's basis and in the phase of broken EW symmetry it has the form 
\begin{equation}
O_{D^2}\sim\bar{\psi}D^2\psi\,,
\end{equation}
where the covariant derivative $D_\mu=\partial_\mu + i e Q A_\mu$ contains only the photon for simplicity. Note, Note that we could have used $\slashed{D}\slashed{D}$ instead of $D^2$. 
The two operators are equivalent, even in the Green's basis, as they are related by a purely algebraic identity \begin{equation}
\slashed{D}\slashed{D}=D^2+\frac{1}{2}e Q_e\sigma_{\mu\nu}F^{\mu\nu}\,.
\label{eq:D2}
\end{equation}
The difference between the two appears only in matrix elements with additional gauge bosons.
The coefficient $c_{D^2}$ of the redundant operator is then given by precisely the 1-loop sized term in the 2-point function proportional to $p^2$ so the last step is to remove the redundant operator.
We find the appropriate field redefiniton to be
\begin{equation}
e\to e+i\frac{c_{D^2}v}{\sqrt{2}}\,\slashed{D}e
\end{equation}
immediately shifting the coefficient of the dipole operator
\begin{equation}
c_{e\gamma}\to c_{e\gamma}-\frac{e Q_e}{2}c_{D^2}.
\end{equation}
By gauge invariance, we expect an operator structure in the $ee\gamma$ vertex function that can not be accounted for by any SM or Warsaw basis operator but instead by the one-photon part of $O_{D^2}$ and is numerically related to the $p^2$ structure we found above. Indeed, we find this exact 1-loop contribution, which serves as another check of our calculations.
For completeness, we quote the additional redundant operator needed in this work, here expressed in the unbroken phase,
\begin{gather}
O^{\text{\tiny{(2)}}}_{D^2}=\left(\bar{\psi}_LD_\mu\psi_R\right)D^\mu H\,.\label{OD2}
\end{gather}
It appeared in the calculation of the dipole operator contributions to $C_{Hud}$.
We refer to \cite{Gherardi:2020det} for the coefficients in Warsaw basis in terms of the ones in the Green's basis.}

Let us note that an alternative to this approach, which avoids the introduction of redundant operators, is to directly compute all reducible diagrams with the desired final states. For our purposes this would correspond to attaching e.g. the 1-loop fermion 2-point function to the tree level dipole vertex. But since we are working in the phase of broken EW symmetry with massive particles a cancellation between the 2-point function and the on-shell propagator connecting the loop to the tree level vertex is not obvious and spurious kinematic divergences appear if not treated with care.

\subsection{Additional cross-checks}

Before moving on to the results of our calculations we want to briefly comment on all the checks performed that give us confidence in their correctness. During the course of arriving at the final results we checked various different aspects of our results. 

First of all we used two different computer programs, the two Mathematica packages Package-X \cite{Patel2015} and the FeynRules/FeynArts/FormCalc \cite{Alloul2013,Hahn2000,Hahn1998} pipeline, finding the same results in both cases. We use the former to obtain explicit analytic expressions of the Passarino-Veltman (PV) loop integrals.

Further, although Feynman gauge is used for the quantum fields in our calculations we explicitly checked gauge invariance by leaving the gauge parameter $\xi$ generic in various subsets of diagrams and confirming analytically that every dependence on $\xi$ drops out. Further, as illustrated in the last section, we used the cancellation of various divergences related by gauge symmetry by the same redundant operator as a further check related to gauge invariance.

By performing a full 1-loop calculation we automatically rederived the RGEs in both the SMEFT \cite{Jenkins2013a,Jenkins2013,Alonso2013} and LEFT \cite{Jenkins2017,Jenkins2017a} and we explicitly verified that our RGEs coincide with the ones in the literature, after performing the respective weak rotations in the case of the SMEFT RGEs.

Finally, the full calculation of all diagrams was performed with two independent implementations, again yielding the same result.

All these aspects collectively give us the confidence to believe that the results presented here are correct and can be reproduced if the same \add{choice of schemes is} employed.

\section{Results and bounds}
\label{sec:res}

Now that we have established all the technical details of our calculation, we will present the results and bounds derived from them in this section. Because the full expressions for all the EDMs are quite long we will not report them here but instead refer the reader to \App{app:expressions}. The results shown there are taken to be at leading order in the external $m/v$, where $m$ is the mass of the external fermion. While this is a good approximation even for the third-generation leptons, this is not applicable for the third generation quarks.
Further, due to the sheer amount of Wilson coefficients appearing we also do not present all the bounds we obtained here, rather we quote them in \App{app:bounds}. Nevertheless, we will discuss the most interesting points in the following. In particular one of the main focuses in this work lies on the inclusion of finite terms, so we are also interested in quantifying the impact these terms have on the final result. To extract the bounds on various Wilson coefficients from any of the experimental EDM bounds, we neglect the SM contributions, that are many orders of magnitude smaller than the experimental constraints, and turn on only one coefficient at a time, rescaling them by the appropriate combination of SM couplings, reflecting the naturally expected to be carried by the corresponding coefficient, see also \Tab{tab:nat_sizes}. Using
this factorization, we expect, in most of the BSM theories, order one rescaled Wilson coefficients, if the parameters of the
UV completion have natural $O(1)$ size. For the new physics scale we assume $\Lambda=5$ TeV. Furthermore, we will also set lower bounds on the new physics scale $\Lambda$, assuming that the Wilson coefficients have the naturally expected size; we will see that EDMs push $\Lambda$ to be
very large, of the order of $10^3$ TeV.

\add{In the following section we define \emph{RG running} contributions to be all terms that explicitly contain a scale dependence, i.e. $\log(\Lambda)$. All remaining terms, both rational and non-rational, are collectively called \emph{finite}.}

\begin{table}[t]
	\begin{center}
	\begin{tabular}{|c|}
\hline
$C_{\underset{ii}{\psi B}}\to \left(y_{\psi}\right)_{ii} g'~ C_{\underset{ii}{\psi B}}$  \Tstrut \\[2ex]
$C_{\underset{ii}{\psi W}}\to \left(y_{\psi}\right)_{ii} g ~C_{\underset{ii}{\psi W}}$   \\[2ex]
$C_{\underset{ii}{q G}}\to \left(y_q\right)_{ii} g_s ~C_{\underset{ii}{q G}}$   \\[2ex]
$C_{\underset{ii}{q H}}\to \left(y_q\right)_{ii} ~C_{\underset{ii}{q H}}$   \\[2ex]
$C_{\underset{iijj}{lequ}}^{\text{\tiny (3)}} \to \left(y_{\ell}\right)_{ii}\left(y_u\right)_{jj} ~C_{\underset{iijj}{lequ}}^{\text{\tiny (3)}}$ \\[2ex]
$C_{\underset{ijji}{quqd}}^{\text{\tiny (1,8)}} \to \left(y_d\right)_{ii}\left(y_u\right)_{jj} ~C_{\underset{ijji}{quqd}}^{\text{\tiny (1,8)}}$ \\[2ex]
\hline
	\end{tabular}
	\hspace{10mm}
	\begin{tabular}{|c|}
\hline
$C_{H\widetilde{B}} \to g'^2 ~C_{H\widetilde{B}}$ \Tstrut\\[2ex]
$C_{H\widetilde{W}} \to g^2 ~C_{H\widetilde{W}}$ \\[2ex]
$C_{HW\widetilde{B}} \to gg' ~C_{HW\widetilde{B}}$ \\[2ex]
$C_{H\widetilde{G}} \to g_s^2 ~C_{H\widetilde{G}}$ \\[2ex]

$\mathcal{C} \equiv\left\{C_{Hud},C_{ud}^{(1,8)},C_{qu}^{\text{\tiny (1,8)}},C_{qd}^{\text{\tiny (1,8)}},C_{le}\right\} \to g'^2 ~\mathcal{C}$ \\[2ex]

\hline
	\end{tabular}
	\end{center}
	
	\caption{Rescalings of the Wilson coefficients performed throughout this work to reflect the natural size we expect them to carry. We assume the operators which are built from vector currents and therefore do not involve a chirality flip to be generated by a heavy vector boson exchange and choose the SM $U(1)_Y$ gauge coupling as a representative.} 
	\label{tab:nat_sizes}
\end{table}

\subsection{Lepton EDMs}

\begin{figure}[t]
  \centering
\setkeys{Gin}{width=0.7\textwidth}
  \includegraphics{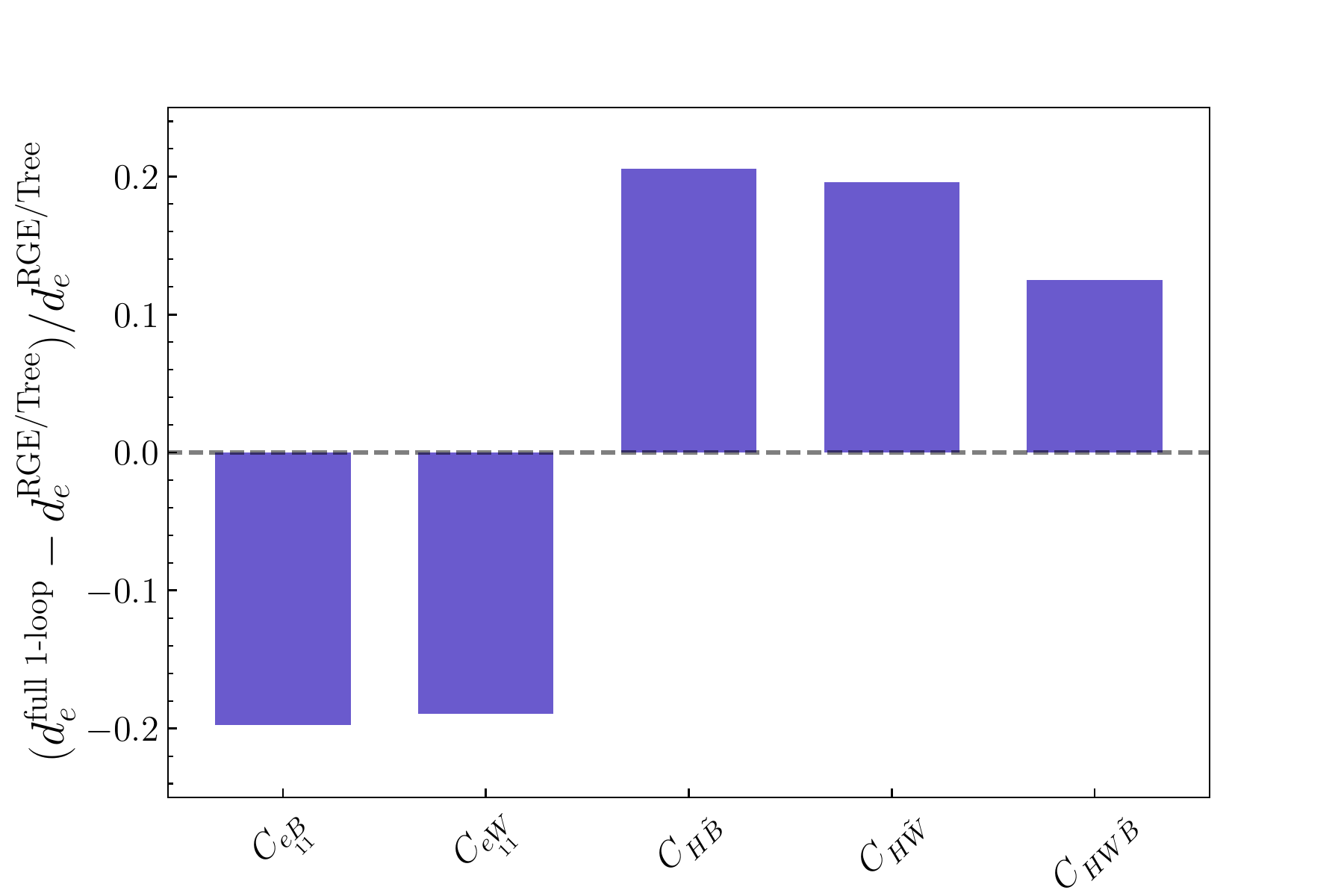}\\
  \includegraphics{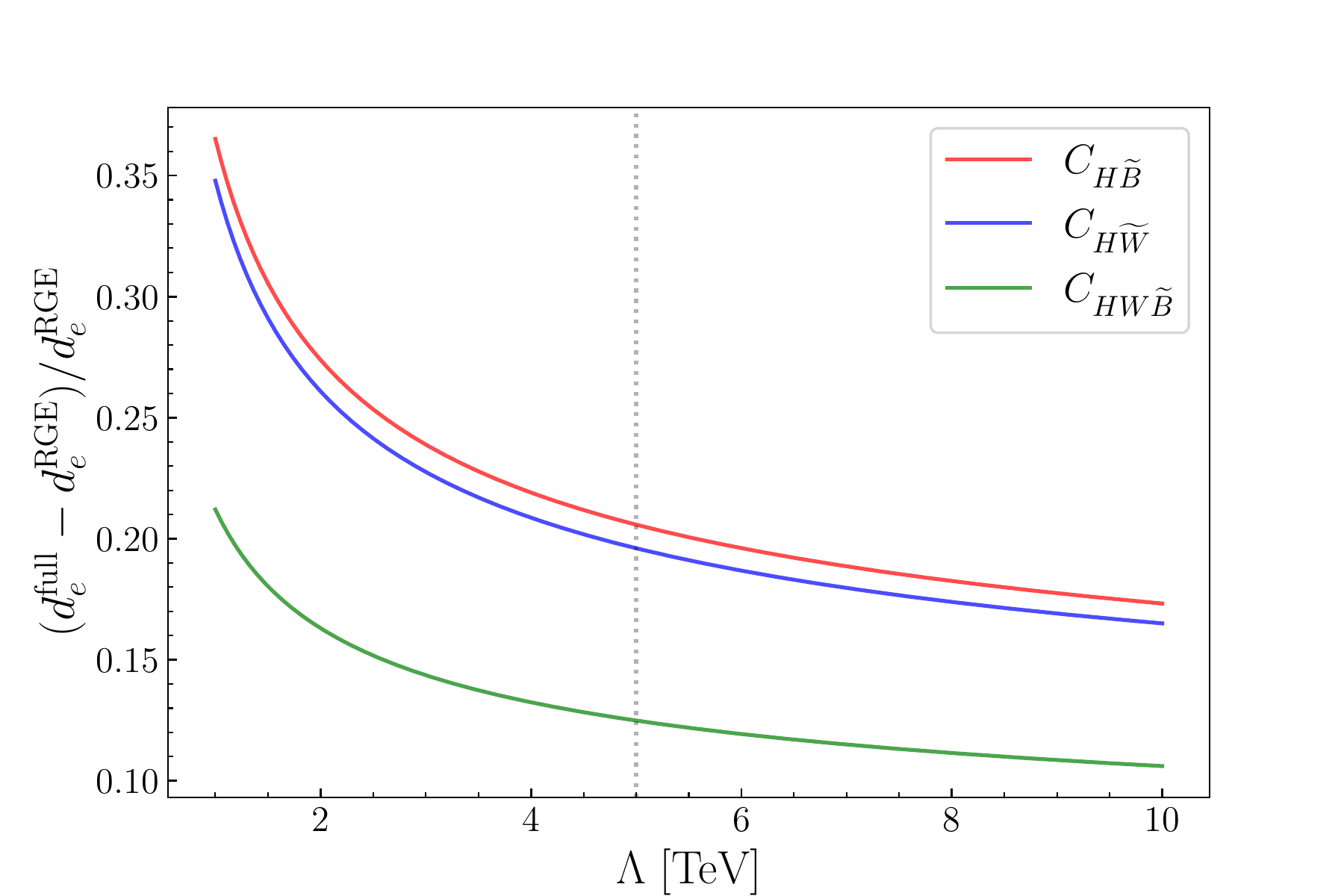}
  \caption{\textit{Upper panel:} Relative change of the electron EDM when using the full 1-loop result compared to only the RG running ($H^2F^2$ operators) and impact of the full 1-loop effects compared to the tree level term ($F\psi^2 H$ operators). \textit{Lower panel:} Dependence of the relative shift in the EDMs as a function of the scale $\Lambda.$ Here the dotted line shows the benchmark value of $\Lambda=5$ TeV used in this paper.}
  \label{fig:RGEvFiniteE} 
\end{figure}

We will start by investigating the lepton EDMs, where less operators appear, compared to the neutron case. In the following, we illustrate the impact of different terms in the contributions to EDMs coming from various class of operators. For the $H^2F^2$ class, we illustrate the impact of finite terms, showing, in the upper panel of \Fig{fig:RGEvFiniteE}, the relative change when using only the RGE versus the full 1-loop result. For illustrative purposes we use the electron EDM, and while the numerics change due to differing masses, the overall pattern is the same for the other lepton flavors.

In fact, these, together with the dipole operators themselves, are the only operators that give both RGE and finite contributions, while operators of the $\psi^4$ class give vanishing rational terms and both the $F^3$ and $\psi^2\bar{\psi}^2$ class operators enter only through purely rational terms. We want to note that, on the other hand, for the dipole operators finite terms play a negligible role affecting the result by $\lesssim 1\%$, but this is simply because they enter the EDMs also at tree level, completely dominating over corrections to higher order terms. This is why, in this case, we do not show the impact of the 1-loop finite terms but rather of the full 1-loop result compared to the tree level term for these operators only, in \Fig{fig:RGEvFiniteE}. We see that these higher order effects add to the destructively to the tree level piece, therefore actually lowering the bound on the scale $\Lambda$.

On the contrary, for the $H^2F^2$ class operators any tree level contribution is obviously absent, which presents a great opportunity to study the size of finite terms. Indeed, we find that the finite terms change the bound by $\sim10-20\%$, however, due to positive relative signs they interfere constructively and consequently increase the bound compared to when using the RG running only. By looking at the corresponding expression we can also easily explain why the effects of the two operators with only one kind of gauge field appearing are very similar but on the other hand quite different from the mixed one. The operators $O_{H\widetilde{B}}$ and $O_{H\widetilde{W}}$ do only get contributions from the photon and Z components of the weak bosons, meaning apart from numerical prefactors coming from different couplings they give the same contributions. On top of that the mixed operator, $O_{HW\widetilde{B}}$, also receives contributions from its W component and it turns out that this piece has the opposite sign of the neutral ones, again reducing the total impact on the lepton EDMs.

Of course, these statements are depend on the scale $\Lambda$, as this changes the energy regime that needs to be swept by the RGE logs. This implies that for new physics sectors well above the TeV the finite terms will be completely subdominant compared to the huge logarithms appearing. On the other hand, the closer the new sector lies to EW scale the smaller the logs and therefore finite terms can have an increasingly big effect. We illustrate this in the lower panel of \Fig{fig:RGEvFiniteE}, where we show the dependence on $\Lambda$ of the relative shift in the electron EDM for the $H^2F^2$-class operators. We see that, due to the slow logarithmic growth, the effect of finite terms does not deteriorate tremendously for e.g. $\Lambda\sim10$ TeV, while it almost doubles for $\Lambda$ approaching $\sim 1$ TeV.

\begin{figure}[t]
  	\centering
	\setkeys{Gin}{width=0.7\linewidth}
		\includegraphics{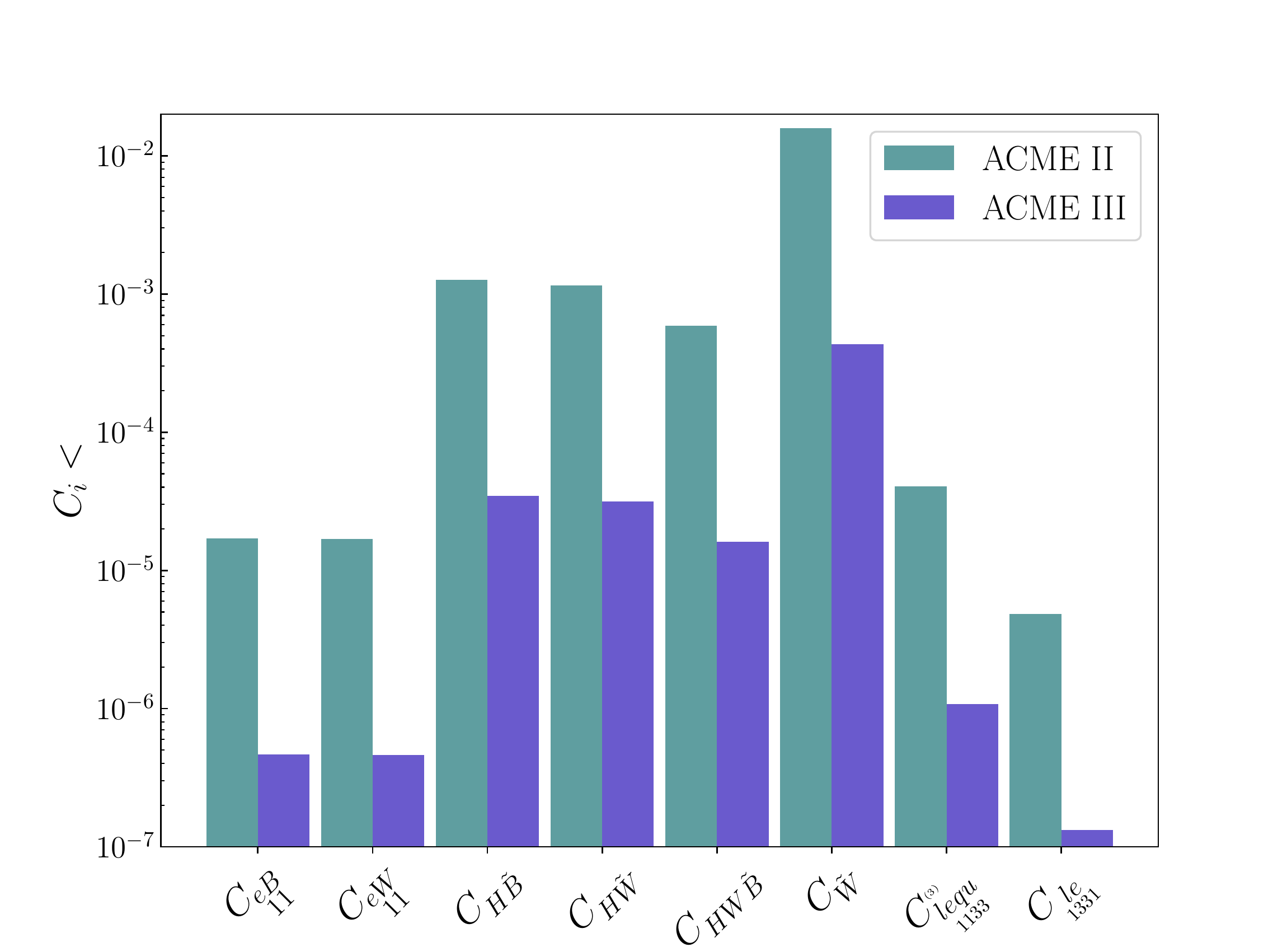}\\		
		\includegraphics{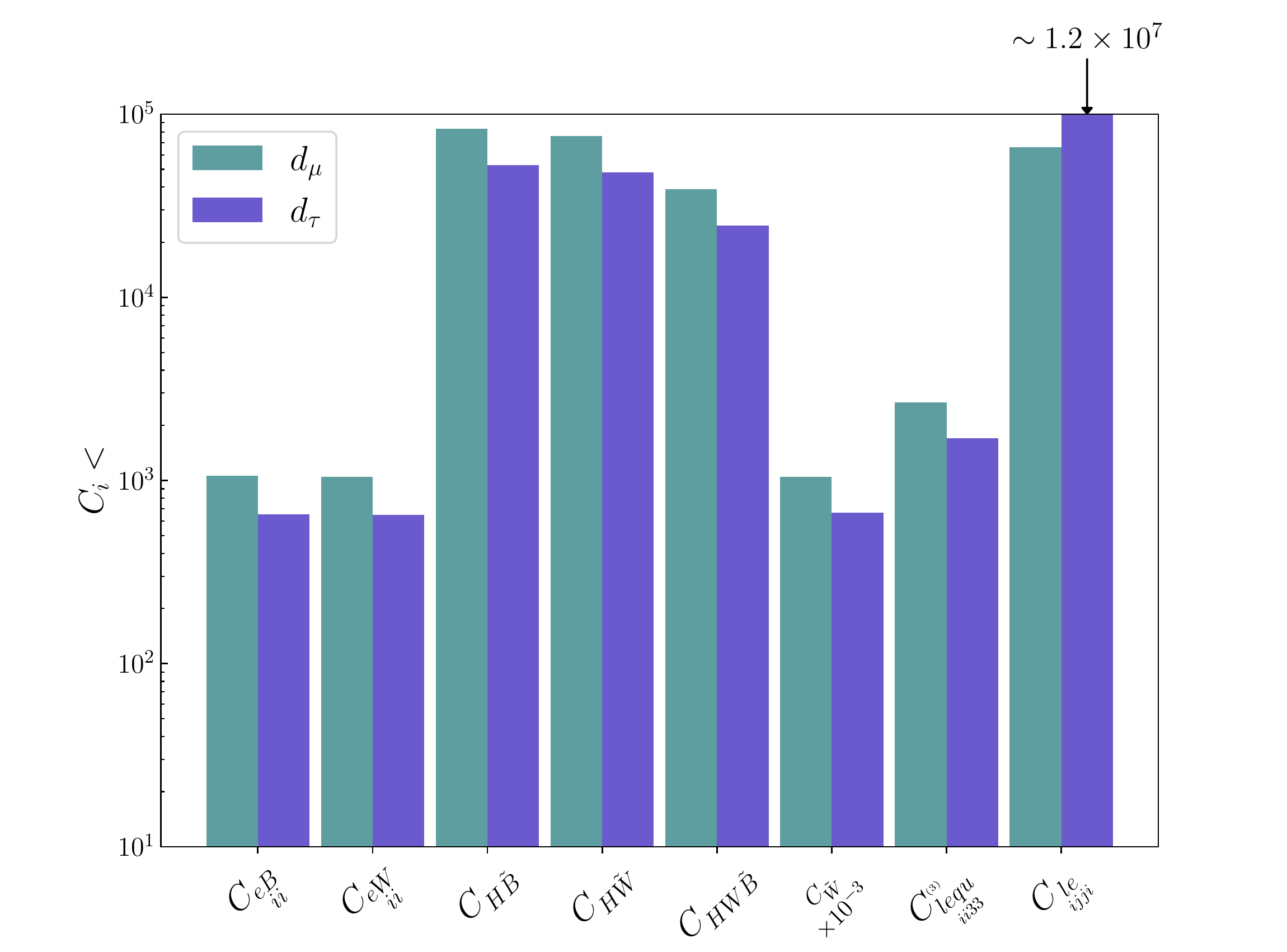}
  		\caption{Upper bounds on the Wilson coefficients, assuming $\Lambda=5$ TeV and applying the rescalings shown in Table~\ref{tab:nat_sizes}, obtained including the full 1-loop expressions, from the experimental bounds  on the different lepton EDMs. \textit{Upper panel:} The current constraints (ACME II) coming from the best bounds on the electron EDM, compared to the ones from the projected future bounds (ACME III). \textit{Lower panel:} We compare the bounds from the two heavy lepton flavors with each other. Here  $i=2(3)$ stands for the muon (tau) EDM and $j$ denotes the heavier of the two lepton flavors different from $i$ in the operator $O_{le}$. }
  		\label{fig:FlavorEDMBounds}
\end{figure}

Finally, let us briefly discuss the bounds on the Wilson coefficients from the electron, muon and tau EDM, summarized in \Fig{fig:FlavorEDMBounds} and computed assuming $\Lambda=5$ TeV and applying the rescalings shown in Table~\ref{tab:nat_sizes}. Here we show the full tree plus loop level result, i.e. including both the RG running and finite terms; in the case of the electron EDM the prospected future bounds are shown as well. Note that for the 4-fermion operators, we chose to show only the component with the most stringent bound for each of the operators. The bounds on other components can easily be obtained from the ones shown in \Fig{fig:FlavorEDMBounds} by rescaling them with the appropriate ratio of fermion masses.
The most obvious conclusion that can be drawn from this figure, by comparing the upper panel with the lower one (and with the values in \Tab{tab:eEDMBoundsWC}) is that the supreme precision of the eEDM measurement gives by far the most stringent bounds from any of the lepton flavors. One can notice that, for $\Lambda=5$ TeV, the constraints from the electron EDM can set bounds of order $10^{-5}$ on the Wilson coefficients of operators with fermions and of $10^{-3}\div 10^{-2}$ in the case of purely bosonic operators. These bounds will further improve of one or two orders of magnitude at ACME III.

Nevertheless, we can make another interesting observation. Even though the experimental sensitivity to the muon EDM is roughly one order of magnitude higher than for the tau EDM, it still happens to be the case that the tau lepton is slightly more constraining than its lighter cousin. Speaking of the different masses of these leptons, this is exactly the reason why this happens. For every operator the contribution is proportional to the lepton Yukawa, either through our rescaling of the Wilson coefficients to their natural size or because the contribution itself is directly proportional to the lepton mass. So it turns out that with the current sensitivities the mass difference between the muon and tau lepton barely overcompensates the lower experimental reach for the latter such that the tau EDM is indeed more constraining than its muonic counterpart. This argument, however, does not hold for the operator $O_{le}$. For this operator we see the inverted situation, where the tau EDM is less constraining that the muon EDM. But this is readily explained by closer examining the corresponding expression in Eq.~\eqref{eq:ole}. Here we see that it is in fact not proportional to mass of the external lepton but of the lepton inside the loop instead. Because we chose the most constraining component of each Wilson coefficient, this mass is the tau mass for the muon EDM and vice versa, such that the reasoning here is exactly inverted with respect to all the other operators and on top of the weaker experimental bound, the constraint from the tau EDM is further suppressed by the muon mass, contrary to the tau mass in the muon EDM.
From this point of view, the phenomenal constraining power of the electron EDM is even more impressive, as the mass gap between the electron mass and the other lepton masses spans multiple orders of magnitude, but still the electron bounds by far overshadow the other ones.

As mentioned before, we also set lower bounds on the new physics scale $\Lambda$, assuming that the Wilson coefficients have values corresponding to the natural size indicated in Table~\ref{tab:nat_sizes}. Turning on one operator at a time, the strongest constraints come from the dipole and the $O^{\tiny(3)}_{\underset{1133}{lequ}}$ and $O_{\underset{1331}{le}}$ contributions and are of the order of $10^3$ TeV.

\subsection{Neutron EDM}

\begin{figure}[b]
  	\centering
	\includegraphics[width=0.85\textwidth]{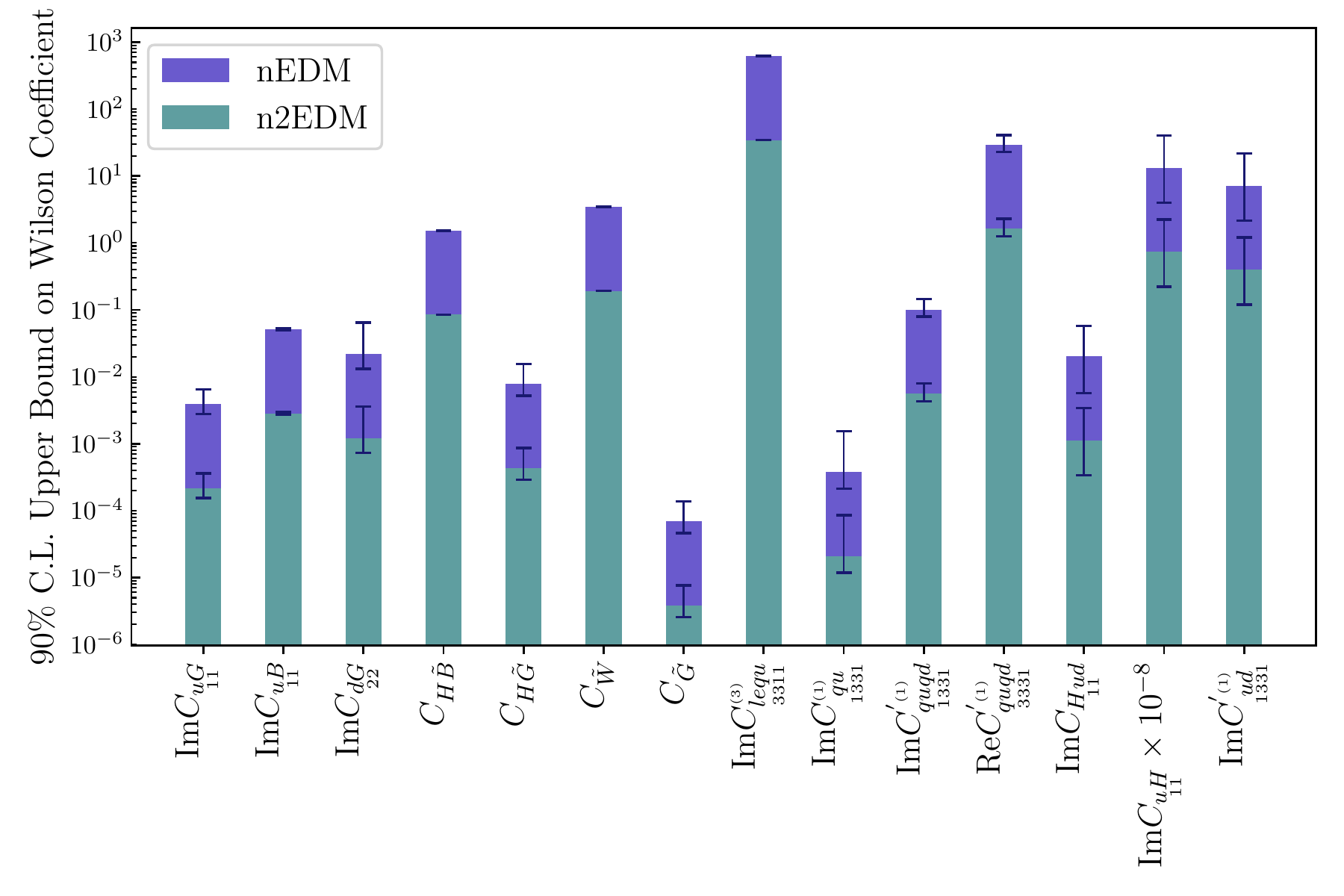}
  	\caption{Selected upper bounds on the Wilson coefficients, assuming $\Lambda=5$ TeV and applying the rescalings shown in Table~\ref{tab:nat_sizes}, obtained including the full 1-loop expressions, from the experimental bounds on the neutron EDM. In addition to the bounds from the central values, we also show the influence of the uncertainties in the determination of the chromo-dipole and Weinberg operator matrix elements. We also show bounds on the Wilson coefficients for the projected accuracy of the n2EDM experiment. Notice that the last two Wilson coefficients are in the up-quark gauge basis, while the others in the mass basis.}
  	\label{fig:nEDMWCBounds}
\end{figure}
We proceed with the neutron EDM which is composed of the (chromo-)EDMs of the quarks and gluons as well as the operators $O_{\underset{11}{Hud}}$ and $O_{\underset{1111}{quqd}}^{(1,8)}$ which can be matched to operators which have a non-vanishing matrix element on the neutron EDM. There are several differences with respect to the lepton EDMs as we can now have cancellations between the 1-loop contributions of the EDMs and chromo-EDMs of the light quarks, more flavor components of the Wilson coefficients are contributing to the dipole amplitudes (this is all the more true in the gauge basis, due to the non-trivial rotation between gauge and mass basis, see Sec.~\ref{sec:basis}) and in general more operators due to the presence of QCD degrees of freedom. 
We show a selection of bounds in Fig.~\ref{fig:nEDMWCBounds} where we have also included a conservative estimate of the influence of the uncertainties in the determination of the matrix elements of all contributing effective operators in the expression of the neutron EDM and a projection for the expected accuracy of the n2EDM experiment \cite{Ayres:2021hoq}. The full set of bounds can be found in App.s~\ref{sec:nEDMwoFlavor} and \ref{sec:nEDMwFlavor}. 

Starting with the dipole operators, in addition to the electroweak dipole operators also the gluonic dipole operators contribute to the neutron EDM. There, the effects of including finite terms are much larger than for the electroweak dipoles. This is due to the large rational terms in the wave function renormalization of the gluon. In addition, we can also probe more flavor components of the dipole operators through the appearance of the strange quark dipole in the neutron EDM as well as the appearance of all flavor components of the quark dipole Wilson coefficients in the 3-gluon 1-loop amplitudes. The bounds on these flavor components are suppressed with respect to the dominant up and down quark chromo-dipole operators, since the matrix elements in the expression for the neutron EDM are smaller and some of the flavor elements only enter through loop corrections. Note also, that the contribution of the dipole operators through effective operators other than dipole operators in the expression of the neutron EDM is negligible, since these contributions are suppressed by the much smaller matrix elements of the effective operators and the common loop factor that all dipole contributions receive that are sourced by these additional effective operators. One exception to this is the contribution through the Weinberg operator as those loop contributions are enhanced by an inverse quark mass. This can be seen in particular in the bounds on the coefficients in the spurionic expansion of the different flavor symmetries as we will see later.

For the $H^2F^2$ type operators we also have to differentiate between the operators with field strengths of electroweak and strong gauge bosons. The bounds on the electroweak operators are less stringent, by around three orders of magnitude, than the ones obtained from the electron EDM, as is expected due to the experimental bound on the neutron EDM being so much weaker. Interestingly, for all three electroweak operators there is a constructive interference between the terms from the different quark EDMs, enhancing the contribution to the neutron EDM, together with the enhancement from the quark Yukawas with respect to the electron case. Therefore with an experimental bound on the neutron EDM with the same constraining power as the current electron EDM sensitivity, the bounds on the Wilson coefficients would actually be stronger than those obtained from the electron EDM. The neutron EDM receives, through the quark chromo-EDMs, contributions also from the gluonic $H^2G^2$ operator. Such terms are additionally enhanced by the strong coupling and for this reason the bound on the corresponding Wilson coefficient is stronger than the constraints obtained for the Wilson coefficients of the electroweak bosonic operators by more than two orders of magnitude, as shown in Fig.~\ref{fig:nEDMWCBounds}.

For the 4-fermion operators we have the same situation as for the lepton EDMs, only now there are more operators including quarks contributing to the EDM. As for the lepton EDMs, the 4-fermion operators either enter only via RG running or only via rational terms to the dominant contributions that are given by the (chromo-)dipole operators. They can also enter directly with a small hadronic matrix element in the neutron EDM. What is interesting for these 4-fermion operators made from quarks is that the change of basis from the gauge to the mass basis is non-trivial, as discussed in Sec.~\ref{sec:basis}. 
Starting, for example, from an up- or down-quark gauge basis, in the rotation to the mass basis a CKM matrix appears for the down or up component of the operators, respectively. As mentioned above, whenever we use expressions in terms of Wilson coefficients in the gauge basis, we choose the up-basis since more operators with up quarks appear in the final expression of the neutron EDM. In fact, with this choice, a larger number of operators is left unchanged by the basis transformation; for example, this is the case for the $O_{lequ}^{\tiny(3)}$ operator already considered in the previous section in the discussion of the lepton EDMs.
However, since both the up and down type dipole appear in the neutron EDM it is inevitable that CKM matrix elements appear somewhere. Since the CKM matrix contains a CP violating phase this also enables us to probe the real part of some of the Wilson coefficients in the gauge basis, in particular of some of the flavor off-diagonal ones (see the rightmost column in Fig.~\ref{fig:nEDMWCBounds}). In fact, these real parts contribute to the imaginary parts of the Wilson coefficients in the mass basis, that enter the EDMs expressions. Those constraints are of the same order as the bounds on the corresponding imaginary parts, since the imaginary part of the very off-diagonal part of the CKM matrix is of the same order as its real part. 

Another interesting contribution appears through the Weinberg operator. Unlike $O_{\tilde{W}}$, it can also contribute with RG running and in addition to its appearance through the quark chromo-dipoles, it also enters directly in the expression of the neutron EDM, interpreted as the chromo-dipole of the gluon. As can be seen in the analytical expressions of the dipoles in combination with how they enter in the neutron EDM, the interference between the different chromo-EDMs is constructive and all effects proportional to the Weinberg Wilson coefficient add up to the comparably strong bound. This, together with the strong coupling enhancement for this contributions, leads to the most stringent among the constraints imposed by the neutron EDM experimental bound, being of order $10^{-4}$ for $\Lambda=5$ TeV and for a $C_{\tilde{G}}$ rescaled as in Table~\ref{tab:nat_sizes}. In addition, there are large finite terms in the self 1-loop contributions of the Weinberg operator which give corrections of $\sim45\%$ with respect to only including RG running at the considered scale. 

Furthermore, there can be direct contributions of the 4-fermion operators $O_{quqd}^{(1,8)}$ which are however largely suppressed by their small matrix element in the neutron EDM. This leads to an interesting interference where loop suppressed contributions of these 4-fermion operators to the dipole operators, which are further suppressed by small Yukawa couplings, are of the same order as the direct tree level contributions of those operators (see App.\ref{app:expressions} and Eq.~\eqref{eq:nEDM}). The dipole contributions to those 4-fermion operators are suppressed by small matrix elements and loop factors as discussed before.

Finally, there is a small direct contribution to the neutron EDM of the operator $O_{Hud}$ which also contributes with a finite term to the dipole operators. As can be seen in Fig.~\ref{fig:nEDMWCBounds}, the Wilson coefficient of this operators gets a significant bound from the neutron EDM mostly due to the tree level contribution to the neutron EDM. The Yukawa-like operators $O_{uH,dH}$ which appear in the 1-loop contribution to this operator on the other hand are largely suppressed by a loop-factor and small Yukawas and therefore only get bounds beyond the perturbative unitarity limit. As mentioned previously, the dipole contributions which also enter in this 1-loop expression are negligible when compared to the dominant direct contributions to the neutron EDM. Lastly, there is another 4-fermion operator which enters in the 1-loop expression of the operator $O_{Hud}$, $O_{ud}^{(1,8)}$, which also only receives a bound around the perturbative unitarity limit.

We also show in Fig.~\ref{fig:nEDMWCBounds} the error bars associated to the 50\% uncertainties of the matrix elements of the quark and gluon chromo-EDMs. Wherever the Wilson coefficients of the chromo-dipole operators enter at tree level, the uncertainties translate direcly to the bound. In the case of the electroweak operators, which can only enter at loop level in the chromo-EDMs, the dependence on the uncertainties is much smaller.

Furthermore, we also estimate the bounds on all Wilson coefficients with the projected experimental bound of the n2EDM experiment \cite{Ayres:2021hoq}. With the projected experimental bound of $\sim 10^{-27}e$ cm, we expect an improvement of about one order of magnitude for all Wilson coefficients.
\begin{figure}[t]
  	\centering
	\includegraphics[width=0.85\textwidth]{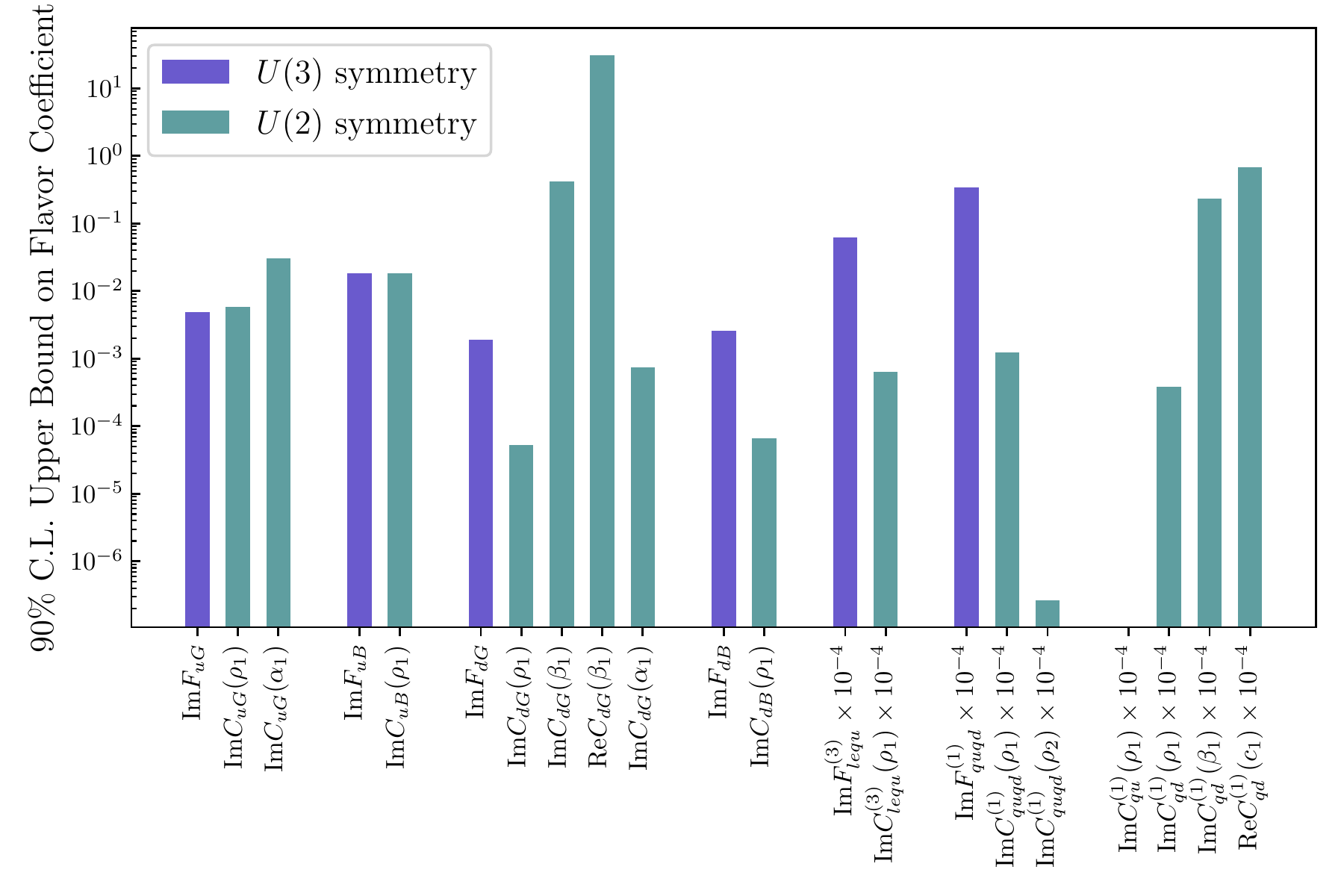}
  	\caption{Selected bounds on coefficients in spurionic expansion assuming the different flavor symmetries and for $\Lambda=5$ {TeV}.}
  	\label{fig:nEDMWCBoundsFlavor}
\end{figure}

As mentioned before, see Sec.~\ref{sec:neutronflavor}, it is also interesting to consider the expression of the neutron EDM under flavor symmetries relating the different flavor components that appear in the neutron EDM with some minimal assumptions (for the notation we refer to \App{app:flavorsym}). The bounds on the coefficients in the spurionic expansion of the Wilson coefficients, as discussed above, can be found in Fig.~\ref{fig:nEDMWCBoundsFlavor} (notice that this expansion is performed for the Wilson coefficients in the up-quark gauge basis). 
The key feature of the different flavor symmetric scenarios, namely the correlation among components of flavor tensors, leads to the combination in a single bound of various contributions, that would have been separated for a generic flavor structure. Then, the constraints on flavor blind coefficients of the spurionic expansion are dominated by the strongest among the bounds on the various flavor components. For example, the up-quark dipole receives contributions from all the ${\rm Im}[C^{\tiny{(3)}}_{{lequ}}]_{ii11}$ components, which, if taken as independent among each others, have very different constraints: ${\rm Im}[C^{\tiny{(3)}}_{{lequ}}]_{1111}< 7.54 \cdot 10^9\lambda_e\lambda_u$ and ${\rm Im}[C^{\tiny{(3)}}_{{lequ}}]_{3311}< 6.21 \cdot 10^2\lambda_\tau\lambda_u$, where the $\lambda$s are entries of the diagonalized Yukawa matrices. On the other hand, if a $U(3)^5$ flavor symmetry is imposed, exactly the same Yukawa dependence as above is assigned to each component, but with a unique coefficient in front, whose bound reads ${\rm Im}F^{\tiny{(3)}}_{lequ}< 6.19 \cdot 10^2$: it is of the same order, but even slightly stronger, as the previous bound on the $\tau$ matrix element, which was the most severe. Similarly, the limit on the down-type flavor coefficients is particularly interesting because it combines the contributions of the down and strange quark dipoles into one bound. In addition, the $U(2)$ flavor symmetry disentangles the contributions from the third and first two generations which is visible in the bounds on $\text{Im}C_{uG}(\rho_1)$ and $\text{Im}C_{uG}(\alpha_1)$, where the $\alpha_1$ component only receives contributions from the contributions of the top dipole operator to the three-gluon amplitude and, thus, has weaker constraints. 

Most of the bounds on the flavor coefficients of the 4-fermion operators are just around or beyond the perturbative unitarity limit, still allowing the flavor symmetries as a valid symmetry of UV physics, but not setting any significant constraint on the parameter space. As for the dipole operators, the difference between the $U(2)$ and $U(3)$ flavor symmetry is apparent in the expansion of the Wilson coefficient $C_{lequ}^{\tiny{(3)}}$. In the $U(3)$ spurionic expansion all lepton flavors contribute in the loop but they are all suppressed with the respective small lepton Yukawa. For the $U(2)$ symmetry on the other hand, only the third generation of the leptons is allowed at the considered accuracy. However, since the third generation is excluded from the flavor group, it is completely unsuppressed apart from the small up-quark Yukawa that is also present in the $U(3)$ spurionic expansion. 

What is also worth noting are the Wilson coefficients $C_{qu,qd}^{\tiny{(1)}}$ which are completely forbidden by the $U(3)$ flavor symmetry at the considered order. Some elements of the flavor tensor are allowed in the $U(2)$ expansion, giving however fairly lose bounds. As we saw earlier in the discussion of the neutron bounds without flavor symmetries, we can also probe the real parts of flavor coefficient, if other phases are present. This is the case here, where the CKM phase can also appear through the $V_q$ spurion in the expansion of these Wilson coefficients. 

One should notice that, in the $U(2)$ case, the different independent terms in the spurionic expansion of a certain Wilson coefficient have to be of the same order, in order to allow the parameters $\rho_{1,2}$, $\alpha_1$, $\beta_1$ and $c_1$ (see \App{app:flavorsym}) to be of order 1, such that the flavor symmetry breaking pattern is respected. However, as we can see from Table~\ref{tab:nEDMBoundsU2WC}, this is usually not the case.

Importantly, we notice that, assuming the Wilson coefficients are of the natural size shown in Table~\ref{tab:nat_sizes}, the experimental constraint on the neutron EDM sets a lower bound on the new physics scale of order $10^3$ TeV, coming from the Weinberg operator $G^3$ contributions. All the bounds imposed when any of the other operators is instead turned on are at least one order of magnitude weaker.

\section{Conclusions} \label{sec:conclusions}

In this paper, we perform the analysis at 1-loop level of the lepton and neutron electric dipole moments, using the model independent EFT approach. We provide, at this accuracy, the complete expressions of these CP violating low energy observables as a function of the dimension-6 SMEFT Wilson coefficients in the Warsaw basis, including the RG running effects as well as finite terms. The latter play a fundamental role in the cases of operators that do not renormalize the dipoles, but there are also classes of operators for which they provide an important fraction, $10-20\%$, of the total 1-loop contribution, if the NP scale is around $\Lambda=5$ TeV. In presenting these results, we also discuss the various loop contributions to the EDMs under the light of selection rules, based on helicity, angular momentum and CP arguments. 

Furthermore, we compute the full set of bounds that the current and prospected experimental constraints impose on the Wilson coefficients, with one single operator turned on at a time, for a fixed SMEFT cut-off scale. On the other hand, we provide also the lower bounds on the scale of new physics, obtained assuming that the Wilson coefficients values are given by the natural sizes that we expect them to carry. The analysis of the neutron EDM is performed both in scenarios with generic flavor structure and in presence of $U(3)^5$ and $U(2)^5$ flavor symmetries for the SMEFT.
One can see that EDMs provide a very powerful probe for deviations from the SM, since the computed bounds are very strong and can push the scale of new physics above $10^3$ TeV, with the mentioned natural values for the Wilson coefficients. This means that any UV completion of the SM, for which the operators responsible for these strong bounds are generated, should accidentally have a very suppressed CP violation, similar to the SM one, unless some fine-tuning mechanism is present. 


\section*{Acknowledgments}

The authors thank Emanuele Mereghetti for helpful exchanges and discussions. E.V. would like to thank Pietro Baratella for useful discussions and comments on the draft. This work has been partially funded by the Deutsche Forschungs-gemeinschaft (DFG, German Research Foundation) under Germany's Excellence Strategy- EXC-2094 - 390783311, by the Collaborative Research Center SFB1258, and the BMBF grant 05H18WOCA1. We warmly thank the Munich Institute for Astro- and Particle Physics (MIAPP) for hospitality, which is also funded under Germany’s Excellence Strategy - EXC-2094 - 390783311.

\newpage

\appendix

\section{Relevant diagrams} \label{app:diags}

In this appendix we present all the relevant diagrams needed to calculate all the 1-loop contributions to the (c)EDMs. Since the diagrams contributing to the EDMs are the same for any fermion species and flavor we collectively denote them as $f$. Additionally, because of possible $W$ bosons in the loop the fermion running in the loop is not necessarily of the same flavor as the external ones, hence they are denoted by $f'$

In all diagrams the black dot denotes the insertion of any of the dimension-6 operators possible at any given position. Since we are performing all calculations in the phase of broken EW symmetry we denote Higgs fields that are set to their VEV by scalar legs ending in a cross. 

Notice that we do not show diagrams that could in principle contribute but vanish due to scaleless integrals or antisymmetry. Additionally, we do not show diagrams of 3-point functions with only scalars in the loop, as these would contribute only at higher orders in the fermion masses.

\begin{figure}[H]
  \centering
\setkeys{Gin}{width=0.8\textwidth}
  \includegraphics{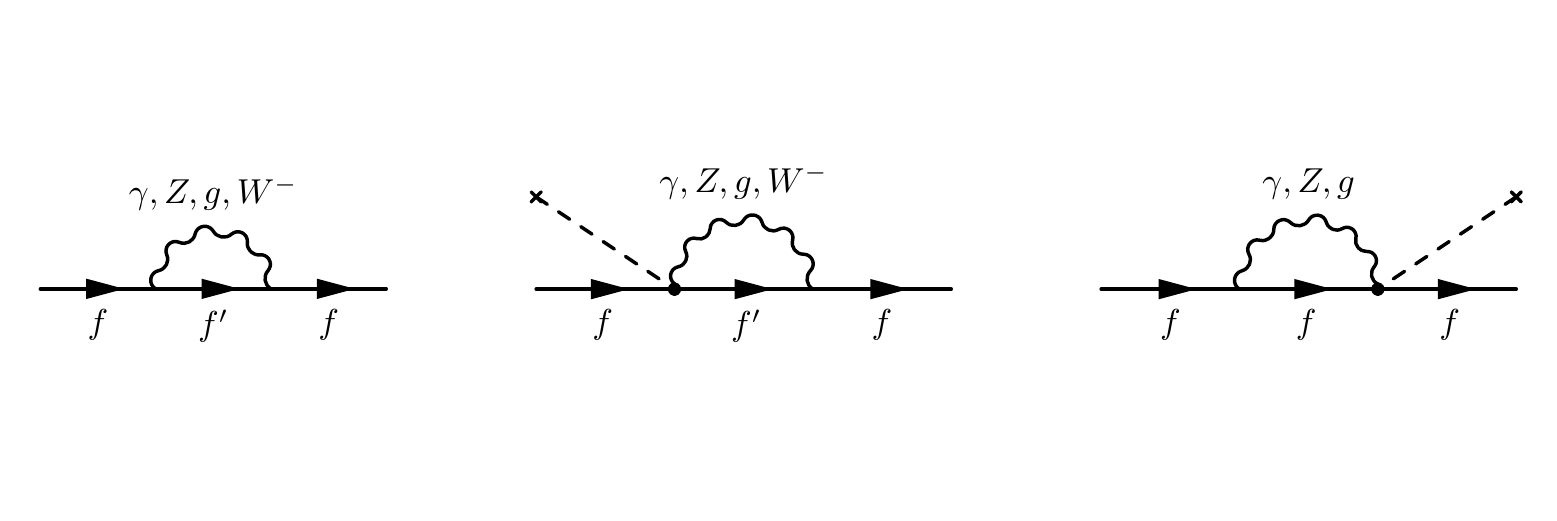}
  \caption{Diagrams contributing to the fermion 2-point function.}
  \label{fig:Fermion_2pt}
\end{figure}
 
\begin{figure}[H]
  \centering
\setkeys{Gin}{width=0.8\textwidth}
  \includegraphics{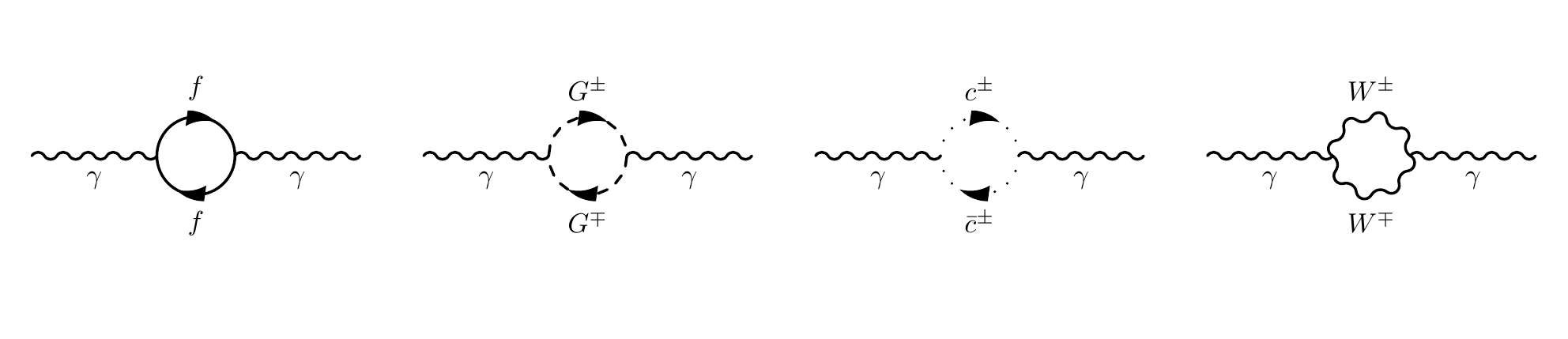}\hspace{0.5cm}%
\setkeys{Gin}{width=0.7\textwidth}
  \includegraphics{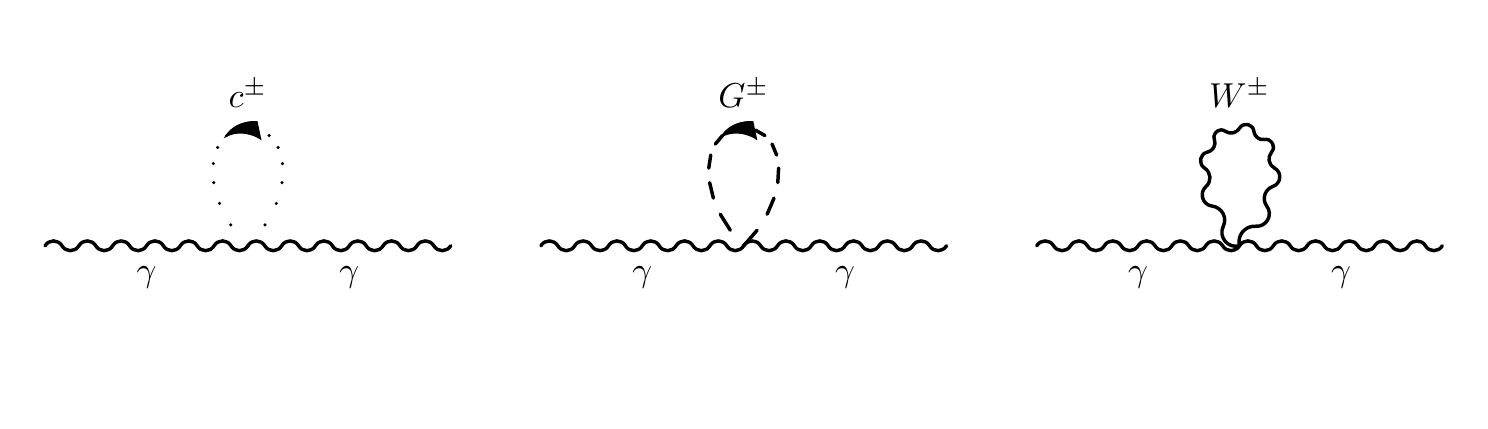}\hspace{0.5cm}%

  \caption{Diagrams contributing to the photon 2-point function.}
  \label{fig:Photon_2pt}
\end{figure}

\begin{figure}[H]
  \centering
\setkeys{Gin}{width=0.8\textwidth}
  \includegraphics{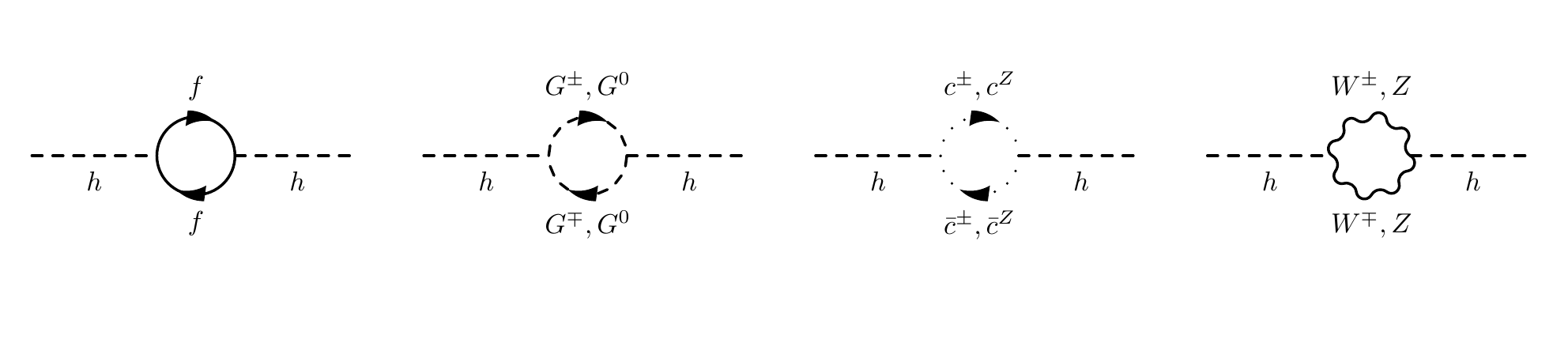}\hspace{0.5cm}%
  \includegraphics{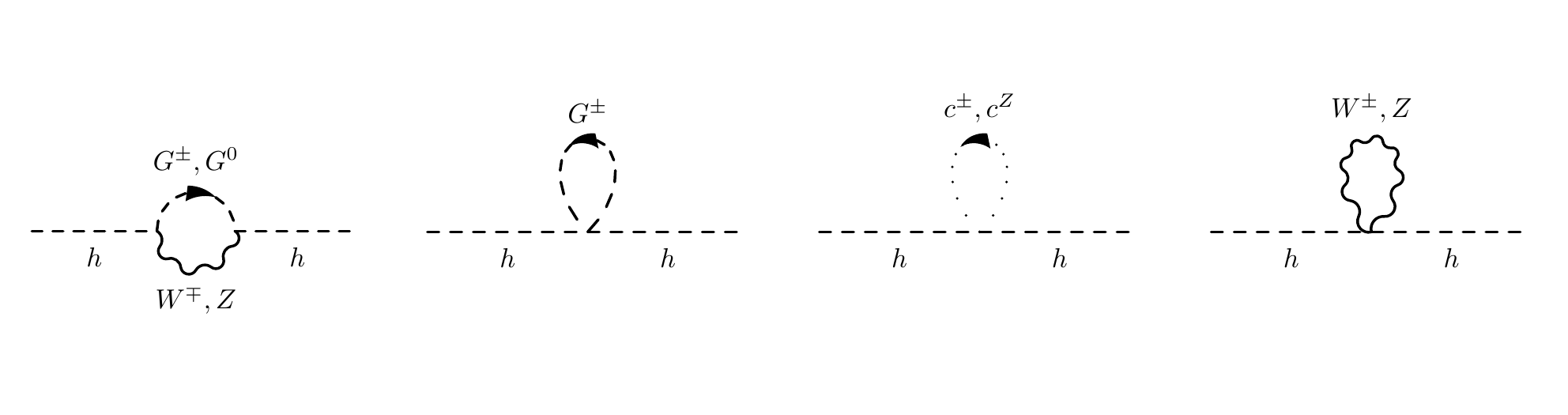}\hspace{0.5cm}%

  \caption{Diagrams contributing to the Higgs 2-point function.}
  \label{fig:Higgs_2pt}
\end{figure}

\begin{figure}[H]
  \centering
\setkeys{Gin}{width=0.8\textwidth}
  \includegraphics{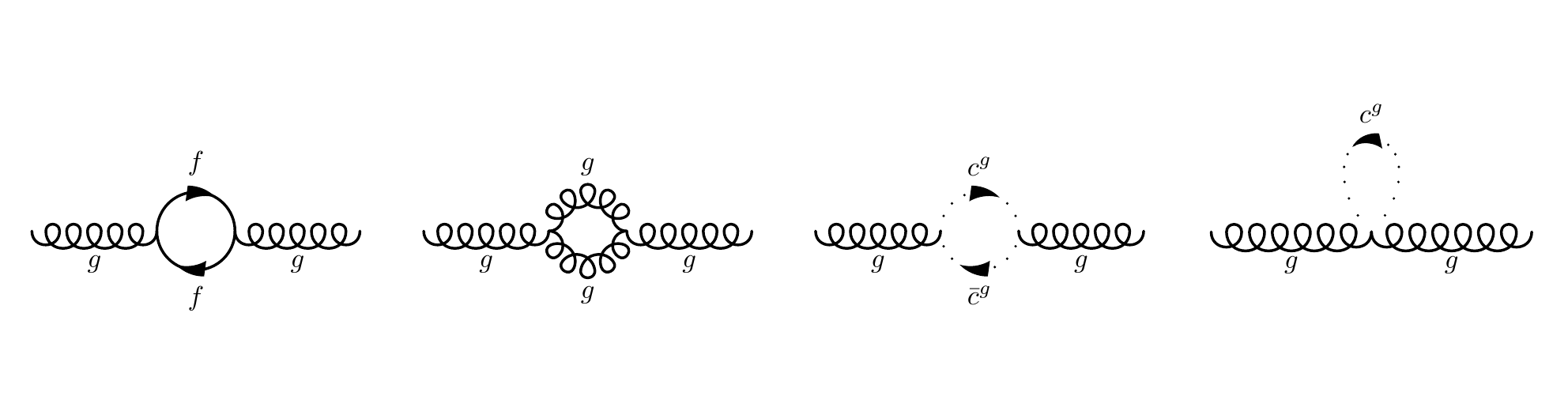}\hspace{0.5cm}%

  \caption{Diagrams contributing to the gluon 2-point function.}
  \label{fig:Gluon_2pt}
\end{figure}

\begin{figure}[H]
  \centering
\setkeys{Gin}{width=0.25\linewidth}
  \includegraphics{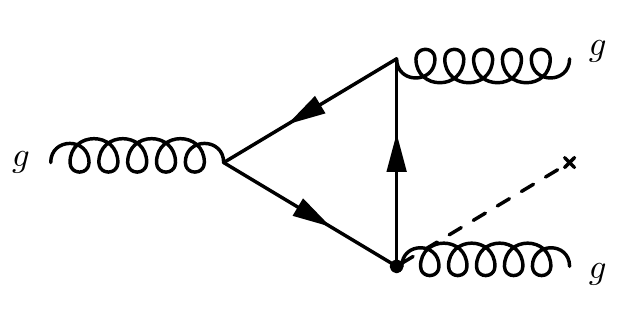}\hspace{0.5cm}%
  \includegraphics{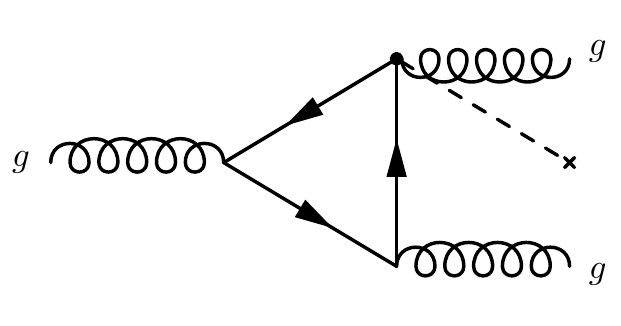}\hspace{0.5cm}%
  \includegraphics{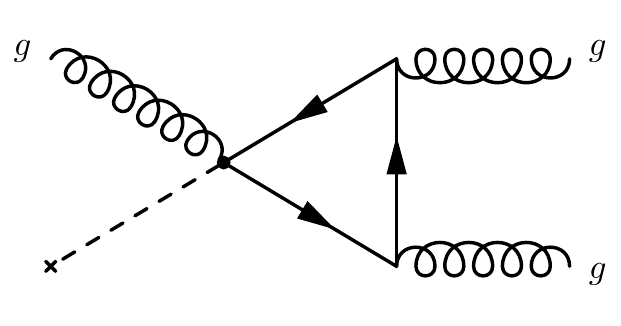}

  \vspace{.5cm}

  \includegraphics{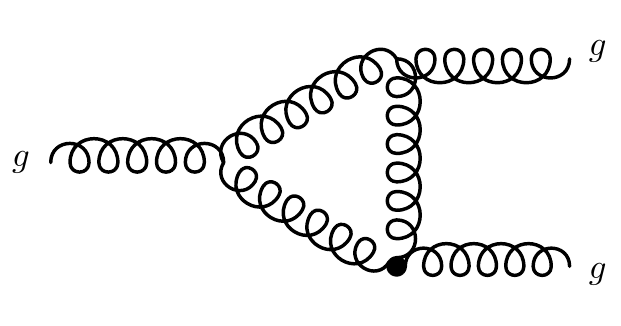}\hspace{0.5cm}%
  \includegraphics{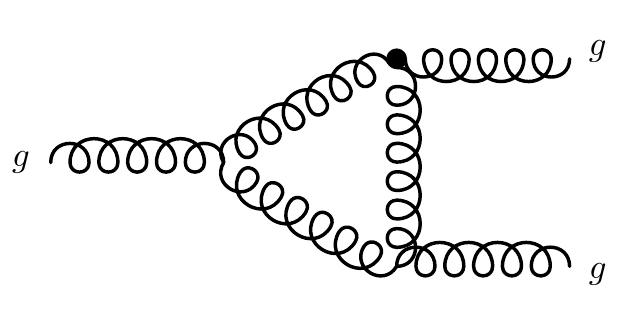}\hspace{0.5cm}%
  \includegraphics{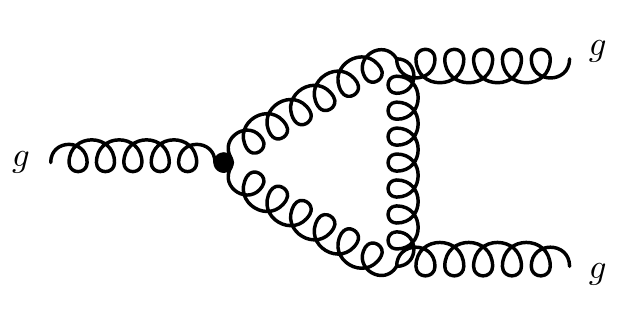}
  
  \caption{1PI diagrams contributing to the $ggg$ 3-point function. Of course, there are also diagrams with only two propagators for both the insertion of the Weinberg and the dipole operator, but we find that these vanish, so we do not display them here. We also do not show diagrams with the external gluons attached to an SM vertex crossed.}
  \label{fig:Gluon_3pt}
\end{figure}

\begin{figure}[H]
  \centering
\setkeys{Gin}{width=0.7\textwidth}
  \includegraphics{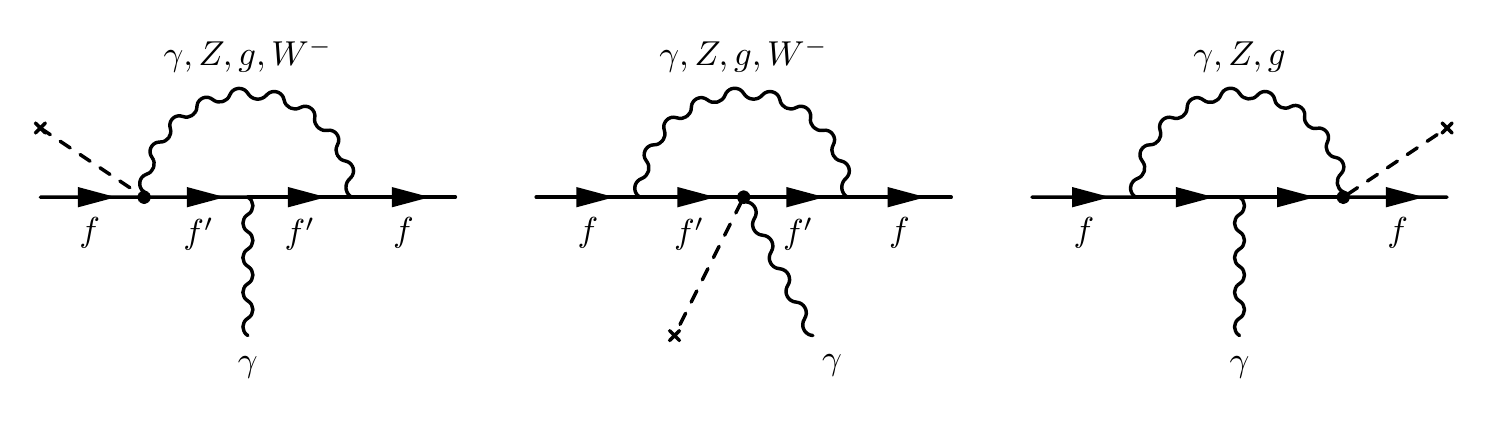}\hspace{0.5cm}%
  \includegraphics{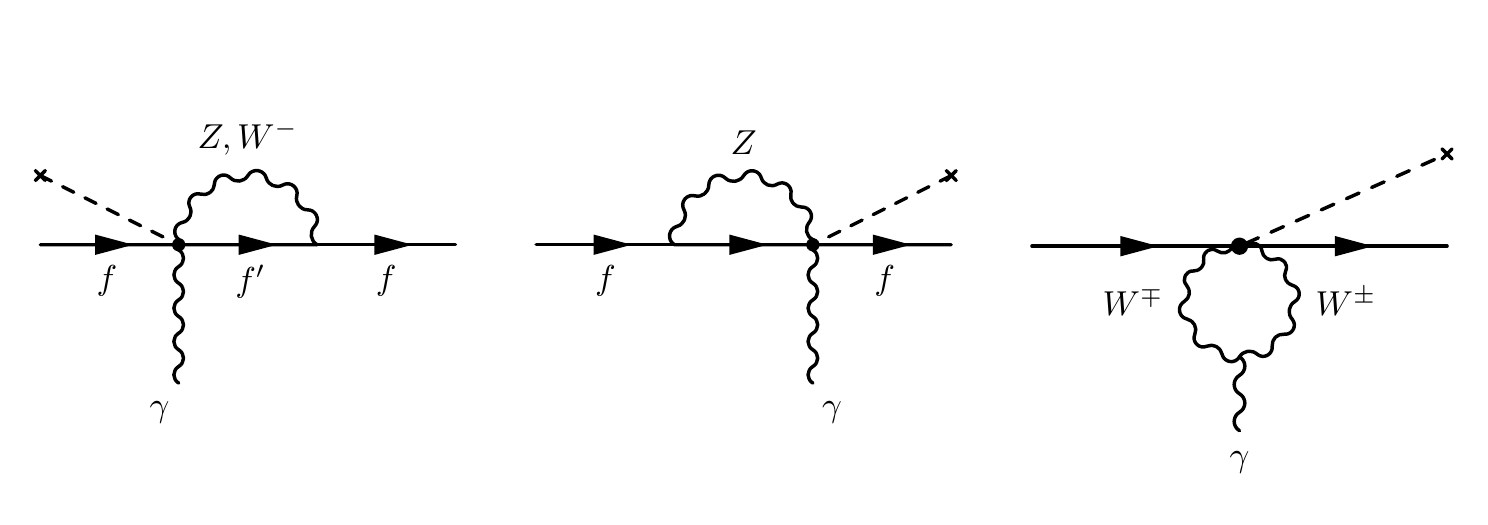}\hspace{0.5cm}%
\setkeys{Gin}{width=0.5\textwidth}
  \includegraphics{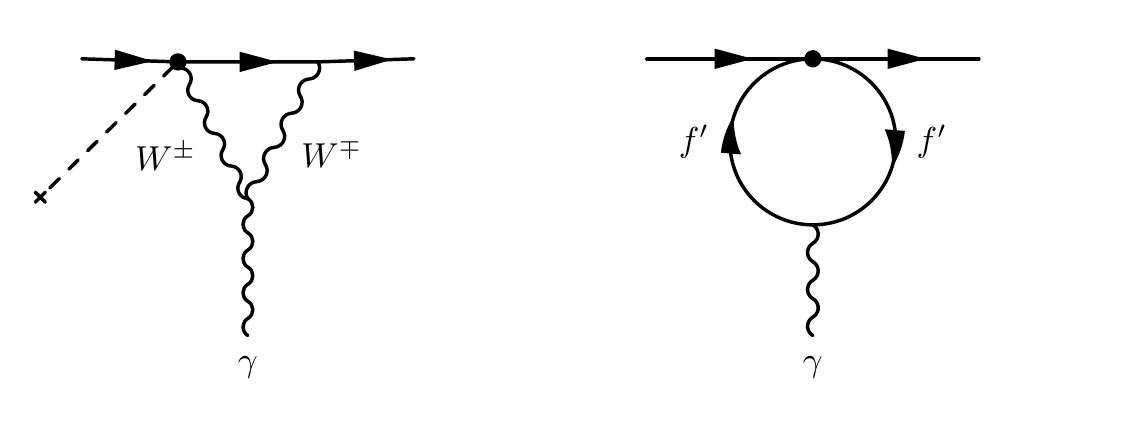}\hspace{0.5cm}%
\setkeys{Gin}{width=0.7\textwidth}
  \includegraphics{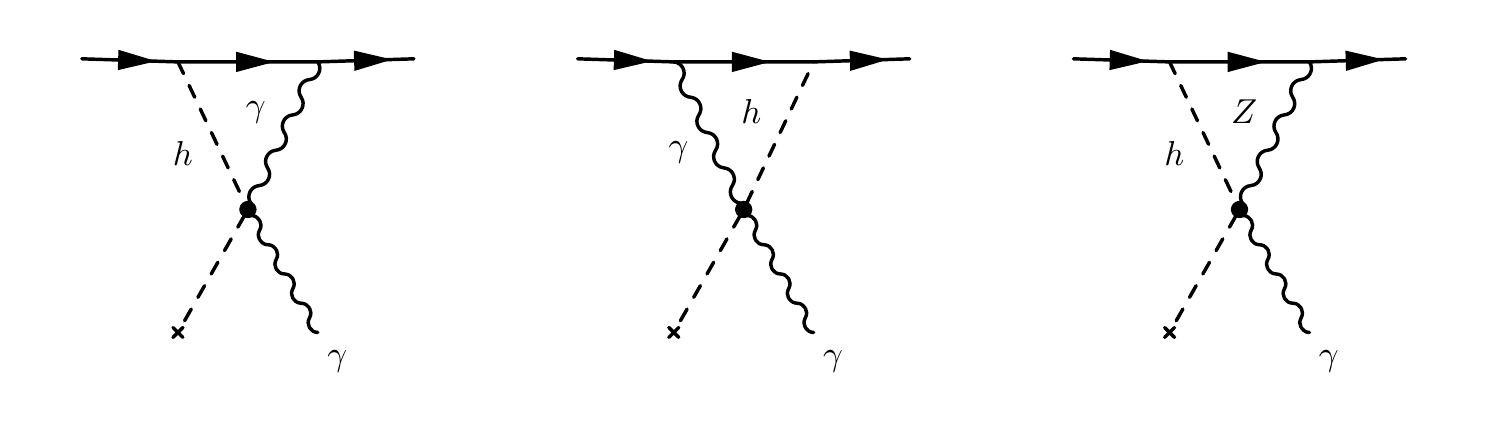}\hspace{0.5cm}%
  \includegraphics{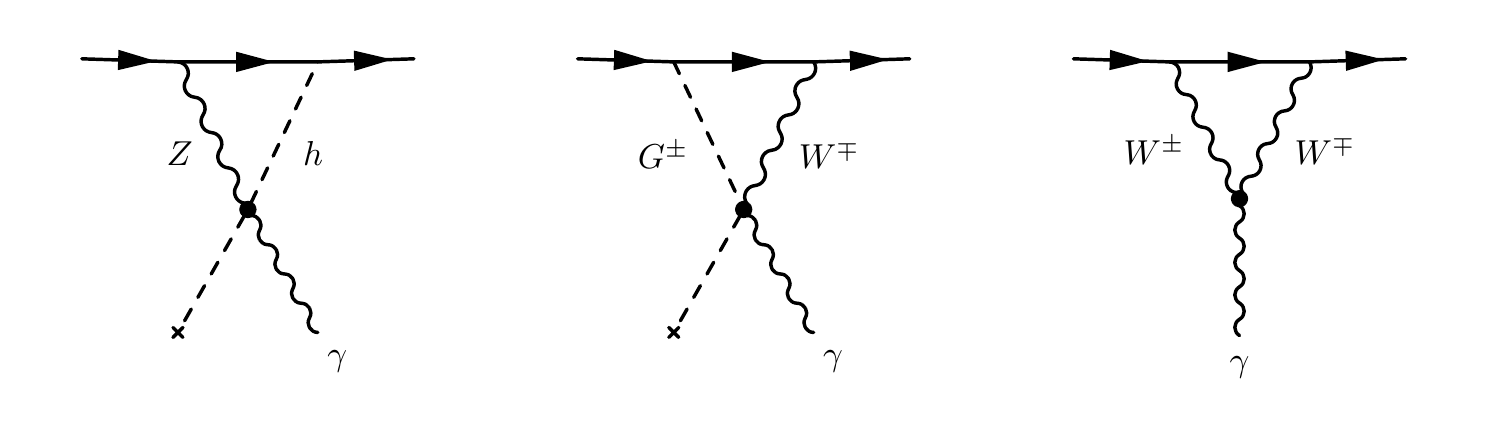}\hspace{0.5cm}%

  \caption{1PI diagrams contributing to the $\bar{\psi}\psi\gamma$ 3-point function. Notice that the diagram with the lepton loop exists only for external up-type quarks and leptons.}
  \label{fig:Electron_3pt}
\end{figure}

\begin{figure}[H]
  \centering
\setkeys{Gin}{width=0.7\textwidth}
  \includegraphics{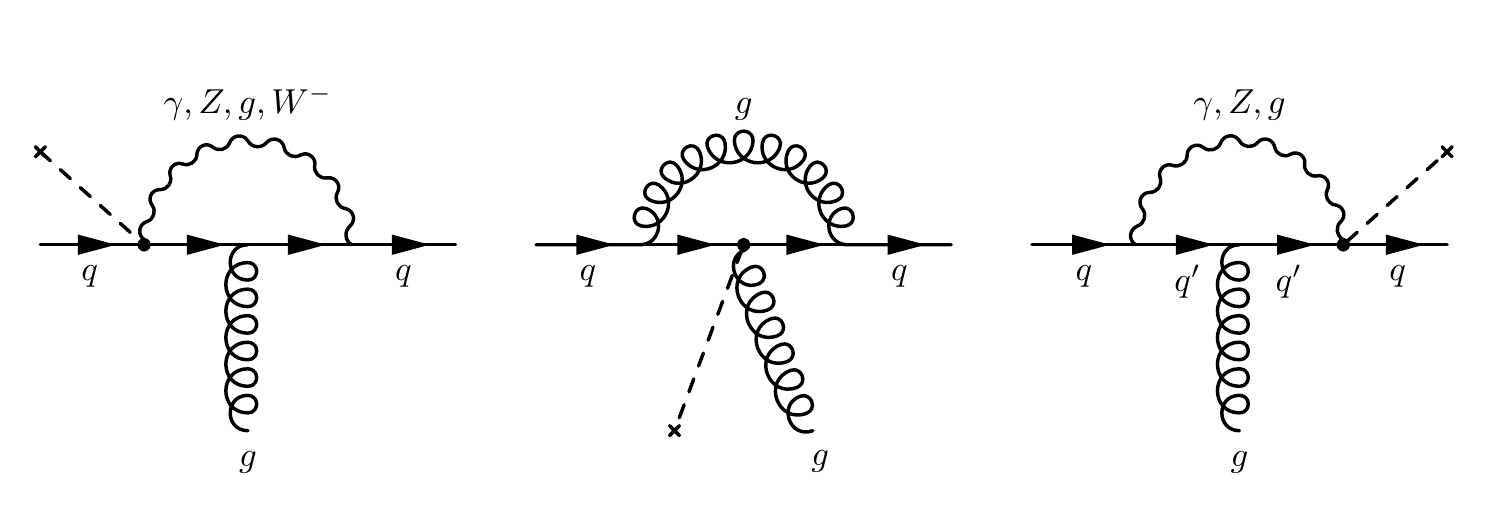}\hspace{0.5cm}%
\setkeys{Gin}{width=0.5\textwidth}
  \includegraphics{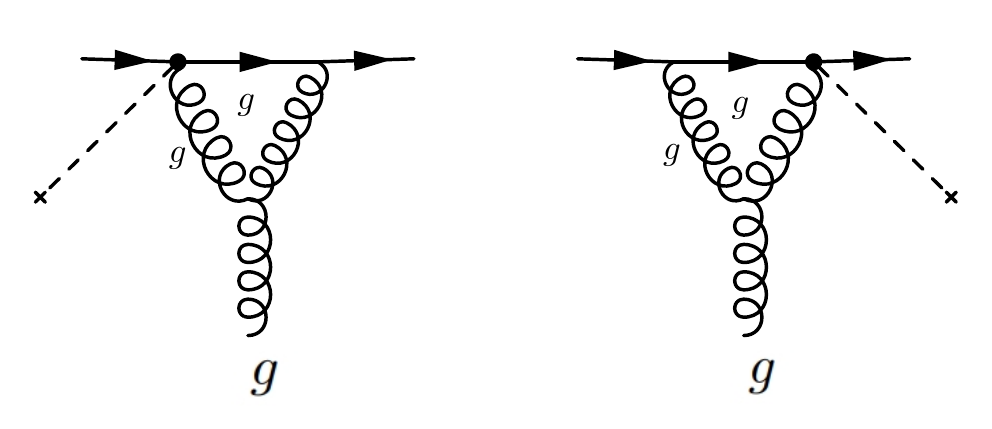}\hspace{0.5cm}%
  \includegraphics{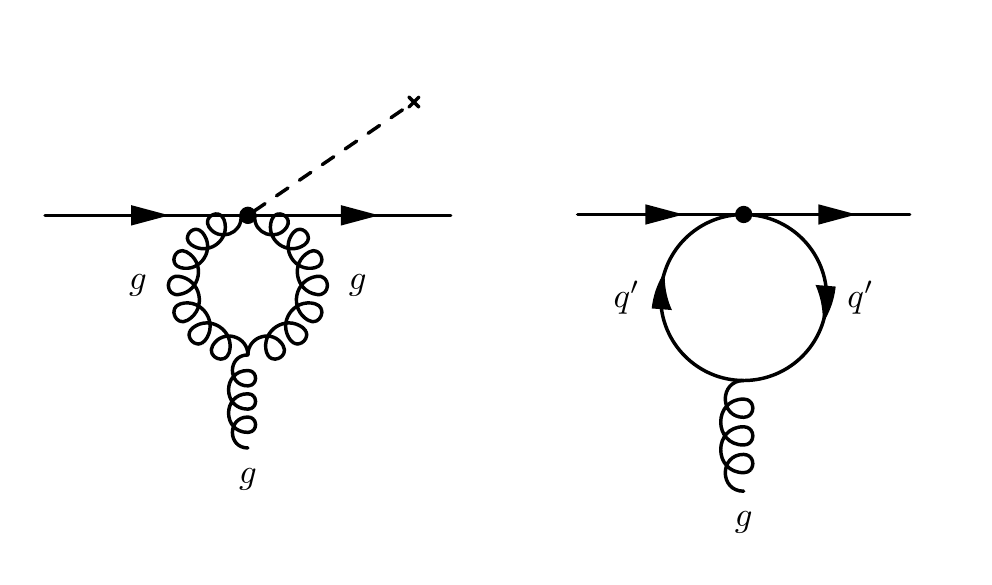}\hspace{0.5cm}%
\setkeys{Gin}{width=0.7\textwidth}
  \includegraphics{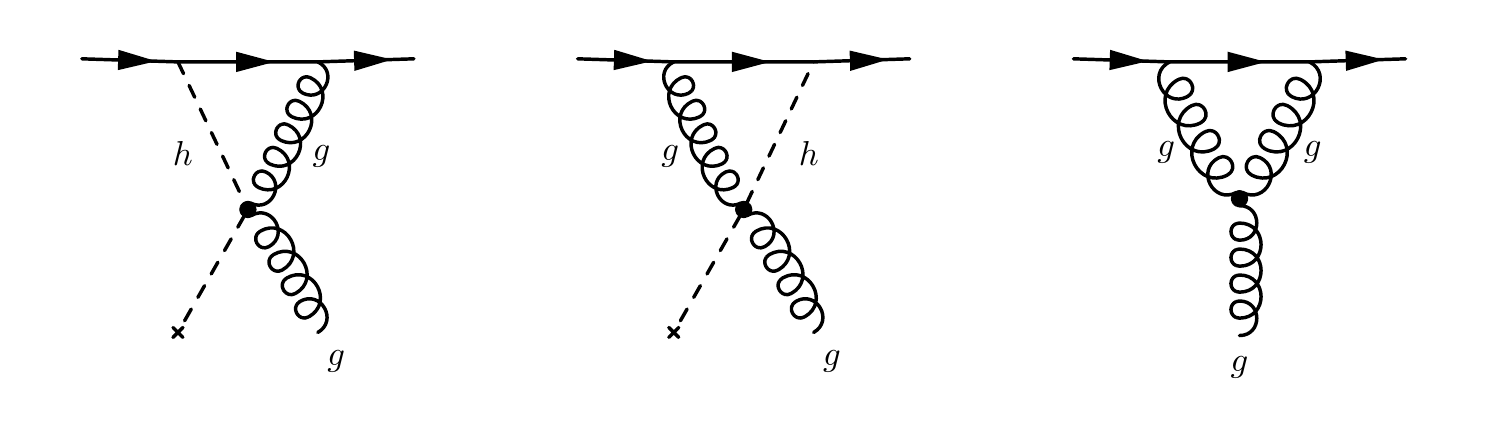}\hspace{0.5cm}%

  \caption{1PI diagrams contributing to the $\bar{q}qg$ 3-point function.}
  \label{fig:Quark_3pt}
\end{figure}

\begin{figure}[H]
    \centering
  \setkeys{Gin}{width=0.23\linewidth}
    \includegraphics{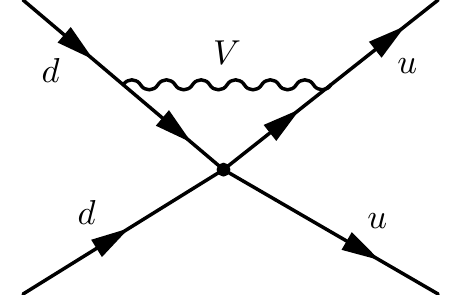}\hspace{0.5cm}
    \includegraphics{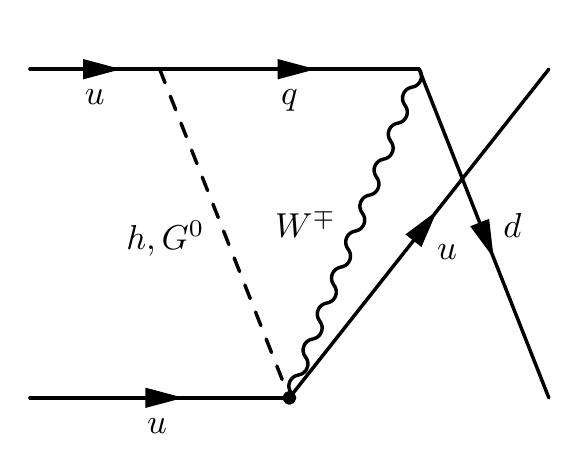}
    \includegraphics{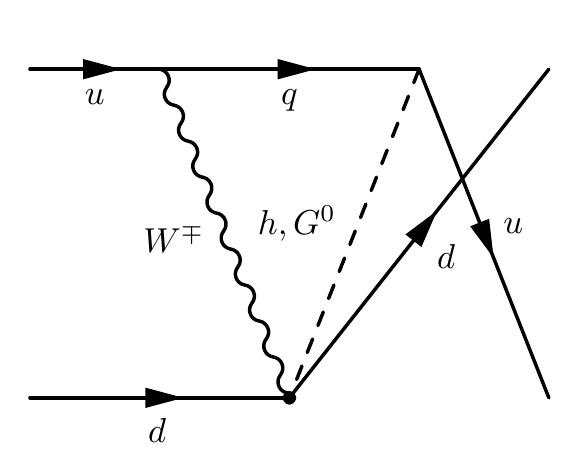}

    \includegraphics{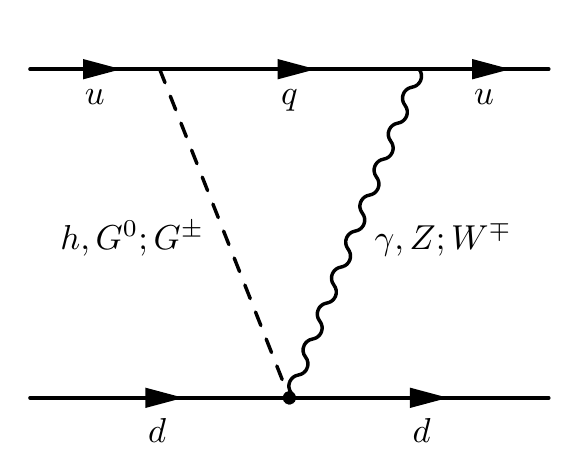}%
    \includegraphics{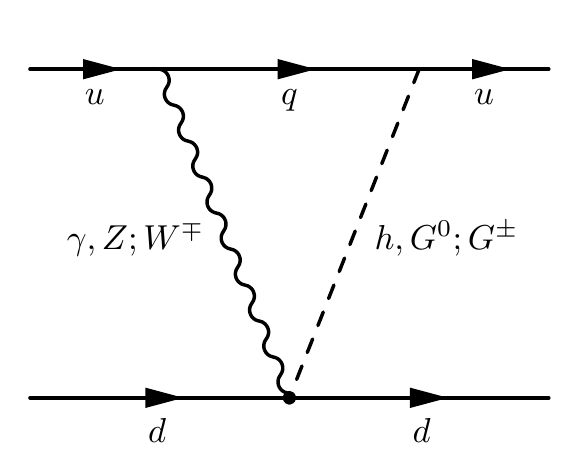}%

    \caption{1PI diagrams contributing to the $udud$ 4-point function. For diagrams contributing to the self-renormalization, we show only one representative diagram, all the others can be obtained by connecting all possible pairs of external fermions with the internal vector. The other diagrams show the contribution of the down-type dipole operators. The corresponding up-dipole diagrams can be obtained by just exchanging up and down quarks. }
\end{figure}

\begin{figure}[H]
    \centering
    \setkeys{Gin}{width=0.9\textwidth}
    \includegraphics{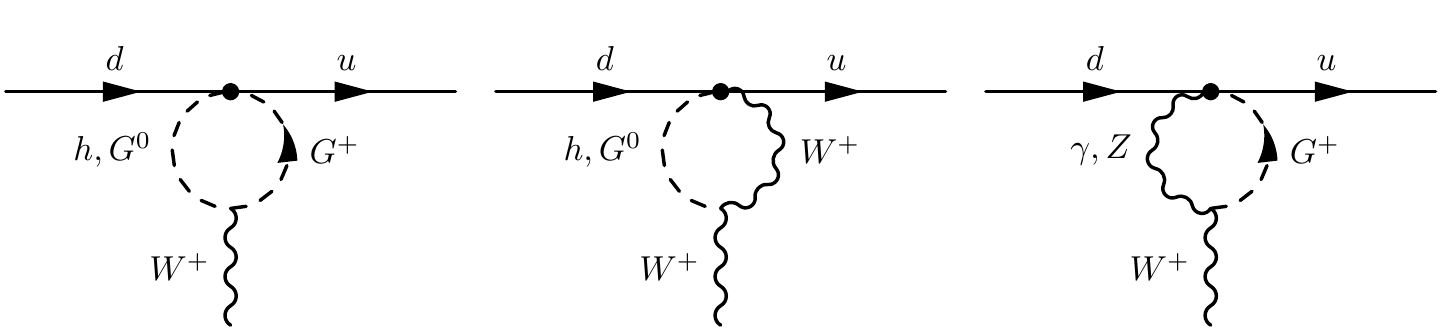}\vspace{0.5cm}
    \includegraphics{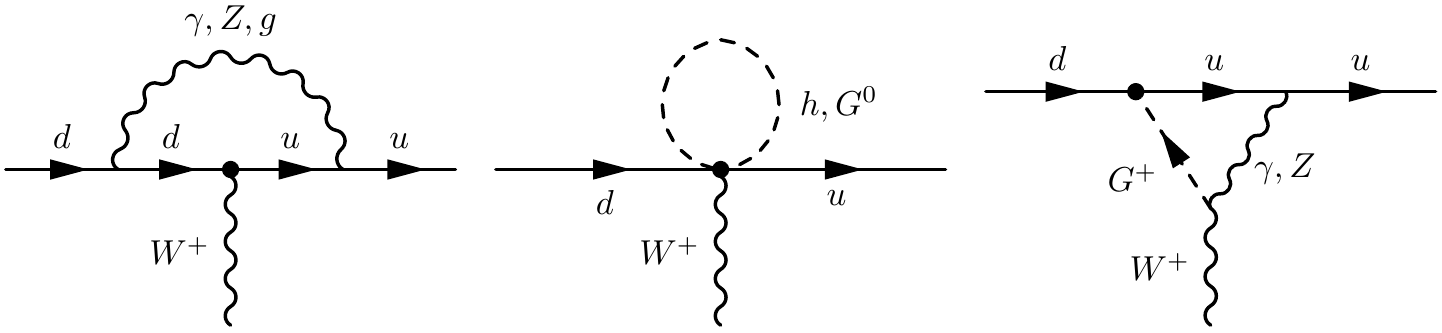}\vspace{0.5cm}
    \includegraphics{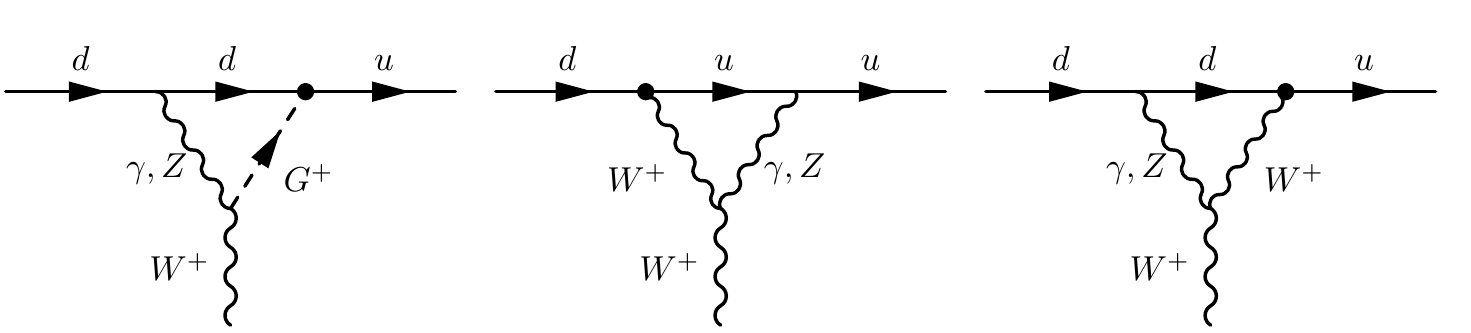}
    \caption{1PI diagrams contributing to the $udW^+$ 4-point function which were used to calculate the self-renormalization of $O_{Hud}$.}
\end{figure}

\begin{figure}[H]
    \centering
    \setkeys{Gin}{width=0.9\textwidth}
    \includegraphics{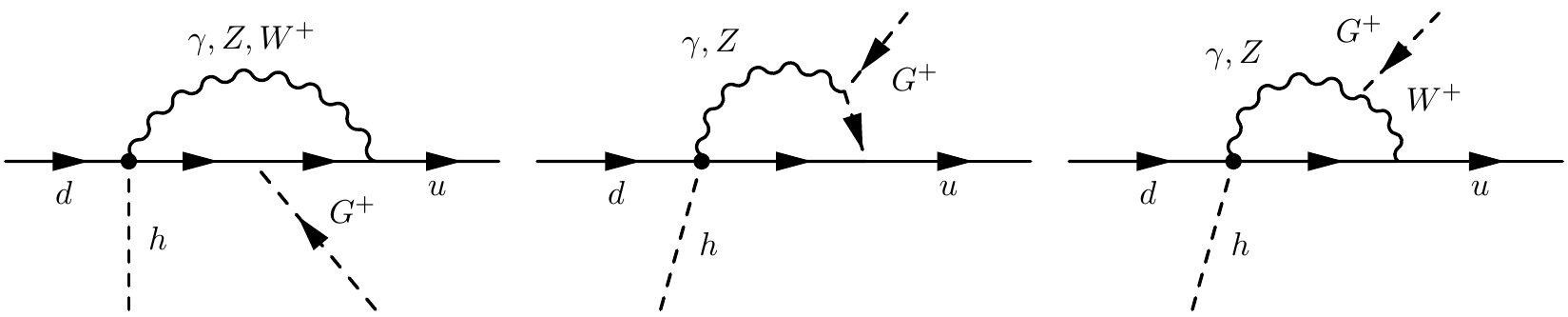}\vspace{0.5cm}
    \includegraphics{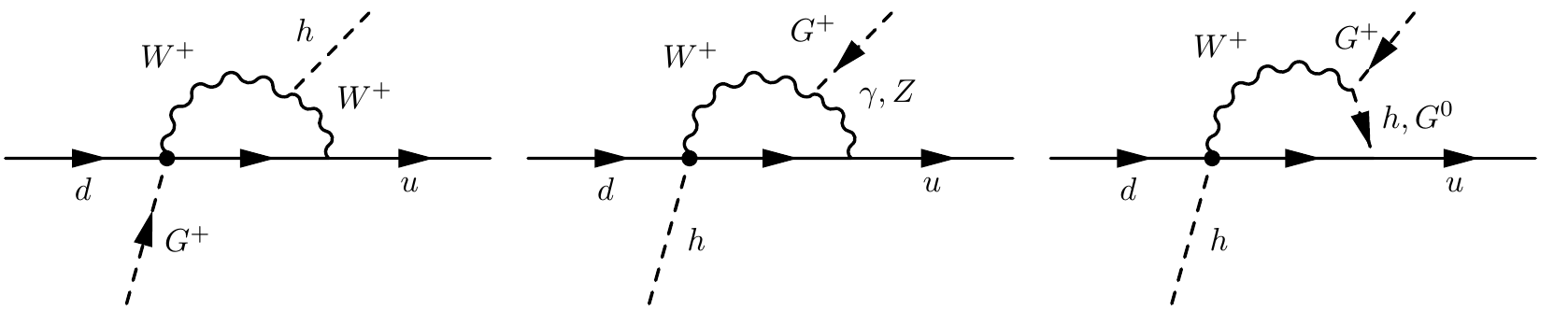}
    \caption{1PI diagrams contributing to the $udhG^+$ 4-point function. We only show the contribution of the down-type dipole. Furthermore, additional diagrams can be generated by exchanging $h\leftrightarrow G^+$.}
\end{figure}

\clearpage

\section{Analytic expressions of various EDMs} \label{app:expressions}

In this appendix we report the analytic expressions computed in this work. To improve readability we divide the full expressions into categories defined by the field content of the operators contributing to the dipole. Because we give the expression of the observable EDM we repeat here its relation to the Wilson coefficient $c_{f\gamma}$, of the operator $\bar{f_L}\sigma^{\mu\nu}f_R F_{\mu\nu}$,
\begin{equation}
d_f = -\frac{2}{\Lambda^2}\text{Im}\,c_{f\gamma},
\end{equation}
and similar for the chromo-dipoles.

\subsection{Universal contributions}

Since the full expression of the fermion (c)EDMs is rather long, we will start by providing their universal parts first. Apart from the term proportional to $g_s$, \Eq{eq:FermionWFR_Strong}, which is present only for quark dipoles, these are universal in the sense that they correspond to pure SM loops on the external particle 2-point functions and are independent of the fermion species and therefore enter all dipoles in the same way. This includes both the renormalization of the Higgs VEV, which in this work is given by just the loops in the physical Higgs 2-point function, as well as the mixing of the neutral gauge bosons at 1-loop.

All these contributions are:

\begin{itemize}
  \item Loops on external left-handed (LH) and right-handed (RH) fermions: 
  \begin{subequations}
    \begin{align}
    \hspace{-.5cm}
    16\pi^2\times\text{(LH Fermion 2-pt.)}_f &= 2\,e^2Q_f^2-\frac{e^2}{4s_w^2}-\frac{e^2}{2s_w^2c_w^2}(T^3_f-Q_f s_w^2)^2 \label{eq:FermionWFR_L_Weak}\\[2ex]
    &+2\,e^2 Q_f^2 \log\left(\frac{\Lambda}{m_f}\right)+\frac{e^2}{2s_w^2}\log\left(\frac{\Lambda}{m_W}\right)\nonumber\\[2ex]
    &+\frac{e^2}{s_w^2c_w^2}(T^3_f-Q_f s_w^2)^2\log\left(\frac{\Lambda}{m_Z}\right)\nonumber\\[2ex]
    &+2\,c_{F,3}g_s^2+2\,c_{F,3} g_s^2\log\left(\frac{\Lambda}{m_f}\right)\label{eq:FermionWFR_L_Strong}
    \end{align}
    \begin{align}
    \hspace{-.5cm}
    16\pi^2\times\text{(RH Fermion 2-pt.)}_f &= 2\,e^2Q_f^2-\frac{e^2Q_f^2t_w^2}{2}+2\,e^2 Q_f^2 \log\left(\frac{\Lambda}{m_f}\right)\label{eq:FermionWFR_R_Weak}\\[2ex]
    &+e^2Q_f^2t_w^2\log\left(\frac{\Lambda}{m_Z}\right)\nonumber\\[2ex]
    &+2\,c_{F,3} g_s^2 + 2\,c_{F,3} g_s^2\log\left(\frac{\Lambda}{m_f}\right)\label{eq:FermionWFR_R_Strong}
    \end{align}
  \end{subequations}
  \item Loop contributions to the Higgs VEV:  
    \begin{align}
    \hspace{-.5cm}
    16\pi^2\times\text{Higgs 2-pt.}&=\frac{4\,N_c m_t^2}{v^2}-\frac{4\,e^2}{s_w^2}-\frac{2\,e^2}{s_w^2c_w^2}\label{eq:HiggsWFR}\\[2ex]
    &+\frac{4\,N_c m_t^2}{v^2}\log\left(\frac{\Lambda}{m_t}\right)-\frac{4\,e^2}{s_w^2}\log\left(\frac{\Lambda}{m_W}\right)\nonumber\\[2ex]
    &-\frac{2\,e^2}{s_w^2c_w^2}\log\left(\frac{\Lambda}{m_Z}\right)\nonumber\\[2ex]
    &-\frac{2N_c\,m_t^2}{v^2}\sqrt{\frac{4m_t^2-m_h^2}{m_h^2}}\arctan\left(m_h\sqrt{\frac{4m_t^2-m_h^2}{\left(2m_t^2-m_h^2\right)^2}}\right)\nonumber\\[2ex]
    &+\frac{2\,e^2}{s_w^2}\sqrt{\frac{4m_W^2-m_h^2}{m_h^2}}\left[\arctan\left(m_h\sqrt{\frac{4m_W^2-m_h^2}{\left(2m_W^2-m_h^2\right)^2}}\right)+\pi\right]\nonumber\\[2ex]
    &\left.+\frac{e^2}{s_w^2c_w^2}\sqrt{\frac{4m_Z^2-m_h^2}{m_h^2}}\arctan\left(m_h\sqrt{\frac{4m_Z^2-m_h^2}{\left(2m_Z^2-m_h^2\right)^2}}\right)\right\}\nonumber
    \end{align}
  \item Loops on external photons: 
    \begin{align}
    \hspace{-.5cm}
    16\pi^2\times\text{Photon 2-pt.}=&-\frac{2\,e^2}{3}-14\,e^2\log\left(\frac{\Lambda}{m_W}\right)\label{eq:PhotonWFR}\\[2ex]
    &+\frac{8}{3}\sum_{\text{fermions}}\left(\delta_{i\ell}+N_c \delta_{iq}\right)e^2Q_i^2\log\left(\frac{\Lambda}{m_i}\right)\nonumber
    \end{align}
  \item Photon-Z mixing:
    \begin{align}
    16\pi^2\times\text{Photon-Z Mixing}&=-\frac{2e^2}{3t_w}-\frac{1+42c_{w}^2}{6s_wc_w}e^2\log\left(\frac{\Lambda}{m_Z}\right)\label{eq:Mixing}\\[2ex]
    &+\frac{4}{3}\frac{e^2}{s_wc_w}\sum_{i\neq t}\left(\delta_{i\ell}+N_c \delta_{iq}\right)Q_i (T^3_i-2 Q_is_w^2)\log\left(\frac{\Lambda}{m_Z}\right)\nonumber\\[1ex]
    &+\frac{4}{3}\frac{N_c e^2}{s_wc_w} Q_u (T^3_u-2 Q_us_w^2)\log\left(\frac{\Lambda}{m_t}\right)\nonumber
    \end{align}
  \item Loops on external gluons: 
    \begin{align}
    \hspace{-.5cm}
    16\pi^2\times\text{Gluon 2-pt.}&=-\frac{67N_c}{9}g_s^2-\frac{22}{9}N_c g_s^2\log\left(\frac{\Lambda}{\mu_H}\right)+\frac{4}{3}g_s^2 \sum_{q=u,d,s,c}\log\left(\frac{\Lambda}{\mu_H}\right)\label{eq:GluonWFR}\\[2ex]
    &\left.+\frac{4}{3}g_s^2\log\left(\frac{\Lambda}{m_b}\right)+\frac{4}{3}g_s^2\log\left(\frac{\Lambda}{m_t}\right)\nonumber\right\}
    \end{align}
\end{itemize}

\subsection{Lepton EDMs}

We start with showing the results for lepton EDMs. Note that the logs arising from the divergent terms of the photon wave function renormalization do not necessarily run down to the mass of the fermion running in the loop but only to the mass of the external lepton if the latter is heavier than the former.

\paragraph{Contributions from $\psi^2HF$ operators} 

\begin{subequations}
\begin{align}
\hspace{-2cm}\frac{d_{\ell}}{e} \times (4\pi\Lambda)^2 \supset \Im{c_w C_{\underset{11}{eB}}+2~T_{\ell}^3s_w C_{\underset{11}{eW}}}&\left\{-\frac{16\sqrt{2}\,\pi^2\,v}{e}+4\sqrt{2}\,eQ_{\ell}^2v+8\sqrt{2}\,eQ_{\ell}^2v\log\left(\frac{\Lambda}{m_{\ell}}\right)\right. \label{eq:eEDM_Tree1PI}\\[2ex]
&\hspace{-1.5cm}+\frac{v}{\sqrt{2}e}\Big(\Eq{eq:FermionWFR_L_Weak}_{\ell}+\Eq{eq:FermionWFR_R_Weak}_{\ell}+\Eq{eq:HiggsWFR}+\Eq{eq:PhotonWFR}\Big)\Bigg\}\nonumber\\[2ex]
+\Im{-s_w C_{\underset{11}{eB}}+2~T_{\ell}^3c_wC_{\underset{11}{eW}}}&\left\{\sqrt{2}\,eQ_{\ell}\,v\frac{T_{\ell}^3-2Q_{\ell}s_w^2}{s_wc_w}\left[\frac{1}{2}+\log\left(\frac{\Lambda}{m_Z}\right)\right]\right.\label{eq:eEDM_Z1PI}\\[2ex]
&+\frac{v}{\sqrt{2}e}\Big(\Eq{eq:Mixing}\Big)\Bigg\}\nonumber\\[2ex]
-\Im{C_{\underset{11}{eW}}}&\left\{\frac{ev(5+Q_{\ell})}{2\sqrt{2}s_w}-\frac{\sqrt{2}\,ev(3Q_{\ell}-1)}{s_w}\log\left(\frac{\Lambda}{m_W}\right)\right\}\label{eq:eEDM_W1PI}.
\end{align}
\end{subequations}

\paragraph{Contributions from $H^2F^2$ operators} 

\begin{subequations}
\begin{align}
\frac{d_{\ell}}{e}\times (4\pi\Lambda)^2 \supset & - m_{\ell}\left\{3\,(2\,Q_{\ell}-T_{\ell}^3)+(8\,Q_{\ell}-4T_{\ell}^3)\log\left(\frac{\Lambda}{m_h}\right)\right.\label{eq:eH2X2_1}\\[2ex]
&\hspace{0.8cm}\left.+(2 s_w^2 Q_{\ell}-T_{\ell}^3)\frac{4m_Z^2}{m_Z^2-m_h^2}\log\left(\frac{m_h}{m_Z}\right)\right\}\,C_{H\widetilde{B}}\nonumber\\[2ex]
&-m_{\ell}\left\{3~ T_{\ell}^3+4T_{\ell}^3\log\left(\frac{\Lambda}{m_h}\right)\right.\label{eq:eH2X2_2}\\[2ex]
&\hspace{0.8cm}\left.-(2 s_w^2 Q_{\ell}-T_{\ell}^3)\frac{4m_Z^2}{m_Z^2-m_h^2}\log\left(\frac{m_h}{m_Z}\right)\right\}\,C_{H\widetilde{W}}\nonumber\\[2ex]
&+m_{\ell}\left\{\frac{6~Q_{\ell} s_w^2-c_w^2 + 3~ T_{\ell}^3 c_{2w}}{2 c_w s_w}\right.\label{eq:eH2X2_3}\\[2ex]
&\hspace{0.8cm}\left.+\frac{4 s_w^2 Q_{\ell}+2T_{\ell}^3c_{2w}}{c_w s_w}\log\left(\frac{\Lambda}{m_h}\right)-\frac{2}{t_w}\log\left(\frac{\Lambda}{m_W}\right)\right.\nonumber\\[2ex]
&\hspace{0.8cm}\left.-(2 s_w^2 Q_{\ell}-T_{\ell}^3)\frac{c_{2w}}{c_w s_w}\frac{2m_Z^2}{m_Z^2-m_h^2}\log\left(\frac{m_h}{m_Z}\right)\right\}\,C_{HW\widetilde{B}}\nonumber
\end{align}
\end{subequations}

\paragraph{Contributions from $F^3$ operators} 
\begin{equation}
\frac{d_{\ell}}{e}\times (4\pi\Lambda)^2 \supset -\frac{3}{2}\frac{e\, m_{\ell}}{s_w} C_{\widetilde{W}}
\end{equation}

\paragraph{Contributions from $\psi^4$ operators} 
\begin{equation}
\frac{d_{\ell}}{e}\times (4\pi\Lambda)^2 \supset 16N_c Q_u \sum_{i\in\{1,2,3\}}m_{u,i} \,\log\left(\frac{\Lambda}{m_{u,i}}\right) \Im{C^{\text{\tiny{(3)}}}_{\underset{11 ii}{lequ}}}
\end{equation}

\paragraph{Contributions from $\psi^2\bar{\psi}^2$ operators} 
\begin{equation}
\frac{d_{\ell}}{e}\times (4\pi\Lambda)^2 \supset -2Q_e\sum_{i \in \{2,3\} }m_{\ell,i}\,\Im{C_{\underset{1 ii 1}{le}}}
\label{eq:ole}
\end{equation}

\subsection{Quark EDMs}

We show here the results for the quark EDMs; for the scale in the logs of the photon 2-point function, the same discussion as in the case of the lepton EDMs applies. \add{Furthermore, $\mu_H \sim \mathcal{O}(\text{GeV})$ denotes the hadronic scale. We define the following frequently used combination}
\begin{equation}\label{eq:FermionWFR_LR_StrongWeak}
    \text{(LH+RH quark WFR)}_q = \Eq{eq:FermionWFR_L_Weak}_q+\Eq{eq:FermionWFR_L_Strong}_q+\Eq{eq:FermionWFR_R_Weak}_q+\Eq{eq:FermionWFR_R_Strong}_q
\end{equation}
where the subscript $q=u,d$ denotes the type of quark.

\paragraph{Contributions from $\psi^2HF$ operators} 

\begin{subequations}
\begin{align}
\hspace{-1cm}\frac{d_q}{e}\times (4\pi\Lambda)^2 \supset \Im{c_w C_{\underset{11}{qW}}+2~T_q^3s_w C_{\underset{11}{qW}}}&\left\{-\frac{16\sqrt{2}\,\pi^2\,v}{e}+4\sqrt{2}\,eQ_{q}^2v+8\sqrt{2}\,eQ_{q}^2v\log\left(\frac{\Lambda}{\mu_H}\right)\right. \label{eq:qEDM_Tree1PI}\\[2ex]
&\hspace{-2cm}+\frac{v}{\sqrt{2}e}\Big(\Eq{eq:FermionWFR_LR_StrongWeak}_q+\Eq{eq:HiggsWFR}+\Eq{eq:PhotonWFR}\Big)\nonumber\\[2ex]
+\Im{-s_w C_{\underset{11}{qB}}+2~T_q^3c_wC_{\underset{11}{qW}}}&\left\{\sqrt{2}eQ_qv\,\frac{T_q^3-2Q_qs_w^2}{s_wc_w}\left[\frac{1}{2}+\log\left(\frac{\Lambda}{m_Z}\right)\right]\right.\label{eq:qEDM_Z1PI}\\[2ex]
&+\frac{v}{\sqrt{2}e}\Big(\Eq{eq:Mixing}\Big)\nonumber\\[2ex]
+2T_q^3\Im{C_{\underset{11}{qW}}}&\left\{\frac{ev(5+Q_q)}{2\sqrt{2}s_w}-\frac{\sqrt{2}\,ev(3Q_q-1)}{s_w}\log\left(\frac{\Lambda}{m_W}\right)\right\}\label{eq:qEDM_W1PI}\\[2ex]
+\Im{C_{\underset{11}{qG}}}&\left\{2\sqrt{2}v\, c_{F,3} Q_q g_s+4\sqrt{2}v\, c_{F,3} Q_q g_s\log\left(\frac{\Lambda}{\mu_H}\right)\right\}.
\end{align}
\end{subequations}

\paragraph{Contributions from $H^2F^2$ operators}

\begin{subequations}
\begin{align}
\frac{d_q}{e}\times (4\pi\Lambda)^2 \supset & - m_q\left\{(6\,Q_q-3T_q^3)+(8\,Q_q-4T_q^3)\log\left(\frac{\Lambda}{m_h}\right)\right.\\[2ex]
&\hspace{0.8cm}\left.+(2 s_w^2 Q_q-T_q^3)\frac{4m_Z^2}{m_Z^2-m_h^2}\log\left(\frac{m_h}{m_Z}\right)\right\}\,C_{H\widetilde{B}}\nonumber\\[2ex]
&-m_q\left\{3\,T_q^3+4\,T_q^3\log\left(\frac{\Lambda}{m_h}\right)\right.\\[2ex]
&\hspace{0.8cm}\left.-(2 s_w^2 Q_q-T_q^3)\frac{4m_Z^2}{m_Z^2-m_h^2}\log\left(\frac{m_h}{m_Z}\right)\right\}\,C_{H\widetilde{W}}\nonumber\\[2ex]
&+m_q\left\{\frac{12~Q_q s_w^2-2 ~c_w^2 + 6~ T_q^3 c_{2w}}{4 c_w s_w}\right.\\[2ex]
&\hspace{0.8cm}\left.+\frac{4 s_w^2 Q_q+2T_q^3c_{2w}}{c_w s_w}\log\left(\frac{\Lambda}{m_h}\right)-\frac{2}{t_w}\log\left(\frac{\Lambda}{m_W}\right)\right.\nonumber\\[2ex]
&\hspace{0.8cm}\left.-(2 s_w^2 Q_q-T_q^3)\frac{c_{2w}}{c_w s_w}\frac{2m_Z^2}{m_Z^2-m_h^2}\log\left(\frac{m_h}{m_Z}\right)\right\}\,C_{HW\widetilde{B}}\nonumber
\end{align}
\end{subequations}

\paragraph{Contributions from $F^3$ operators} 
\begin{equation}
\frac{d_q}{e}\times (4\pi\Lambda)^2 \supset 2T_q^3 \frac{3}{2}\frac{e\, m_q}{s_w}~  C_{\widetilde{W}}
\end{equation}

\paragraph{Contributions from $\psi^4$ operators}

\begin{align}
\frac{d_d}{e}\times (4\pi\Lambda)^2 \supset 2\sum_{i\in\{1,2\}}m_{u,i} Q_u\,\log\left(\frac{\Lambda}{\mu_H}\right) \Im{C^{\text{\tiny{(1)}}}_{\underset{1ii1}{quqd}}+c_{F,3}\,C^{\text{\tiny{(8)}}}_{\underset{1ii1}{quqd}}} \\[2ex]
+2 m_t Q_u \,\log\left(\frac{\Lambda}{m_t}\right) \Im{C^{\text{\tiny{(1)}}}_{\underset{1331}{quqd}}+c_{F,3}\,C^{\text{\tiny{(8)}}}_{\underset{1331}{quqd}}} \nonumber
\end{align}
\begin{align}
\frac{d_u}{e}\times (4\pi\Lambda)^2 \supset&\,8\sum_{i\in\{1,2,3\}}m_{\ell,i} Q_{e}\,\log\left(\frac{\Lambda}{\mu_H}\right) \Im{C^{\text{\tiny{(3)}}}_{\underset{ii11}{lequ}}}
\\[2ex]
+&\,2\sum_{i\in\{1,2\}}m_{d,i} Q_d\,\log\left(\frac{\Lambda}{\mu_H}\right) \Im{C^{\text{\tiny{(1)}}}_{\underset{i11i}{quqd}}+c_{F,3}\,C^{\text{\tiny{(8)}}}_{\underset{i11i}{quqd}}}\nonumber \\[2ex]
+&\,2 m_{b} Q_d\,\log\left(\frac{\Lambda}{m_b}\right) \Im{C^{\text{\tiny{(1)}}}_{\underset{3113}{quqd}}+c_{F,3}\,C^{\text{\tiny{(8)}}}_{\underset{3113}{quqd}}}\nonumber
\end{align}

\paragraph{Contributions from $\psi\bar{\psi}H^2D$ operators} 

\begin{subequations}
\begin{align}
\frac{d_d}{e}\times (4\pi\Lambda)^2 \supset \sum_{i\in\{1,2\}}\frac{4m_i}{\sqrt{2}v}&(1+Q_u)\Im{C_{\underset{i1}{Hud}}
}\\[2ex]
+\frac{m_t}{\sqrt{2}v}&\left[\frac{m_t^4-11m_t^2m_W^2+4m_W^4}{(m_t^2-m_W^2)^2}\right.\nonumber\\[2ex]
+&Q_u\frac{m_t^4+m_t^2m_W^2+4m_W^4}{(m_t^2-m_W^2)^2}\nonumber\\[2ex]
+&\left.6m_t^2m_W^2\frac{m_t^2-Q_u m_W^2}{(m_t^2-m_W^2)^3}\log\left(\frac{m_t^2}{m_W^2}\right)\right]\, \Im{C_{\underset{31}{Hud}}}
\end{align}
\end{subequations}

\begin{equation}
\frac{d_u}{e}\times (4\pi\Lambda)^2 \supset - \frac{2\sqrt{2}}{v}\sum_{i\in\{1,2,3\}} m_{d,i} \, \Im{C_{\underset{1i}{Hud}}^{\dagger}}
\end{equation}

\paragraph{Contributions from $\psi^2\bar{\psi}^2$ operators} 
\begin{equation}
\frac{d_d}{e}\times (4\pi\Lambda)^2 \supset -2\sum_{i\in\{2,3\}} m_{d,i} Q_d\,\Im{C^{\text{\tiny{(1)}}}_{\underset{1ii1}{qd}}+c_{F,3}\,C^{\text{\tiny{(8)}}}_{\underset{1ii1}{qd}}}
\end{equation}

\begin{equation}
\frac{d_u}{e}\times (4\pi\Lambda)^2 \supset -2\sum_{i\in\{2,3\}}m_{u,i} Q_u\,\Im{C^{\text{\tiny{(1)}}}_{\underset{1ii1}{qu}}+c_{F,3}\,C^{\text{\tiny{(8)}}}_{\underset{1ii1}{qu}}}
\end{equation}
\subsection{Quark cEDM}

\paragraph{Contributions from $\psi^2HF$ operators} 

\begin{subequations}
\begin{align}
\hspace{-1.5cm}\frac{\hat{d}_q}{e}\times (4\pi\Lambda)^2 \supset \Im{c_w C_{\underset{11}{qB}}+2T_q^3s_w C_{\underset{11}{qW}}}&\left\{6\sqrt{2}\,g_s Q_q v+8\sqrt{2}\,g_s Q_q v\log\left(\frac{\Lambda}{\mu_H}\right)\right\}\\[2ex]
 +\Im{-s_w C_{\underset{11}{qB}}+2T_q^3c_w C_{\underset{11}{qW}}}&\left\{\sqrt{2}\frac{(T_q^3-2Q_qs_w^2)}{s_wc_w}g_s v+4\sqrt{2}\,\frac{(T_q^3-2Q_qs_w^2)}{s_wc_w}g_s v\log\left(\frac{\Lambda}{m_Z}\right)\right\}\\[2ex]
-2T_q^3\Im{C_{\underset{11}{qW}}}&\left\{\frac{\sqrt{2}\,g_sv}{s_w}+\frac{4\sqrt{2}\,g_sv}{s_w}\log\left(\frac{\Lambda}{m_W}\right)\right\}\\[2ex]
+\Im{C_{\underset{11}{qG}}}&\left\{-\frac{16\sqrt{2}\pi^2v}{e}-2\sqrt{2}eQ_q^2 v-\frac{v}{\sqrt{2}}\frac{4+3 N_c^2}{N_c}\frac{g_s^2}{e}\right.\\[2ex]
&-\frac{4\sqrt{2}\,v}{N_c}\frac{g_s^2}{e}\log\left(\frac{\Lambda}{\mu_H}\right)\nonumber\\[2ex]
&\left.+\frac{v}{\sqrt{2}e}\Big(\Eq{eq:FermionWFR_LR_StrongWeak}_q+\Eq{eq:HiggsWFR}+\Eq{eq:GluonWFR}\Big)\right\}
\end{align}
\end{subequations}

\paragraph{Contributions from $H^2F^2$ operators}

\begin{equation}
\frac{\hat{d}_q}{e}\times (4\pi\Lambda)^2 \supset -C_{H\widetilde{G}}\left\{\frac{6m_qg_s}{e}+\frac{8m_qg_s}{e}\log\left(\frac{\Lambda}{m_h}\right)\right\}
 \end{equation}

\paragraph{Contributions from $F^3$ operators} 

\begin{equation}
\frac{\hat{d}_q}{e}\times (4\pi\Lambda)^2 \supset C_{\widetilde{G}}\left\{8N_cm_q\frac{g_s^2}{e}+6N_cm_q\frac{g_s^2}{e}\log\left(\frac{\Lambda}{\mu_H}\right)\right\}
\end{equation}

\paragraph{Contributions from $\psi^4$ operators} 
\begin{align}
\frac{\hat{d}_d}{e}\times (4\pi\Lambda)^2 &\supset -2\sum_{i\in\{1,2\}}\frac{m_{u,i} g_s}{e}\,\log\left(\frac{\Lambda}{\mu_H}\right) \Im{C^{\text{\tiny{(1)}}}_{\underset{1ii1}{quqd}}-\frac{1}{2 N_c}\,C^{\text{\tiny{(8)}}}_{\underset{1ii1}{quqd}}}\\[2ex]
&\hspace{0.4cm}-\frac{2m_tg_s}{e}\,\log\left(\frac{\Lambda}{m_t}\right) \Im{C^{\text{\tiny{(1)}}}_{\underset{1331}{quqd}}-\frac{1}{2 N_c}\,C^{\text{\tiny{(8)}}}_{\underset{1331}{quqd}}}\nonumber\\[2ex]\nonumber\\[2ex]
\frac{\hat{d}_u}{e}\times (4\pi\Lambda)^2 &\supset -2\sum_{i\in\{1,2\}}\frac{m_{d,i} g_s}{e}\,\log\left(\frac{\Lambda}{\mu_H}\right) \Im{C^{\text{\tiny{(1)}}}_{\underset{i11i}{quqd}}-\frac{1}{2 N_c}\,C^{\text{\tiny{(8)}}}_{\underset{i11i}{quqd}}}\\[2ex]
&\hspace{0.4cm}-\frac{2m_bg_s}{e}\,\log\left(\frac{\Lambda}{m_b}\right) \Im{C^{\text{\tiny{(1)}}}_{\underset{3113}{quqd}}-\frac{1}{2 N_c}\,C^{\text{\tiny{(8)}}}_{\underset{3113}{quqd}}}\nonumber
\end{align}

\paragraph{Contributions from $\psi\bar{\psi}H^2D$ operators} 

\begin{subequations}
\begin{align}
\frac{\hat{d}_d}{e}\times (4\pi\Lambda)^2 \supset \frac{g_s}{e}&\sum_{i\in\{1,2\}}\frac{4m_i}{\sqrt{2}v}\, \Im{C_{\underset{i1}{Hud}}}\\[2ex]
+&\frac{g_s m_t}{e\sqrt{2}v}\left[\frac{m_t^4+m_t^2m_W^2+4m_W^4}{(m_t^2-m_W^2)^2}\right.\nonumber\\[2ex]
-&\left.\frac{6m_t^2m_W^4}{(m_t^2-m_W^2)^3}\log\left(\frac{m_t^2}{m_W^2}\right)\right]\, \Im{C_{\underset{31}{Hud}}}
\end{align}
\end{subequations}

\begin{equation}
\frac{\hat{d}_u}{e}\times (4\pi\Lambda)^2 \supset 0 \times \Im{C_{\underset{ij}{Hud}}^{\dagger}}
\end{equation}

\paragraph{Contributions from $\psi^2\bar{\psi}^2$ operators} 
\begin{align}
\frac{\hat{d}_d}{e}\times (4\pi\Lambda)^2 \supset -2\sum_{i\in\{2,3\}}\frac{m_ig_s}{e}\,\Im{C^{\text{\tiny{(1)}}}_{\underset{1ii1}{qd}}-\frac{1}{2 N_c}\,C^{\text{\tiny{(8)}}}_{\underset{1ii1}{qd}}}\\[2ex]
\frac{\hat{d}_u}{e}\times (4\pi\Lambda)^2 \supset -2\sum_{i\in\{2,3\}}\frac{m_ig_s}{e}\,\Im{C^{\text{\tiny{(1)}}}_{\underset{1ii1}{qu}}-\frac{1}{2 N_c}\,C^{\text{\tiny{(8)}}}_{\underset{1ii1}{qu}}}
\end{align}

\subsection{Gluon cEDM}

\paragraph{Contributions from $F^3$ operators}

\begin{align}
C_{\widetilde{G}}\times (4\pi\Lambda)^2 \supset &\left\{2\sqrt{3}\pi^2 N_c-\frac{7 N_c}{2}-\left[8+N_c\right]\log\left(\frac{\Lambda}{\mu_H}\right)\right.\nonumber\\[2ex]
&\left.-2\log\left(\frac{\Lambda}{m_b}\right)-2\log\left(\frac{\Lambda}{m_t}\right)\right\} g_s^2C_{\widetilde{G}}
\end{align}

\paragraph{Contributions from $\psi^2HF$ operators}

\begin{equation}
C_{\widetilde{G}}\times (4\pi\Lambda)^2 \supset \frac{\sqrt{2}\,g_s^2 v}{3}\sum_{\substack{q\in\{u,d\} \\ i\in\{1,2,3\}}}\frac{\Im{C_{\underset{ii}{qG}}}}{m_{q,i}}
\end{equation}

\subsection{$O^{\text{\tiny{(S1/8, RR)}}}_{ud}$}
\paragraph{Contributions from $\psi^4$ operators}
\begin{subequations}
\begin{align}
\Im{c^{\text{\tiny{(S1,RR)}}}_{\underset{1111}{ud}}} \times(4\pi\Lambda)^2&\supset\:-\frac{c_{F,3}g_s^2}{N_c}\left\{3+4\log\left(\frac{\Lambda}{\mu_H}\right)\right\}\Im{C^{\text{\tiny{(8)}}}_{\underset{1111}{quqd}}}\\[2ex]
+&\;\Big\{(4\pi)^2-\frac{1}{2}\Big(\Eq{eq:FermionWFR_LR_StrongWeak}_u+\Eq{eq:FermionWFR_LR_StrongWeak}_d\Big)\\[2ex]
&+4c_{F,3}g_s^2+2e^2(Q_d^2-3Q_dQ_u+Q_u^2)+\frac{5e^2}{4s_w^2}\nonumber\\[2ex]
&+2e^2t_w(Q_d^2-3Q_dQ_u+Q_u^2)-\frac{5}{2}\frac{T_d^3T_u^3}{c_w^2s_w^2}-\frac{5e^2}{c_w^2}\nonumber\\[2ex]
&+8\left[2c_{F,3}g_s^2+e^2(Q_d^2-Q_dQ_u+Q_u^2)\right]\log\left(\frac{\Lambda}{\mu_H}\right)+2\frac{e^2}{s_w^2}\log\left(\frac{\Lambda}{m_W}\right)\nonumber\\[2ex]
&\left.+\frac{4e^2}{s_w^2c_w^2}\left[2(Q_d^2-Q_dQ_u+Q_u^2)s_w^4-3T_u^3s_w^2-T_d^3T_u^3\right]\right\}\Im{C^{\text{\tiny{(1)}}}_{\underset{1111}{quqd}}}\nonumber\\[2ex]
+&\;c_{F,3}\left\{3\,g_s^2\,\frac{N_c^2-2}{N_c^2}+3\,\frac{e^2(Q_d+Q_u)^2}{N_c}-\frac{3\,e^2}{2N_cs_w^2}\right.\\[2ex]
&+\frac{3\,e^2}{N_cs_w^2c_w^2}\left[(Q_d+Q_u)s_w^2-T_d^3\right]\left[(Q_d+Q_u)s_w^2-T_u^3\right]\nonumber\\[2ex]
&+\left[4\,g_s^2\,\frac{N_c^2-2}{N_c^2}-4\,\frac{e^2(Q_d+Q_u)^2}{N_c}\right]\log\left(\frac{\Lambda}{\mu_H}\right)-\frac{2\,e^2}{N_cs_w^2}\log\left(\frac{\Lambda}{m_W}\right)\nonumber\\[2ex]
&\left.+\frac{4e^2}{N_cs_w^2c_w^2}\left[(Q_d+Q_u)s_w^2-T_d^3\right]\left[(Q_d+Q_u)s_w^2-T_u^3\right]\log\left(\frac{\Lambda}{m_Z}\right)\right\}\Im{C^{\text{\tiny{(8)}}}_{\underset{1111}{quqd}}}\nonumber\\[2ex]
+&\;\left\{\frac{12c_{F,3}g_s^2}{N_c}+3\frac{e^2(Q_d+Q_u)^2}{N_c}-\frac{3e^2}{2N_c\,s_w^2}\right.\\[2ex]
&+3\frac{e^2}{N_c\,s_w^2c_w^2}\left[(Q_d+Q_u)s_w^2-T_d^3\right]\left[(Q_d+Q_u)s_w^2-T_u^3\right]\nonumber\\[2ex]
&+\frac{4}{N_c}\left[4c_{F,3}g_s^2+e^2(Q_d+Q_u)^2\right]\log\left(\frac{\Lambda}{\mu_H}\right)-\frac{2e^2}{s_w^2}\log\left(\frac{\Lambda}{m_W}\right)\nonumber\\[2ex]
&\left.+\frac{4e^2}{N_c\,s_w^2c_w^2}\left[(Q_d+Q_u)s_w^2-T_d^3\right]\left[(Q_d+Q_u)s_w^2-T_u^3\right]\log\left(\frac{\Lambda}{m_Z}\right)\right\}\Im{C^{\text{\tiny{(1)}}}_{\underset{1111}{quqd}}}\nonumber
\end{align}
\end{subequations}

\begin{subequations}
\begin{align}
\Im{c^{\text{\tiny{(S8,RR)}}}_{\underset{1111}{ud}}} \times(4\pi\Lambda)^2 &\supset \:\Big\{(4\pi)^2-\frac{1}{2}\Big(\Eq{eq:FermionWFR_LR_StrongWeak}_u+\Eq{eq:FermionWFR_LR_StrongWeak}_d\Big) \nonumber \\[2ex]
&+\frac{g_s^2(8-N_c^2)}{2N_c}+2e^2(Q_d^2-3Q_dQ_u+Q_u^2)+\frac{5e^2}{4s_w^2}\nonumber\\[2ex]
&+2e^2t_w(Q_d^2-3Q_dQ_u+Q_u^2)-\frac{5}{2}\frac{T_d^3T_u^3}{c_w^2s_w^2}-\frac{5e^2}{c_w^2}\nonumber\\[2ex]
&+8e^2(Q_d^2-Q_dQ_u+Q_u^2)\log\left(\frac{\Lambda}{\mu_H}\right)+2\frac{e^2}{s_w^2}\log\left(\frac{\Lambda}{m_W}\right)\nonumber\\[2ex]
&\left.+\frac{4e^2}{s_w^2c_w^2}\left[2(Q_d^2-Q_dQ_u+Q_u^2)s_w^4-3T_u^3s_w^2-T_d^3T_u^3\right]\right\}\Im{C^{\text{\tiny{(8)}}}_{\underset{1111}{quqd}}}\nonumber\\[2ex]
-&\;\left\{6 g_s^2+8g_s^2\log\left(\frac{\Lambda}{\mu_H}\right)\right\}\Im{C^{\text{\tiny{(1)}}}_{\underset{1111}{quqd}}}\\[2ex]
+&\;\left\{3g_s^2\,\frac{2+N_c^2}{N_c^2}-3\,\frac{e^2(Q_d+Q_u)^2}{N_c}+\frac{3\,e^2}{2N_cs_w^2}\right.\\[2ex]
&-\frac{3e^2}{N_cs_w^2c_w^2}\left[(Q_d+Q_u)s_w^2-T_d^3\right]\left[(Q_d+Q_u)s_w^2-T_u^3\right]\nonumber\\[2ex]
&+\left[4g_s^2\,\frac{2+N_c^2}{N_c^2}-4\,\frac{e^2(Q_d+Q_u)^2}{N_c}\right]\log\left(\frac{\Lambda}{\mu_H}\right)+\frac{2e^2}{N_cs_w^2}\log\left(\frac{\Lambda}{m_W}\right)\nonumber\\[2ex]
&\left.-\frac{4e^2}{N_cs_w^2c_w^2}\left[(Q_d+Q_u)s_w^2-T_d^3\right]\left[(Q_d+Q_u)s_w^2-T_u^3\right]\log\left(\frac{\Lambda}{m_Z}\right)\right\}\Im{C^{\text{\tiny{(8)}}}_{\underset{1111}{quqd}}}\nonumber\\[2ex]
+&\;\left\{-\frac{12g_s^2}{N_c}+6e^2(Q_d+Q_u)^2-\frac{3e^2}{s_w^2}\right.\\[2ex]
&+6\frac{e^2}{s_w^2c_w^2}\left[(Q_d+Q_u)s_w^2-T_d^3\right]\left[(Q_d+Q_u)s_w^2-T_u^3\right]\nonumber\\[2ex]
&-8\left[\frac{2g_s^2}{N_c}-e^2(Q_d+Q_u)^2\right]\log\left(\frac{\Lambda}{\mu_H}\right)-\frac{4e^2}{s_w^2}\log\left(\frac{\Lambda}{m_W}\right)\nonumber\\[2ex]
&\left.+8\frac{e^2}{s_w^2c_w^2}\left[(Q_d+Q_u)s_w^2-T_d^3\right]\left[(Q_d+Q_u)s_w^2-T_u^3\right]\log\left(\frac{\Lambda}{m_Z}\right)\right\}\Im{C^{\text{\tiny{(1)}}}_{\underset{1111}{quqd}}}\nonumber
\end{align}
\end{subequations}

\paragraph{Contributions from $\psi^2HF$ operators}

\begin{subequations}
\begin{align}
\Im{c^{\text{\tiny{(S1,RR)}}}_{\underset{1111}{ud}}} \times(4\pi\Lambda)^2&\supset\:\frac{\sqrt{2}e\,m_u}{v}\;\Im{c_w C_{\underset{11}{dB}}-s_w C_{\underset{11}{dW}}}\left\{\frac{6}{N_c}[Q_d+Q_u-N_cQ_u]\right.\\[2ex]
&-\left.\frac{8(Q_d+Q_u)}{N_c}\log\left(\frac{\Lambda}{m_W}\right)-4Q_u\left[\log\left(\frac{\Lambda}{m_Z}\right)+\log\left(\frac{\Lambda}{m_h}\right)\right]\right\}\nonumber\\[2ex]
+&\:\frac{\sqrt{2}e\,m_u}{v\,c_w s_w}\;\Im{s_w C_{\underset{11}{dB}}+c_w C_{\underset{11}{dW}}}\left\{\frac{2}{N_c}(N_c T_u^3-3T_d^3)\right.\\[2ex]
&+\frac{2s_w^2}{N_c}(3Q_d+3Q_u-2N_cQ_u)+\frac{2}{N_c}(N_c T_u^3-4T_d^3)\log\left(\frac{\Lambda}{m_Z}\right)\nonumber\\[2ex]
&+\frac{2s_w^2}{N_c}(4Q_d+4Q_u-2N_cQ_u)\log\left(\frac{\Lambda}{m_Z}\right)+2(T_u^3-2Q_us_w^2)\log\left(\frac{\Lambda}{m_h}\right)\nonumber\\[2ex]
&-2(T_u^3-2Q_us_w^2)\frac{m_Z^2}{m_h^2-m_Z^2}\log\left(\frac{m_h}{m_Z}\right)+\nonumber\\[2ex]
&\left.+8(T_d^3-(Q_u+Q_d)s_w^2)\frac{m_W^2}{m_Z^2-m_W^2}\log\left(\frac{m_Z}{m_W}\right)\right\}\nonumber\\[2ex]
+&\:\frac{\sqrt{2}e\,m_u}{v\, s_w}\;\Im{C_{\underset{11}{dW}}}\left\{1-6\frac{1-2T_d^3}{N_c}\right.\\[2ex]
&-\frac{16 s_w^2\, (Q_d+Q_u)}{N_c}\frac{m_Z^2}{m_Z^2-m_W^2}\nonumber\\[2ex]
&+4\left[1+\frac{4(Q_d+Q_u)s_w^2}{N_c}\right]\log\left(\frac{\Lambda}{m_W}\right)\nonumber\\[2ex]
&-4\left[\frac{1-4T_d^3}{N_c}+\frac{4(Q_d+Q_u)s_w^2}{N_c}\right]\log\left(\frac{\Lambda}{m_Z}\right)-\frac{4}{N_c}\log\left(\frac{\Lambda}{m_h}\right)\nonumber\\[2ex]
&+4\left[1-4T_d^3+4(Q_d+Q_u)s_w^2\right]\frac{m_W^2}{m_Z^2-m_W^2}\log\left(\frac{m_Z}{m_W}\right)\nonumber\\[2ex]
&+\frac{32\,s_w^2\,m_W^2}{N_c(m_Z^2-m_W^2)}\left[\frac{T_u^3}{s_w^4}+Q_u+(Q_d+Q_u)\frac{m_Z^2}{m_Z^2-m_W^2}\right]\log\left(\frac{m_Z}{m_W}\right)\nonumber\\[2ex]
&\left.+\frac{4}{N_c}\frac{m_W^2}{m_h^2-m_W^2}\log\left(\frac{m_h}{m_W}\right)\right\}\nonumber\\[2ex]
+&\:\frac{4\sqrt{2}g_sm_u}{v}\frac{c_{F,3}}{N_c}\;\Im{C_{\underset{11}{dG}}}\left\{3+4\log\left(\frac{\Lambda}{m_W}\right)\right\}\\[2ex]
&+\left(C_{\underset{11}{dW}} \rightarrow -C_{\underset{11}{uW}}, d\leftrightarrow u\right)
\end{align}
\end{subequations}

\begin{subequations}
\begin{align}
\Im{c^{\text{\tiny{(S8,RR)}}}_{\underset{1111}{ud}}}  \times(4\pi\Lambda)^2&\supset\:\frac{\sqrt{2}e\,m_u}{v}\;\Im{c_w C_{\underset{11}{dB}}-s_w C_{\underset{11}{dW}}}\left\{4(Q_d+Q_u)\left[3+4\log\left(\frac{\Lambda}{m_W}\right)\right]\right\}\\[2ex]
+&\:\frac{\sqrt{2}e\,m_u}{v\,c_w s_w}\;\Im{s_w C_{\underset{11}{dB}}+c_w C_{\underset{11}{dW}}}\\[2ex]
&\times\left\{4\left[(Q_d+Q_u)s_w^2-T_d^3\right]\left[3-\frac{4m_W^2}{m_Z^2-m_W^2}\log\left(\frac{m_Z}{m_W}\right)+4\log\left(\frac{\Lambda}{m_W}\right)\right]\right\}\nonumber\\[2ex]
+&\:\frac{\sqrt{2}e\,m_u}{v\, s_w}\;\Im{C_{\underset{11}{dW}}}\left\{12(2T_d^3-1)\right.\\[2ex]
&-32 s_w^2\, (Q_d+Q_u)\frac{m_Z^2}{m_Z^2-m_W^2}\nonumber\\[2ex]
&+32(Q_d+Q_u)s_w^2\log\left(\frac{\Lambda}{m_W}\right)-8\log\left(\frac{\Lambda}{m_h}\right)\nonumber\\[2ex]
&-8\left[1-4T_d^3+4(Q_d+Q_u)s_w^2\right]\log\left(\frac{\Lambda}{m_Z}\right)+\frac{8m_W^2}{m_h^2-m_W^2}\log\left(\frac{m_h}{m_W}\right)\nonumber\\[2ex]
&+\left[1-4T_d^3+4(Q_d+Q_u)s_w^2\right]\frac{8m_W^2}{m_Z^2-m_W^2}\log\left(\frac{m_Z}{m_W}\right)\nonumber\\[2ex]
&\left.+\frac{64\,s_w^2\,m_W^2}{m_Z^2-m_W^2}\left[\frac{T_u^3}{s_w^2}+Q_u +(Q_d+Q_u)\frac{m_Z^2}{m_Z^2-m_W^2}\right]\log\left(\frac{m_Z}{m_W}\right)\right\}\nonumber\\[2ex]
+&\:\frac{2\sqrt{2}g_sm_u}{v}\;\Im{C_{\underset{11}{dG}}}\left\{-\frac{3(2+N_c)}{N_c}-\frac{8}{N_c}\log\left(\frac{\Lambda}{m_W}\right)\right.\\[2ex]
&\left.-2\log\left(\frac{\Lambda}{m_Z}\right)-2\log\left(\frac{\Lambda}{m_h}\right)\right\}\nonumber\\[2ex]
&+\left(C_{\underset{11}{dW}} \rightarrow -C_{\underset{11}{uW}}, d\leftrightarrow u\right)
\end{align}
\end{subequations}

\subsection{$O^{\text{\tiny{(S1/8, RR)}}}_{duud}$}

\begin{subequations}
\paragraph{Contributions from $\psi^4$ operators}
\begin{align}
\Im{c^{\text{\tiny{(S8,RR)}}}_{\underset{1111}{duud}}}\times(4\pi\Lambda)^2 &\supset \left\{3g_s^2\frac{2+N_c^2}{N_c^2}-\frac{12e^2Q_dQ_u}{N_c}+\frac{3e^2}{2s_w^2}\right.\\[2ex]
&-\frac{3e^2}{N_c\,s_w^2c_w^2}(2Q_ds_w^2-T_d^3)(2Q_us_w^2-T_u^3)\nonumber\\[2ex]
&+\frac{4}{N_c^2}\left[g_s^2(2+N_c^2)-4e^2N_cQ_dQ_u\right]\log\left(\frac{\Lambda}{\mu_H}\right)+\frac{2e^2}{N_c\,s_w^2}\log\left(\frac{\Lambda}{m_W}\right)\nonumber\\[2ex]
&\left.-\frac{4e^2}{N_c\,s_w^2c_w^2}(2Q_ds_w^2-T_d^3)(2Q_us_w^2-T_u^3)\log\left(\frac{\Lambda}{m_Z}\right)\right\}\Im{C^{\text{\tiny{(8)}}}_{\underset{1111}{quqd}}}\nonumber\\[2ex]
+&\;\left\{-\frac{12g_s^2}{N_c}+24e^2Q_dQ_u-\frac{3e^2}{s_w^2}\right.\\[2ex]
&+\frac{6e^2}{\,s_w^2c_w^2}(2Q_ds_w^2-T_d^3)(2Q_us_w^2-T_u^3)\nonumber\\[2ex]
&-\frac{16g_s^2}{N_c}+32e^2Q_dQ_u\log\left(\frac{\Lambda}{\mu_H}\right)-\frac{4e^2}{s_w^2}\log\left(\frac{\Lambda}{m_W}\right)\\[2ex]
&\left.+\frac{8e^2}{s_w^2\,c_w^2}(2Q_ds_w^2-T_d^3)(2Q_us_w^2-T_u^3)\log\left(\frac{\Lambda}{m_Z}\right)\right\}\Im{C^{\text{\tiny{(1)}}}_{\underset{1111}{quqd}}}\nonumber\\[2ex]
+&\;\Big\{(4\pi)^2-\frac{1}{2}\Big(\Eq{eq:FermionWFR_LR_StrongWeak}_u+\Eq{eq:FermionWFR_LR_StrongWeak}_d\Big)\\[2ex]
&+\frac{g_s^2(8-N_c^2)}{2N_c}-\frac{e^2(Q_d+Q_u)^2}{2}+\frac{5e^2}{4s_w^2}\nonumber\\[2ex]
&-\frac{e^2}{2c_w^2s_w^2}\left[5T_d^3T_u^3+(Q_d+Q_u)^2s_w^4\right]\nonumber\\[2ex]
&+8e^2Q_dQ_u\log\left(\frac{\Lambda}{\mu_H}\right)+\frac{2e^2}{s_w^2}\log\left(\frac{\Lambda}{m_W}\right)\nonumber\\[2ex]
&+4e^2\left.\left[2Q_dQ_ut_w^2+\frac{T_u^3}{c_w^2}-\frac{T_u^3T_d^3}{s_w^2c_w^2}\right]\log\left(\frac{\Lambda}{m_Z}\right)\right\}\Im{C^{\text{\tiny{(8)}}}_{\underset{1111}{quqd}}}\nonumber\\[2ex]
&\;-2g_s^2\left\{3+4\log\left(\frac{\Lambda}{\mu_H}\right)\right\}\Im{C^{\text{\tiny{(1)}}}_{\underset{1111}{quqd}}}
\end{align}
\end{subequations}

\begin{subequations}
\begin{align}
\Im{c^{\text{\tiny{(S1,RR)}}}_{\underset{1111}{duud}}}\times(4\pi\Lambda)^2 &\supset\,c_{F,3}\left\{3g_s^2\frac{N_c^2-2}{N_c^2}+\frac{12e^2Q_dQ_u}{N_c}-\frac{3e^2}{2s_w^2}\right.\\[2ex]
&+\frac{3e^2}{N_c\,s_w^2c_w^2}(2Q_ds_w^2-T_d^3)(2Q_us_w^2-T_u^3)\nonumber\\[2ex]
&+\frac{4}{N_c^2}\left[g_s^2(N_c^2-2)+4e^2N_cQ_dQ_u\right]\log\left(\frac{\Lambda}{\mu_H}\right)-\frac{2e^2}{N_c\,s_w^2}\log\left(\frac{\Lambda}{m_W}\right)\nonumber\\[2ex]
&\left.+\frac{4e^2}{N_c\,s_w^2c_w^2}(2Q_ds_w^2-T_d^3)(2Q_us_w^2-T_u^3)\log\left(\frac{\Lambda}{m_Z}\right)\right\}\Im{C^{\text{\tiny{(8)}}}_{\underset{1111}{quqd}}}\nonumber\\[2ex]
+&\;\left\{\frac{12c_{F,3}g_s^2}{N_c}+12\frac{e^2Q_dQ_u}{N_c}-\frac{3e^2}{2N_c\,s_w^2}\right.\\[2ex]
&+\frac{3e^2}{N_c\,s_w^2c_w^2}(2Q_ds_w^2-T_d^3)(2Q_us_w^2-T_u^3)\nonumber\\[2ex]
&+\frac{16}{N_c}\left[g_s^2c_{F,3}+e^2Q_dQ_u\right]\log\left(\frac{\Lambda}{\mu_H}\right)-\frac{2e^2}{N_c\,s_w^2}\log\left(\frac{\Lambda}{m_W}\right)\\[2ex]
&\left.+\frac{4e^2}{N_c\,s_w^2c_w^2}(2Q_ds_w^2-T_d^3)(2Q_us_w^2-T_u^3)\log\left(\frac{\Lambda}{m_Z}\right)\right\}\Im{C^{\text{\tiny{(1)}}}_{\underset{1111}{quqd}}}\nonumber\\[2ex]
-&\;\frac{c_{F,3}g_s^2}{N_c}\left\{3+4\log\left(\frac{\Lambda}{\mu_H}\right)\right\}\Im{C^{\text{\tiny{(8)}}}_{\underset{1111}{quqd}}}\\[2ex]
+&\;\Big\{(4\pi)^2-\frac{1}{2}\Big(\Eq{eq:FermionWFR_LR_StrongWeak}_u+\Eq{eq:FermionWFR_LR_StrongWeak}_d\Big)\\[2ex]
&+4c_{F,3}g_s^2-\frac{e^2(Q_d+Q_u)^2}{2}+\frac{5e^2}{4s_w^2}\nonumber\\[2ex]
&-\frac{e^2}{2c_w^2s_w^2}\left[5T_d^3T_u^3+(Q_d+Q_u)^2s_w^4\right]\nonumber\\[2ex]
&+8\left[2c_{F,3}g_s^2+e^2Q_dQ_u\right]\log\left(\frac{\Lambda}{\mu_H}\right)+\frac{2e^2}{s_w^2}\log\left(\frac{\Lambda}{m_W}\right)\nonumber\\[2ex]
&+4e^2\left.\left[2Q_dQ_ut_w^2+\frac{T_u^3}{c_w^2}-\frac{T_u^3T_d^3}{s_w^2c_w^2}\right]\log\left(\frac{\Lambda}{m_Z}\right)\right\}\Im{C^{\text{\tiny{(1)}}}_{\underset{1111}{quqd}}}\nonumber
\end{align}
\end{subequations}

\paragraph{Contributions from $\psi^2HF$ operators}

\begin{subequations}
\begin{align}
\Im{c^{\text{\tiny{(S1,RR)}}}_{\underset{1111}{duud}}}\times(4\pi\Lambda)^2&\supset\:\frac{\sqrt{2}e\,m_u}{v}\;\Im{c_w C_{\underset{11}{dB}}-s_w C_{\underset{11}{dW}}}\left\{\frac{12 Q_u}{N_c}-3(Q_d+Q_u)\right.\\[2ex]
&+2(Q_d+Q_u)\log\left(\frac{\Lambda}{m_W}\right)-\frac{4Q_u}{N_c}\left[\log\left(\frac{\Lambda}{m_Z}\right)+\log\left(\frac{\Lambda}{m_h}\right)\right]\nonumber\\[2ex]
+&\:\frac{\sqrt{2}e\,m_u}{v\,c_w s_w}\;\Im{s_w C_{\underset{11}{dB}}+c_w C_{\underset{11}{dW}}}\left\{3T_d^3-\frac{4T_u^3}{N_c}\right.\\[2ex]
&+\frac{8Q_u-3N_c(Q_d+Q_u)}{N_c}s_w^2\nonumber\\[2ex]
&-2\left[T_d^3-\frac{T_u^3}{N_c}-\frac{N_c(Q_d+Q_u)-2Q_u}{N_c}s_w\right]\log\left(\frac{\Lambda}{m_Z}\right)\nonumber\\[2ex]
&+\frac{2}{N_c}(T_u^3-2Q_us_w^2)\log\left(\frac{\Lambda}{m_h}\right)\nonumber\\[2ex]
&-(T_u^3-2Q_us_w^2)\frac{4m_Z^2}{m_Z^2-m_H^2}\log\left(\frac{m_h}{m_Z}\right)\nonumber\\[2ex]
&-\left.((Q_d+Qu)s_w^2-T_d^3)\frac{4m_W^2}{m_Z^2-m_W^2}\log\left(\frac{m_Z}{m_W}\right)\right\}\nonumber\\[2ex]
+&\:\frac{\sqrt{2}e\,m_u}{v\, s_w}\;\Im{C_{\underset{11}{dW}}}\left\{3-6T_d^3-\frac{2}{N_c}\right.\\[2ex]
&+8 s_w^2\, (Q_d+Q_u)\frac{m_Z^2}{m_Z^2-m_W^2}\nonumber\\[2ex]
&-8\left[\frac{1}{N_c}+(Q_d+Q_u)s_w^2\right]\log\left(\frac{\Lambda}{m_W}\right)\nonumber\\[2ex]
&-2\left[4T_d^3-1-4(Q_d+Q_u)s_w^2\right]\log\left(\frac{\Lambda}{m_Z}\right)+2\log\left(\frac{\Lambda}{m_h}\right)\nonumber\\[2ex]
&+\left[4T_d^3-1-4(Q_d+Q_u)s_w^2\right]\frac{2m_W^2}{m_Z^2-m_W^2}\log\left(\frac{m_Z}{m_W}\right)\nonumber\\[2ex]
&-\frac{16\,s_w^2\,m_W^2}{m_Z^2-m_W^2}\left[\frac{T_u^3}{s_w^2}+Q_u +(Q_d+Q_u)\frac{m_Z^2}{m_Z^2-m_W^2}\right]\log\left(\frac{m_Z}{m_W}\right)\nonumber\\[2ex]
&-\frac{2m_W^2}{m_h^2-m_W^2}\log\left(\frac{m_h}{m_W}\right)\nonumber\\[2ex]
+&\:\frac{4\sqrt{2}g_sm_u}{v}\frac{c_{F,3}}{N_c}\;\Im{C_{\underset{11}{dG}}}\left\{3+2\log\left(\frac{\Lambda}{m_Z}\right)+2\log\left(\frac{\Lambda}{m_h}\right)\right\}\\[1ex]
&+\left(C_{\underset{11}{dW}} \rightarrow -C_{\underset{11}{uW}}, d\leftrightarrow u\right)
\end{align}
\end{subequations}

\begin{subequations}
\begin{align}
\Im{c^{\text{\tiny{(S8,RR)}}}_{\underset{1111}{duud}}} \times(4\pi\Lambda)^2&\supset\:\frac{\sqrt{2}e\,m_u}{v}\;\Im{c_w C_{\underset{11}{dB}}-s_w C_{\underset{11}{dW}}}\Big\{24 Q_u \\[2ex]
&\left.+16Q_u\log\left(\frac{\Lambda}{m_Z}\right)+16Q_u\log\left(\frac{\Lambda}{m_h}\right)\right\}\nonumber\\[2ex]
+&\:\frac{\sqrt{2}e\,m_u}{v\,c_w s_w}\;\Im{s_w C_{\underset{11}{dB}}+c_w C_{\underset{11}{dW}}}\Bigg\{8(2Q_us_w^2-T_u^3)\\[2ex]
&+8(2Q_us_w^2-T_u^3)\left[\log\left(\frac{\Lambda}{m_Z}\right)+\log\left(\frac{\Lambda}{m_h}\right)\right]\nonumber\\[2ex]
&-\left.8(2Q_us_w^2-T_u^3)\frac{m_Z^2}{m_h^2-m_Z^2}\log\left(\frac{m_h}{m_Z}\right)\right\}\nonumber\\[2ex]
-&\:\frac{4\sqrt{2}e\,m_u}{v\, s_w}\;\Im{C_{\underset{11}{dW}}}\left\{1+4\log\left(\frac{\Lambda}{m_W}\right)\right\}\\[2ex]
-&\:\frac{2\sqrt{2}g_sm_u}{v}\;\Im{C_{\underset{11}{dG}}}\left\{\frac{3(2+N_c)}{N_c}+4\log\left(\frac{\Lambda}{m_W}\right)\right.\\[2ex]
&+\left.\frac{4}{N_c}\log\left(\frac{\Lambda}{m_Z}\right)+\frac{4}{N_c}\log\left(\frac{\Lambda}{m_h}\right)\right\}\nonumber\\[2ex]
&+\left(C_{\underset{11}{dW}} \rightarrow -C_{\underset{11}{uW}}, d\leftrightarrow u\right)
\end{align}
\end{subequations}

\subsection{$O_{Hud}$}

\paragraph{Contributions from $\psi\bar{\psi}H^2D$ operators}

\begin{align}
\Im{C_{\underset{11}{Hud}}}\times (4\pi\Lambda)^2 & \supset \:\Bigg\{(4\pi)^2-\frac{1}{2}\Big(\Eq{eq:FermionWFR_R_Weak}_u+\Eq{eq:FermionWFR_R_Strong}_u\Eq{eq:HiggsWFR}\\
&+\Eq{eq:FermionWFR_R_Weak}_d+\Eq{eq:FermionWFR_R_Strong}_d\Big) -\pi^2e^2 \\[2ex]
&-\frac{2\pi^2}{3}e^2\frac{m_Z^2}{m_W^2}\left[\frac{Q_d-Q_d\,t_w^2}{4}-Q_dQ_u\,t_w^2\frac{m_Z^2}{m_W^2}\right]\nonumber\\[2ex]
&-4e^2-4g^2-2g^2c_{2w}-\frac{g^2}{2c_w^2}+e^2 Q_d\frac{c_{2w}}{c_w^2}+\frac{g^2}{12c_w^4}+2e^2t_w^2\nonumber\\[2ex]
&-g^2t_w^2-2e^2 Q_d Q_u \frac{t_w^2}{c_w^2}+2\frac{m_h^2}{v^2}-\frac{m_h^4}{3m_W^2v^2}\nonumber\\[2ex]
&+2\left[c_{F,3} g_s^2+e^2 Q_dQ_u\right]\log\left(\frac{\Lambda}{\mu_H}\right)-2e^2\log\left(\frac{\Lambda}{m_W}\right)\nonumber\\[2ex]
&-\left[2e^2+\frac{7}{3}g^2+2g^2(c_w^2-2s_w^2)-2e^2(1+Q_dQ_u)t_w^2\right]\log\left(\frac{\Lambda}{m_Z}\right)\nonumber\\[2ex]
&-\frac{5g^2}{3}\log\left(\frac{\Lambda}{m_h}\right)-4e^2t_w^2Q_dQ_u\frac{m_Z^4}{m_W^4}\log^2\left(\frac{m_Z}{m_W}\right)\nonumber\\[2ex]
&+g^2\left[\frac{1}{3}+\frac{3m_h^2}{v^2}-\frac{2m_h^4}{m_W^2v^2}+\frac{m_h^6}{3m_W^4v^2}\right]\log\left(\frac{m_h}{m_W}\right)-\frac{g^2}{3}\log\left(\frac{m_h}{m_Z}\right)\nonumber\\[2ex]
&+2e^2\left[Q_u(3-t_w^2+Q_d\,t_w^2)+\frac{c_{2w}}{c_w^2}-(3-Q_dQ_ut_w^2)\frac{m_Z^2}{m_W^2}\right]\log\left(\frac{m_Z}{m_W}\right)\nonumber\\[2ex]
&-g^2\left[\frac{13}{3}+2c_{2w^2}+\frac{1}{2c_w^2}-\left(1+c_{2w}+\frac{1}{2c_w^2}\right)\frac{m_Z^2}{m_W^2}+\frac{1}{12 c_w^2}\frac{m_Z^2}{m_W^2}\right]\log\left(\frac{m_Z}{m_W}\right)\nonumber\\[2ex]
&-\frac{e^2 Q_d}{c_w^2}\frac{m_Z^2}{m_W^2}\left[c_{2w}-2Q_dQ_u s_w^2\frac{m_Z^2}{m_W^2}\right]\text{Li}_2\left(1+\frac{m_Z^2}{m_W^2}\right)\nonumber\\[2ex]
&-\left[g^2+\frac{4m_h^2}{3v^2}-\frac{m_h^4}{3m_W^2v^2}\right]F(m_W^2,m_h,m_W)\nonumber\\[2ex]
&+\left[4e^2-2g^2c_w-\frac{g^2}{3c_w^2}+g^2\frac{m_Z^2}{12m_W^2}\left(\frac{1}{c_w^2}-12\right)\right]F(m_W^2,m_W,m_Z)\nonumber\\[2ex]
&\left.+e^2\left[2-\frac{1}{c_w^2}\right]C_0(m_W^2,m_Z,m_W)\right\} \Im{C_{\underset{11}{Hud}}}\nonumber,
\end{align}

where we defined 
\begin{equation}
F(x,y,z)=\frac{\sqrt{\lambda(x,y^2,z^2)}}{x}\,\log\left(\frac{y^2+z^2-x+\sqrt{\lambda(x,y^2,z^2)}}{2 yz}\right),
\end{equation}
\begin{align}
C_0(x,y,\sqrt{x})=&\frac{\pi^2}{6}+\frac{1}{2}\log\left(\frac{\sqrt{y^4-4xy^2}-y^2}{2x+\sqrt{y^4-4xy^2}-y^2}\right)\\[2ex]
&-\text{Li}_2\left(\frac{2x}{y^2-\sqrt{y^4-4xy^2}}\right)+\text{Li}_2\left(-\frac{2x}{2x+\sqrt{y^4-4xy^2}-y^2}\right), \nonumber
\end{align}

with the usual Kallen $\lambda$-function and $\text{Li}_2\left(x\right)$ denotes the dilogarithm. \\[1cm]

\paragraph{Contributions from $\psi^2 H F$ operators}

\begin{subequations}
\begin{align}
\Im{C_{\underset{11}{Hud}}} \times & (4\pi\Lambda)^2 \supset - \frac{5c_{F,3}g_s m_u}{\sqrt{2}v} \Im{C_{\underset{11}{dG}}} \\
- & \;\frac{e m_u}{\sqrt{2}v} \Im{-s_w C_{\underset{11}{dW}} + c_w C_{\underset{11}{dB}}} \left[-1 + 9 Q_u + 4 Q_u m_h^2 C_0(m_h^2,m_W^2,m_h^2+m_W^2,0,0,0) \right] \\[-2ex]
+ & \; \frac{e m_u}{\sqrt{2} s_w c_w v} \Im{-s_w C_{\underset{11}{dB}} - c_w C_{\underset{11}{dW}}} \left\{ -4 +\frac{9}{2}Q_u + 8 T_u^3  \right. \\[2ex]
& \left. + c_{2w} \left( \frac{1}{2} \left(1-9Q_u\right) -4\frac{m_Z^2}{m_W^2} + \pi^2 \frac{2m_Z^4}{3m_W^4} \right) + 2 Q_u s_w^2 \log\left(\frac{m_Z^2}{m_h^2}\right) \right. \nonumber \\[2ex]
& \left. + 2 Q_u s_w^2 \left[ \frac{m_h^4-m_W^4+m_h^2m_Z^2+3m_W^2m_Z^2-2m_Z^4}{m_h^2(m_h^2+m_W^2)} \log\left(\frac{m_Z^2}{m_h^2+m_W^2-m_Z^2}\right) \right. \right. \nonumber \\[2ex]
& \left. \left. + \frac{(m_W^2-m_Z^2)(m_W^2-2m_Z^2-4m_h^2)}{m_h^2 m_W^2} \log\left(\frac{m_Z^2}{m_Z^2-m_W^2} \right)  \right] \right. \nonumber \\[2ex]
& \left. + 2 c_{2w} \frac{(m_W^2+2m_Z^2)}{m_W^2} \log\left(\frac{m_Z^2}{m_W^2}\right) - 4 c_{2w} \frac{m_Z^4}{m_W^4} \text{Li}_2\left(1-\frac{m_W^2}{m_Z^2}\right) \right. \nonumber \\[2ex]
& \left. + 2 Q_u s_w^2 \left( 2m_h^2 + m_Z^2 \right) C_0(m_h^2,m_W^2,m_h^2+m_W^2,0,0,m_Z)  \Large\right\} \nonumber \\[2ex]\nonumber
\end{align}
\end{subequations}
\addtocounter{equation}{-1}

\begin{subequations}
\setcounter{equation}{3}
\begin{align}
\qquad\qquad +&\; \frac{e m_u}{\sqrt{2} s_w v} \Im{C_{\underset{11}{dW}}} \Bigg\{ 4 \left( -3 + 4 Q_u s_w^2 (3-\pi^2+3s_wt_w^2) + \frac{m_h^2+m_Z^2}{m_W^2} \right) \\[2ex]
& - \frac{12m_W^4 +9m_h^2 m_W^2 - 4m_h^4}{2m_W^2(m_h^2+m_W^2)} \log\left(\frac{m_W^2}{m_h^2}\right) + \frac{2(m_Z^2-4m_W^2)}{m_W^2} \log\left(\frac{m_W^2}{m_Z^2}\right) \nonumber \\[2ex]
& - \frac{4m_h^4 -12 m_h^2 m_W^2 + 8m_W^4}{m_W^4} \log\left(\frac{m_h^2}{m_h^2-m_W^2}\right) - \frac{8 m_W^4 - 12 m_W^2 m_Z^2 + 4m_Z^4}{m_W^4} \log\left(\frac{m_Z^2}{m_Z^2-m_W^2}\right) \nonumber \\[2ex]
& - \frac{5m_W^4 - 4 m_h^2 m_W^2 + m_h^4}{m_W^4} \log^2\left(\frac{m_W^2}{m_h^2-m_W^2}\right) - \frac{5m_W^4 - 4m_W^2 m_Z^2 + m_Z^4}{m_W^4} \log^2\left(\frac{m_W^2}{m_Z^2-m_W^2}\right) \nonumber \\[2ex]
& + \frac{2\sqrt{m_h^4-4m_h^2m_W^2}}{m_h^2} \log\left(\frac{2m_W^2-m_h^2+\sqrt{m_h^4-4m_h^2m_W^2}}{2m_W^2}\right) + 4Q_u s_w^2 \left[ \log\left(\frac{m_Z^2}{m_h^2}\right) \right. \nonumber \\[2ex]
& \left. + \frac{m_h^4-m_W^4+m_h^2m_Z^2+3m_W^2m_Z^2-2m_Z^4}{m_h^2(m_h^2+m_W^2)} \log\left(\frac{m_Z^2}{m_h^2+m_W^2-m_Z^2}\right) \right. \nonumber \\[2ex]
& \left. + \frac{(m_W^2-2m_H^2-2m_Z^2)(m_W^2-m_Z^2)}{m_h^2 m_W^2} \log\left(\frac{m_Z^2}{m_Z^2-m_W^2}\right) - 3 s_w^2 t_w \left( \frac{2m_Z^2-4m_W^2}{m_W^3-m_Wm_Z^2} \log\left(\frac{m_W^2}{m_Z^2}\right) \right. \right. \nonumber \\[2ex]
& \left. \left. + \frac{4m_Z(m_Z^2-m_W^2)}{m_W^3} \log\left(\frac{m_Z^2}{m_Z^2-m_W^2}\right) + \frac{m_Z(m_Z^2-2m_W^2)}{m_W^3} \log^2\left(\frac{m_W^2}{m_Z^2-m_W^2}\right) \right) \right] \nonumber \\[2ex]
& + 48 Q_u s_w^2 \text{Li}_2\left(2\right)  - 2 \frac{m_h^4-4m_h^2m_W^2+5m_W^4}{m_W^4} \left( \text{Li}_2\left(1-\frac{m_h^2}{m_W^2}\right) + \text{Li}_2\left(\frac{m_h^2-2m_W^2}{m_h^2-m_W^2}\right) \right) \nonumber \\[2ex]
& - 2 \frac{m_Z^4-4m_W^2m_Z^2+5m_W^4}{m_W^4} \left( \text{Li}_2\left(1-\frac{m_Z^2}{m_W^2}\right) + \text{Li}_2\left(\frac{m_Z^2-2m_W^2}{m_Z^2-m_W^2}\right) \right) \nonumber \\[2ex]
& - 24 Q_u s_w^4 t_w \frac{m_Z(m_Z^2-2m_W^2)}{m_W^3} \left( \text{Li}_2\left(1-\frac{m_Z^2}{m_W^2}\right) + \text{Li}_2\left(\frac{m_Z^2-2m_W^2}{m_Z^2-m_W^2}\right) \right) \nonumber \\[2ex]
& + \left( 2 m_W^2 - \frac{5}{2} m_h^2 \right) C_0\left(m_h^2,m_W^2,m_h^2+m_W^2,m_W,m_W,0\right) + 4Q_u s_w^2 \Big(\left( 2m_h^2+m_Z^2 \right) \times \nonumber \\[2ex]
& \times C_0\left(m_h^2,m_W^2,m_h^2+m_W^2,0,0,m_Z\right) -2 m_h^2 C_0\left(m_h^2,m_W^2,m_h^2+m_W^2,0,0,0\right) \Big) \Bigg\} \nonumber \\[2ex]
&+\left(C_{\underset{11}{dG}} \rightarrow C^\dagger_{\underset{11}{uG}}, C_{\underset{11}{dW}} \rightarrow -C^\dagger_{\underset{11}{uW}}, C_{\underset{11}{dB}} \rightarrow C^\dagger_{\underset{11}{uB}}, d\leftrightarrow u\right)
\end{align}
\end{subequations}

where $C_0(s_1,s_{12},s_2,m_0,m_1,m_2)$ is the scalar Passarino-Veltman three-point function with kinematic invariants $s_1,s_{12},s_2$ and masses $m_0,m_1,m_2$ which can be evaluated numerically with computer programs like Package-X \cite{Patel2015}.

\paragraph{Contributions from $\psi^2\bar{\psi^2}$ operators}

\begin{align}
\Im{C_{\underset{11}{Hud}}}\times (4\pi\Lambda)^2 \supset & 4\sum_{i,j\in \{1,2,3\}}\frac{m_{d,i} m_{u,j}}{v^2}\Im{C^\text{\tiny{(1)}}_{\underset{1ji1}{ud}}+c_{F,3}C^\text{\tiny{(8)}}_{\underset{1ji1}{ud}}}\left\{1+\frac{1}{4}\left[2\,\text{sgn}_{ij}+\frac{m_{u,j}^2}{m_W^2}\right]\log\left(\frac{m_{d,i}^2}{m_{u,j}^2}\right)\right.\nonumber\\[2ex]
&\left.+2\log\left(\frac{2 \Lambda}{m_{d,i}+m_{u,j}-(m_{d,i}-m_{u,j})\text{sgn}_{ij}}\right)+F(2 m_W^2, m_{d,i}, m_{u,j})\right\}
\end{align}

Here we defined
\begin{gather}
\text{sgn}_{ij}=\text{sgn}(m_{d,i}-m_{u,j}).
\end{gather}

\paragraph{Contributions from $\psi^2H^3$ operators}

\begin{subequations}
\begin{align}
\Im{C_{\underset{11}{Hud}}}\times \left(4\pi\Lambda\right)^2 \supset \,\frac{m_u}{\sqrt{2}\,v}&\left\{2\pi^2-2-2\pi^2\left[\frac{m_h^2}{m_W^2}+\frac{m_h^4}{2m_W^4}\right]\label{eq:Yukawa_type}\right.\\
&+\frac{2\pi^2}{3}\left[\frac{m_Z^2}{m_W^2}+\frac{m_Z^4}{2m_W^4}\right]+\frac{3\,m_h^2-m_Z^2}{m_W^2}+\nonumber\\[2ex]
&+6\left[2+\frac{m_H^2}{m_W^2}\right]\log\left(\frac{m_h}{m_W}\right)\nonumber\\[2ex]
&-2\left[2+\frac{m_Z^2}{m_W^2}\right]\log\left(\frac{m_Z}{m_W}\right)-8\,\text{Li}_2(2)\nonumber\\[2ex]
&+3\left[1+2\frac{m_h^2}{m_W^2}+\frac{m_h^4}{m_W^4}\right]\left[\text{Li}_2\left(1+\frac{m_h^2}{m_W^2}\right)+\frac{1}{2}\log^2\left(\frac{m_h^2}{m_W^2}\right)\right]\nonumber\\[2ex]
&\left.-\left[1+2\frac{m_Z^2}{m_W^2}+\frac{m_Z^4}{m_W^4}\right]\left[\text{Li}_2\left(1+\frac{m_Z^2}{m_W^2}\right)+\frac{1}{2}\log^2\left(\frac{m_Z^2}{m_W^2}\right)\right]\right\} \Im{C_{\underset{11}{dH}}}\nonumber\\[2ex]
&+\left(C_{\underset{11}{dH}} \rightarrow C^\dagger_{\underset{11}{uH}}, d\leftrightarrow u\right)
\end{align}
\end{subequations}

\clearpage

\section{Spurionic Expansion of the Wilson Coefficients and Form of Spurions}\label{app:flavorsym}

In this appendix we show the spurionic expansion of the Wilson coefficients with the different flavor symmetries, to introduce the notation we use to present the bounds. In addition we give expressions of the spurions in terms of Standard Model parameters which is fairly straightforward for the $U(3)$ flavor group and a bit more involved for the $U(2)$ group. 
We start with MFV, where we consider the biggest flavor group that is allowed by the gauge symmetry group of the Standard Model. As mentioned above we can simply identify the spurions with the Yukawa couplings since the full Yukawa matrices are the only source of flavor symmetry breaking. We work in the up-quark gauge basis where we explicitly have
\begin{equation}
y_u = \text{diag}\left(\lambda_u,\lambda_c,\lambda_t \right) \qquad y_d = V_{\text{CKM}}\text{diag}\left(\lambda_d,\lambda_s,\lambda_b \right) \qquad y_u = \text{diag}\left(\lambda_e,\lambda_{\mu},\lambda_{\tau} \right)\, .
\end{equation}
We find the following expansions for the Wilson coefficients appearing in the neutron EDM expression, in the up-basis defined above, keeping terms up to $\mathcal{O}(y_{u,d,e}^2)$
\begin{equation}
C'_{\substack{uB\\pr}}\mathcal{O}'_{\substack{uB\\pr}} = C'_{\substack{uB\\pr}} \left( \bar{q}'_p \sigma^{\mu\nu} u'_r \right) \widetilde{H} B_{\mu\nu} \longrightarrow F_{uB} \left( \bar{q}'_p \sigma^{\mu\nu} u'_r \right) \widetilde{H} B_{\mu\nu} \left( \left( y_u \right)_{pr} + \mathcal{O}( y_i^3) \right)
\end{equation}
\begin{equation}
C'_{\substack{Hud\\pr}}\mathcal{O}'_{\substack{Hud\\pr}} = C'_{\substack{Hud\\pr}} \left( \tilde{H}^{\dagger} i D_{\mu} H \right) \left( \bar{u}'_p \gamma^{\mu} d'_r \right) \longrightarrow F_{Hud} \left( \tilde{H}^{\dagger} i D_{\mu} H \right) \left( \bar{u}'_p \gamma^{\mu} d'_r \right) \left( \left( y_u^{\dagger} y_d \right)_{pr} + \mathcal{O}( y_i^4) \right)
\end{equation}
\begin{equation}
C'_{\substack{uH\\pr}}\mathcal{O}'_{\substack{uH\\pr}} = C'_{\substack{uH\\pr}} |H|^2 \left( \bar{q}'_p \widetilde{H} u'_r \right) \longrightarrow F_{uH} |H|^2 \left( \bar{q}'_p \widetilde{H} u'_r \right) \left( \left( y_u \right)_{pr} + \mathcal{O}( y_i^3) \right)
\end{equation}
\begin{equation}
\begin{split}
C_{\substack{lequ\\prst}}^{'(3)}\mathcal{O}_{\substack{lequ\\prst}}^{'(3)} = C_{\substack{lequ\\prst}}^{'(3)} \left( \bar{l}_p^{'i} \sigma_{\mu\nu} e'_r \right) & \epsilon_{ij} \left(\bar{q}_s^{'j} \sigma^{\mu\nu} u'_t \right) \longrightarrow F_{lequ}^{(3)} \left( \bar{l}_p^{'i} \sigma_{\mu\nu} e'_r \right) \epsilon_{ij} \left(\bar{q}_s^{'j} \sigma^{\mu\nu} u'_t \right)  \times \\
& \times \left[ \left(y_e \right)_{pr} \left( y_u \right)_{st} + \mathcal{O}(y_i^4) \right]
\end{split}
\end{equation}
\begin{equation}
C_{\substack{quqd\\prst}}^{'(1)}\mathcal{O}_{\substack{quqd\\prst}}^{'(1)} = C_{\substack{quqd\\prst}}^{'(1)} \left( \bar{q}_p^{'i} u'_r \right) \epsilon_{ij} \left(\bar{q}_s^{'j} d'_t \right) \longrightarrow F_{quqd}^{(1)} \left( \bar{q}_p^{'i} u'_r \right) \epsilon_{ij} \left(\bar{q}_s^{'j} d'_t \right) \left[ \left(y_u \right)_{pr} \left(y_d \right)_{st} + \mathcal{O}(y_i^4) \right]
\end{equation}
\begin{equation}
\begin{split}
& C_{\substack{qu\\prst}}^{'(1)}\mathcal{O}_{\substack{qu\\prst}}^{'(1)} = C_{\substack{qu\\prst}}^{'(1)} \left( \bar{q}'_p \gamma_{\mu} q'_r \right) \left( \bar{u}'_s \gamma^{\mu} u'_t \right) \longrightarrow F_{qu}^{(1)} \left( \bar{q}'_p \gamma_{\mu} q'_r \right) \left( \bar{u}'_s \gamma^{\mu} u'_t \right) \times \\
& \times \left[ x_1 \delta_{pr} \delta_{st} + x_2 \left( \left( y_d y_d^{\dagger} \right)_{pr} + \left( y_u y_u^{\dagger} \right)_{pr} \right) \delta_{st} + x_3 \delta_{pr} \left( y_u y_u^{\dagger} \right)_{st} + x_4 \left( y_u \right)_{pt} \left( y_u^{\dagger} \right)_{sr}  + \mathcal{O}(y_i^4) \right]
\end{split}
\end{equation}
\begin{equation}
C_{\substack{ud\\prst}}^{'(1)}\mathcal{O}_{\substack{ud\\prst}}^{'(1)} = C_{\substack{ud\\prst}}^{'(1)} \left( \bar{u}'_p \gamma_{\mu} u'_r \right) \left( \bar{d}'_s \gamma^{\mu} d'_t \right) \longrightarrow F_{ud}^{(1)} \left( \bar{u}'_p \gamma_{\mu} u'_r \right) \left( \bar{d}'_s \gamma^{\mu} d'_t \right) \left[ 0 + \mathcal{O}(y_i^4) \right]\, .
\end{equation}
We noticed above that for the last operator to be CP violating, we have to take off-diagonal currents in flavor space. At the considered order those are not present and all the components of this Wilson coefficient are CP even. The expansions of all other Wilson coefficients which are not shown can be trivially obtained from those presented here. 

Assuming the smaller symmetry group $U(2)^5$ we have to consider more spurions to make all Yukawa interactions invariant. The Yukawa matrices can then be parametrized in terms of the spurions as follows
\begin{equation}
Y_u =  \lambda_{t} \begin{pmatrix}
\Delta_u & x_{t} V_q \\
0 & 1 
\end{pmatrix}\qquad
Y_d =  \lambda_{b} \begin{pmatrix}
\Delta_d & x_{b} V_q \\
0 & 1 
\end{pmatrix}\qquad
Y_e =  \lambda_{\tau} \begin{pmatrix}
\Delta_e & x_{\tau} V_l \\
0 & 1 
\end{pmatrix}.
\end{equation}
In the up-quark basis this leaves us with the following expressions for the spurions in terms of Standard Model parameters
\begin{equation}
\begin{split}
\Delta_e = \text{diag}(\delta_e',&\delta_e) \qquad
\Delta_{u} = \text{diag}(\delta_{u}',\delta_{u}) \qquad \Delta_{d} = O_d^T \text{diag}(\delta_{d}',\delta_{d}) \\
& V_{q} = \lambda_t \begin{pmatrix}
V_{ub} \\
V_{cb}
\end{pmatrix}\qquad V_l \in \mathbb{C}^2 \ \text{unconstrained}
\end{split}
\end{equation}
with
\begin{equation}
\begin{split}
(\delta_u',\delta_d',\delta_e') \approx \left( \frac{\lambda_u}{\lambda_t},\frac{\lambda_d}{\lambda_b}, \frac{\lambda_e}{\lambda_{\tau}} \right) \qquad & \qquad \left(\delta_u,\delta_d,\delta_e \right) \approx \left( \frac{\lambda_c}{\lambda_t}, \frac{\lambda_s}{\lambda_b}, \frac{\lambda_{\mu}}{\lambda_{\tau}} \right) \\
O_d = 
\begin{pmatrix}
c_d & s_d \\
-s_d & c_d 
\end{pmatrix} \qquad \frac{s_d}{c_d} & = \frac{|V_{td}^*|}{|V_{ts}^*|} \qquad \alpha_d = \arg\left(\frac{V_{td}^*}{V_{ts}^*}\right)\, .
\end{split}
\end{equation}
\add{However, there are some left-over $U(2)^5$ parameters that can in general not be related to measurable objects. Ignoring the leptonic parameters which are irrelevant to us, the phase of $V_q$ for example can only be fixed in the limit where $x_t \to 0$ \cite{SMEFTFlavorSym2}.} With these definitions we can once again construct all terms appearing in the expression of the neutron EDM. Since we have a consistent power counting now, we choose to work at an accuracy of $\mathcal{O}(10^{-2})$ which means that we have to keep terms up to $\mathcal{O}(V^2,\Delta)$. Following the notation in Ref.~\cite{SMEFTFlavorSym}, we find, showing only the terms that can contribute the neutron EDM,
\begin{equation}
C'_{\substack{uB\\pr}} \mathcal{O}'_{\substack{uB\\pr}} \supset  f_{uB} \left[ \alpha_1 \bar{q}'_3 \sigma^{\mu\nu} u'_3 \widetilde{H} B_{\mu\nu} + \beta_1 \bar{Q}^{'p} V_q^p \sigma^{\mu\nu} u^{'}_3 \widetilde{H} B_{\mu\nu} + \rho_1 \bar{Q}'_p \sigma^{\mu\nu} (\Delta_u)_{pr} U'_r \widetilde{H} B_{\mu\nu} \right]
\end{equation}
\begin{equation}
C'_{\substack{Hud\\pr}} \mathcal{O}'_{\substack{Hud\\pr}} \supset  f_{Hud} \left( \tilde{H}^{\dagger} i D_{\mu} H \right) \left( \alpha_1 \bar{u}'_3 \gamma^{\mu} d'_3 \right)
\end{equation}
\begin{equation}
C'_{\substack{uH\\prst}} \mathcal{O}'_{\substack{uH\\prst}} \supset f_{uH} |H|^2 \left[ \rho_1 \bar{Q}'_p \widetilde{H} (\Delta_u)_{pr} U'_r + \beta_1 \bar{Q}^{'p}V_q^p \widetilde{H}d_3' + \alpha_1 \bar{q}_3' \widetilde{H} u_3' \right]
\end{equation}
\begin{equation}
C_{\substack{lequ\\prst}}^{'(3)} \mathcal{O}_{\substack{lequ\\prst}}^{'(3)} \supset f_{lequ}^{(3)} \rho_1 \left( \bar{l}_3^{'i} \sigma_{\mu\nu} e'_3 \right) \epsilon_{ij} \left(\bar{Q}_s^{'j} \sigma^{\mu\nu} (\Delta_u)_{st} U'_t \right)
\end{equation}
\begin{equation}
C_{\substack{quqd\\prst}}^{'(1)} \mathcal{O}_{\substack{quqd\\prst}}^{'(1)} \supset f_{quqd}^{(1)} \left[ \rho_1 \left( \bar{q}_3^{'i} (\Delta_u)_{pr} U'_r \right) \epsilon_{ij} \left(\bar{Q}_p^{'j} d'_3 \right) + \rho_2 \left( \bar{Q}_p^{'i} (\Delta_d)_{pr} u'_3 \right) \epsilon_{ij} \left(\bar{q}_3^{'j} D'_r \right) \right]
\end{equation}
\begin{equation}
C_{\substack{qu\\prst}}^{'(1)} \mathcal{O}_{\substack{qu\\prst}}^{'(1)} \supset f_{qu}^{(1)} \rho_1 \left( \bar{Q}'_p \gamma_{\mu} q'_3 \right) (\Delta_u)_{pr} \left( \bar{u}'_3\gamma^{\mu} U'_r \right)
\end{equation}
\begin{equation}
\begin{split}
C_{\substack{qd\\prst}}^{'(1)} \mathcal{O}_{\substack{qd\\prst}}^{'(1)} & \supset f_{qd}^{(1)} \left[\rho_1 \left( \bar{Q}'_p \gamma_{\mu} q'_3 \right) (\Delta_d)_{pr} \left( \bar{d}'_3\gamma^{\mu} D'_r \right) + \beta_1 \left( \bar{Q}^{'p} V_q^p \gamma_{\mu} q'_3 \right) \left( \bar{D}'_r \gamma^{\mu} D'_r \right)  \right. \\
& \left. \qquad\quad + c_1 \left( \bar{Q}^{'p} V_q^p \gamma_{\mu} V_q^{\dagger r} Q^{'r} \right) \left( \bar{D}'_s \gamma^{\mu} D'_s \right) \right]\, .
\end{split}
\end{equation}
\begin{equation}
C_{\substack{ud\\prst}}^{'(1)} \mathcal{O}_{\substack{ud\\prst}}^{'(1)} \supset  f_{ud}^{(1)} \left( \alpha_1 \bar{u}'_3 \gamma^{\mu} u'_3 \bar{d}'_3 \gamma^{\mu} d'_3 \right)
\end{equation}
For simplicity we will adopt the notation from Ref.~\cite{SMEFTFlavorSym} and write $C_X(Y) = f_X Y$, i.e. for example $C_{uG}(\rho_1) = f_{uG} \rho_1$.

\clearpage

\section{Bounds on Wilson coefficients and UV scale $\Lambda$}\label{app:bounds}
In this appendix we present the bounds on all Wilson coefficients that appear in the expression of the electron and neutron EDM. To give more meaningful bounds we factor out their naturally expected scaling in the Standard Model couplings. We also obtained bounds on the scale of new physics $\Lambda$ by rescaling the Wilson coefficients by their natural scaling and demanding the remaining Wilson coefficient to be of order 1.

\subsection{Electron EDM}
\renewcommand{\arraystretch}{1.3}
\begin{table}[H]
\begin{center}
\hspace{-20pt}
\begin{minipage}{.45\linewidth}
\begin{center}
\begin{tabular}{ccc}
	\hline
	Operator & Tree & Tree+Loop\\
	\hline
	$\mathrm{Im}\:C_{\substack{eB\\11}}$&$1.37 \cdot 10^{-5}\lambda_eg^{'}$&$1.70 \cdot 10^{-5}\lambda_eg^{'}$\\
	$\mathrm{Im}\:C_{\substack{eW\\11}}$&$1.37 \cdot 10^{-5}\lambda_eg$&$1.68 \cdot 10^{-5}\lambda_eg$\\[5pt]
	\hline
\end{tabular}\\[10pt]
\begin{tabular}{ccc}
	\hline
	Operator & RGE only & RGE + finite\\
	\hline
	$C_{H\wtilde{B}}$&$5.27 \cdot 10^{-3}g^{'2}$&$3.08 \cdot 10^{-3}g^{'2}$\\
	$C_{H\wtilde{W}}$&$1.95 \cdot 10^{-3}g^2$&$1.18 \cdot 10^{-3}g^2$\\
	$C_{HW\wtilde{B}}$&$1.52 \cdot 10^{-3}gg^{'}$&$2.12 \cdot 10^{-3}gg^{'}$\\
	$C_{\wtilde{W}}$&---&$1.59 \cdot 10^{-2}g^3$\\[5pt]
	\hline
\end{tabular}
\end{center}
\end{minipage}
\hspace{20pt}
\begin{minipage}{.45\linewidth}
\begin{tabular}{ccc}
	\hline
	Operator & RGE only & RGE + finite \\
	\hline
	$\mathrm{Im}\:C_{\substack{lequ\\1111}}^{(3)}$&$5.43 \cdot 10^{4}\lambda_e\lambda_u$&---\\
	$\mathrm{Im}\:C_{\substack{lequ\\1122}}^{(3)}$&$1.57 \cdot 10^{-1}\lambda_e\lambda_c$&---\\
	$\mathrm{Im}\:C_{\substack{lequ\\1133}}^{(3)}$&$4.33 \cdot 10^{-5}\lambda_e\lambda_t$&---\\
	$\mathrm{Im}\:C_{\substack{le\\1221}}$&---&$8.15 \cdot 10^{-5}g^{'2}$\\
	$\mathrm{Im}\:C_{\substack{le\\1331}}$&---&$4.85 \cdot 10^{-6}g^{'2}$\\[5pt]
	\hline
\end{tabular}
\end{minipage}
\end{center}
\caption{Upper bounds on the Wilson coefficients contributing to the EDM of the electron assuming $\Lambda=5$ TeV and no further assumptions. In the upper left table the Wilson coefficients which can enter at tree level are presented. The column 'Tree+Loop' presents bounds including the tree level contribution, the RG running and all finite terms. In the other tables one can find all other Wilson coefficients which cannot enter at tree level. The left column shows only RG running, while the right column shows both RG running and finite terms. Above, the parameter $\lambda_i$ is the $i^{\rm th}$ diagonal entry of the lepton Yukawa matrix.}
\label{tab:eEDMBoundsWC}
\end{table}
\renewcommand{\arraystretch}{1.0}
\renewcommand{\arraystretch}{1.3}
\begin{table}[H]
\begin{center}
\begin{minipage}{.45\linewidth}
\begin{center}
\begin{tabular}{ccc}
	\hline
	Operator & Tree & Tree+Loop\\
	\hline
	$\mathrm{Im}\:C_{\substack{eB\\11}}$&$1.35 \cdot 10^{3}$&$1.11 \cdot 10^{3}$\\
	$\mathrm{Im}\:C_{\substack{eW\\11}}$&$1.35 \cdot 10^{3}$&$1.13 \cdot 10^{3}$\\[5pt]
	\hline
\end{tabular}\\[10pt]
\begin{tabular}{ccc}
	\hline
	Operator & RGE only & RGE + finite\\
	\hline
	$C_{H\wtilde{B}}$&$1.03 \cdot 10^{2}$&$1.2 \cdot 10^{2}$\\
	$C_{H\wtilde{W}}$&$1.1 \cdot 10^{2}$&$1.27 \cdot 10^{2}$\\
	$C_{HW\wtilde{B}}$&$1.62 \cdot 10^{2}$&$1.46 \cdot 10^{2}$\\
	$C_{\wtilde{W}}$&---&$3.96 \cdot 10^{1}$\\[5pt]
	\hline
\end{tabular}
\end{center}
\end{minipage}
\hspace{10pt}
\begin{minipage}{.45\linewidth}
\begin{tabular}{ccc}
	\hline
	Operator & RGE only & RGE + finite \\
	\hline
	$\mathrm{Im}\:C_{\substack{lequ\\1111}}^{(3)}$&$1.73 \cdot 10^{-2}$&---\\
	$\mathrm{Im}\:C_{\substack{lequ\\1122}}^{(3)}$&$1.30 \cdot 10^{1}$&---\\
	$\mathrm{Im}\:C_{\substack{lequ\\1133}}^{(3)}$&$1.16 \cdot 10^{3}$&---\\
	$\mathrm{Im}\:C_{\substack{le\\1221}}$&---&$5.54 \cdot 10^{2}$\\
	\hline
\end{tabular}
\end{minipage}
\end{center}
\caption{Lower bounds on the UV scale $\Lambda$ in TeV assuming the natural scaling for all Wilson coefficients as given in the previous table and no further assumptions. The labeling of the tables is the same as for the bounds on the Wilson coefficients.}
\label{tab:eEDMBoundsLam}
\end{table}
\renewcommand{\arraystretch}{1.0}

\subsection{Neutron EDM}
\subsubsection{Bounds without Flavor Symmetries}
\label{sec:nEDMwoFlavor}
\renewcommand{\arraystretch}{1.3}
\begin{table}[H]
\begin{center}
\hspace{-20pt}
\begin{minipage}{.45\linewidth}
\begin{tabular}{ccc}
	\hline
	Operator & Tree & Tree+Loop\\
	\hline
	$\mathrm{Im}\:C_{\underset{11}{uG}}$&$1.61 \cdot 10^{-2}\lambda_u g_s$&$3.91 \cdot 10^{-3}\lambda_u g_s$\\
	$\mathrm{Im}\:C_{\underset{11}{uB}}$&$2.59 \cdot 10^{-2}\lambda_u g'$&$5.12 \cdot 10^{-2}\lambda_u g'$\\
	$\mathrm{Im}\:C_{\underset{11}{uW}}$&$2.59 \cdot 10^{-2}\lambda_u g$&$4.19 \cdot 10^{-2}\lambda_u g$\\
	$\mathrm{Im}\:C_{\underset{11}{dG}}$&$3.73 \cdot 10^{-3}\lambda_d g_s$&$1.11\cdot 10^{-3}\lambda_d g_s$\\
	$\mathrm{Im}\:C_{\underset{11}{dB}}$&$3.11 \cdot 10^{-3}\lambda_d g'$&$6.48 \cdot 10^{-3}\lambda_d g'$\\
	$\mathrm{Im}\:C_{\underset{11}{dW}}$&$3.11 \cdot 10^{-3}\lambda_d g$&$5.44 \cdot 10^{-3}\lambda_d g$\\
	$\mathrm{Im}\:C_{\underset{22}{dB}}$&$4.54 \cdot 10^{-2}\lambda_s g'$&$9.62 \cdot 10^{-2}\lambda_s g'$\\
	$\mathrm{Im}\:C_{\underset{22}{dW}}$&$4.54 \cdot 10^{-2}\lambda_s g$&$8.95 \cdot 10^{-2}\lambda_s g$\\[5pt]
	\hline
\end{tabular}
\end{minipage}
\hspace{20pt}
\begin{minipage}{.45\linewidth}
\begin{tabular}{ccc}
	\hline
	Operator & RGE only & RGE + finite \\
	\hline
	$\mathrm{Im}\:C_{\underset{22}{dG}}$&$7.42 \cdot 10^{-2}\lambda_s g_s$&$2.19\cdot 10^{-2}\lambda_s g_s$\\
	$\mathrm{Im}\:C_{\underset{22}{uG}}$&---&$1.65 \cdot 10^{-2}\lambda_c g_s$\\
	$\mathrm{Im}\:C_{\underset{33}{dG}}$&---&$1.65 \cdot 10^{-2}\lambda_b g_s$\\
	$\mathrm{Im}\:C_{\underset{33}{uG}}$&---&$1.65 \cdot 10^{-2}\lambda_t g_s$\\[5pt]
	\hline
\end{tabular}
\end{minipage}
\end{center}
\caption{Upper bounds on the Wilson coefficients of the dipole operators assuming $\Lambda=5$ TeV and no further assumptions. On the left-hand side the coefficients are presented which can enter at tree level. The column 'Tree+Loop' presents bounds including the tree level contribution, the RG running and all finite terms. On the right-hand side one can find all elements which cannot enter at tree level. The left column shows only RG running, while the right column shows both RG running and finite terms. Above, the parameter $\lambda_i$ is the $i^{\rm th}$ diagonal entry of the corresponding diagonalized quark Yukawa matrix here and in all tables that follow.}
\label{tab:nEDMBoundsDipWC}
\end{table}
\renewcommand{\arraystretch}{1.0}
\renewcommand{\arraystretch}{1.3}
\begin{table}[H]
\begin{center}
\hspace{-20pt}
\begin{minipage}{.45\linewidth}
\begin{tabular}{ccc}
	\hline
	Operator & RGE only & RGE + finite\\
	\hline
	$C_{H\wtilde{G}}$&$9.40 \cdot 10^{-3}g_s^2$&$7.81 \cdot 10^{-3}g_s^2$\\
	$C_{H\wtilde{B}}$&$2.04 \cdot 10^{0}g'^{2}$&$ 1.53\cdot 10^{0}g'^{2}$\\
	$C_{H\wtilde{W}}$&$2.99 \cdot 10^{-1}g^2$&$2.62 \cdot 10^{-1}g^2$\\
	$C_{HW\wtilde{B}}$&$1.76 \cdot 10^{-1}gg'$&$1.61 \cdot 10^{-1}gg'$\\
	$C_{\wtilde{W}}$&---&$3.46 \cdot 10^{0}g^{3}$\\
	$C_{\wtilde{G}}$&$4.74 \cdot 10^{-5}g_s^3$&$6.91 \cdot 10^{-5}g_s^3$\\
	\hline
\end{tabular}
\end{minipage}
\hspace{20pt}
\begin{minipage}{.45\linewidth}
\begin{tabular}{ccc}
	\hline
	Operator & RGE only & RGE + finite\\
	\hline
	$\mathrm{Im}\:C_{\underset{11}{Hud}}$&$1.87 \cdot 10^{-2}g^{'2}$&$2.03 \cdot 10^{-2}g^{'2}$\\
	$\mathrm{Im}\:C_{\underset{31}{Hud}}$&---&$1.03 \cdot 10^{-2}g^{'2}$\\
	$\mathrm{Re}\:C_{\underset{31}{Hud}}$&---&$3.53 \cdot 10^{-3}g^{'2}$\\
	$\mathrm{Im}\:C_{\underset{11}{uH}}$&---& $1.33 \cdot 10^{9}\lambda_u$\\
	$\mathrm{Im}\:C_{\underset{11}{dH}}$&---&$1.33 \cdot 10^{9}\lambda_d$\\
	\hline
\end{tabular}
\end{minipage}
\end{center}
\caption{Upper bounds on the Wilson coefficients of the bosonic operators on the left and the $\psi\bar\psi H^2D$ and $\psi^2H^3$ type operators on the right assuming $\Lambda=5$ TeV and no further assumptions. The 'RGE + finite' column for $C_{\underset{11}{Hud}}$ also includes the tree level contribution.}
\label{tab:nEDMBoundsBosonWC}
\end{table}
\renewcommand{\arraystretch}{1}
\renewcommand{\arraystretch}{1.3}
\begin{table}[H]
\begin{center}
\begin{minipage}{.45\linewidth}
\begin{tabular}{ccc}
	\hline
	Operator & RGE only & RGE + finite\\
	\hline
	$\mathrm{Im}\:C_{\underset{1111}{lequ}}^{'(3)}$&$7.54 \cdot 10^{9}\lambda_e\lambda_u$&--- \\
	$\mathrm{Im}\:C_{\underset{2211}{lequ}}^{'(3)}$&$1.76 \cdot 10^{5}\lambda_{\mu}\lambda_u$&--- \\
	$\mathrm{Im}\:C_{\underset{2211}{lequ}}^{('3)}$&$6.21 \cdot 10^{2}\lambda_{\tau}\lambda_u$&--- \\
	$\mathrm{Im}\:V_{1i}^{\dagger}C_{\underset{i111}{quqd}}^{'(1)}$&$1.88 \cdot 10^{7}\lambda_u\lambda_d$&$1.84 \cdot 10^{7}\lambda_u\lambda_d$ \\
	$\mathrm{Im}\:V_{1i}^{\dagger}C_{\underset{i111}{quqd}}^{'(8)}$&$3.82 \cdot 10^{7}\lambda_u\lambda_d$&$3.73 \cdot 10^{7}\lambda_u\lambda_d$ \\
	$\mathrm{Im}\:V_{1i}^{\dagger}C_{\underset{i221}{quqd}}^{'(1)}$&$7.79 \cdot 10^{2}\lambda_c\lambda_d$&--- \\
	$\mathrm{Im}\:V_{1i}^{\dagger}C_{\underset{i221}{quqd}}^{'(8)}$&$1.56 \cdot 10^{3}\lambda_c\lambda_d$&--- \\
	$\mathrm{Im}\:V_{1i}^{\dagger}C_{\underset{i331}{quqd}}^{'(1)}$&$9.86 \cdot 10^{-2}\lambda_t\lambda_d$&--- \\
	$\mathrm{Im}\:V_{1i}^{\dagger}C_{\underset{i331}{quqd}}^{'(8)}$&$1.98 \cdot 10^{-1}\lambda_t\lambda_d$&--- \\	
	
	$\mathrm{Im}\:V_{2i}^{\dagger}C_{\underset{i112}{quqd}}^{'(1)}$&$9.35 \cdot 10^{5}\lambda_u\lambda_s$&--- \\
	$\mathrm{Im}\:V_{2i}^{\dagger}C_{\underset{i111}{quqd}}^{'(8)}$&$1.03 \cdot 10^{7}\lambda_u\lambda_s$&--- \\
	$\mathrm{Im}\:V_{2i}^{\dagger}C_{\underset{i222}{quqd}}^{'(1)}$&$2.56 \cdot 10^{4}\lambda_c\lambda_s$&--- \\
	$\mathrm{Im}\:V_{2i}^{\dagger}C_{\underset{i222}{quqd}}^{'(8)}$&$1.92 \cdot 10^{4}\lambda_c\lambda_s$&--- \\
	$\mathrm{Im}\:V_{2i}^{\dagger}C_{\underset{i332}{quqd}}^{'(1)}$&$3.24 \cdot 10^{0}\lambda_t\lambda_s$&--- \\
	$\mathrm{Im}\:V_{2i}^{\dagger}C_{\underset{i332}{quqd}}^{'(8)}$&$2.43 \cdot 10^{0}\lambda_t\lambda_s$&--- \\
	$\mathrm{Im}\:V_{3i}^{\dagger}C_{\underset{i113}{quqd}}^{'(1)}$&$5.15 \cdot 10^{2}\lambda_u\lambda_b$&--- \\
	$\mathrm{Im}\:V_{3i}^{\dagger}C_{\underset{i113}{quqd}}^{'(8)}$&$5.65 \cdot 10^{3}\lambda_u\lambda_b$&--- \\[10pt]
	\hline
\end{tabular}
\end{minipage}
\hspace{35pt}
\begin{minipage}{.45\linewidth}
\begin{tabular}{ccc}
	\hline
	Operator & RGE only & RGE + finite \\
	\hline
	$\mathrm{Im}\:C_{\underset{1221}{qu}}^{'(1)}$&---&$5.79 \cdot 10^{-2}g'^{2}$ \\
	$\mathrm{Im}\:C_{\underset{1221}{qu}}^{'(8)}$&---&$4.70 \cdot 10^{-2}g'^{2}$ \\
	$\mathrm{Im}\:C_{\underset{1331}{qu}}^{'(1)}$&---&$3.76 \cdot 10^{-4}g'^{2}$ \\
	$\mathrm{Im}\:C_{\underset{1331}{qu}}^{'(8)}$&---&$4.03 \cdot 10^{-4}g'^{2}$ \\	
	$\mathrm{Im}\:V_{1i}^{\dagger}V_{j1}C_{\underset{ij11}{qd}}^{'(1)}$&---&$7.66 \cdot 10^{-1}g'^{2}$ \\
	$\mathrm{Im}\:V_{1i}^{\dagger}V_{j1}C_{\underset{ij11}{qd}}^{'(8)}$&---&$6.60 \cdot 10^{-1}g'^{2}$ \\
	$\mathrm{Im}\:V_{1i}^{\dagger}V_{j2}C_{\underset{ij21}{qd}}^{'(1)}$&---&$3.85 \cdot 10^{-1}g'^{2}$ \\
	$\mathrm{Im}\:V_{1i}^{\dagger}V_{j2}C_{\underset{ij21}{qd}}^{'(8)}$&---&$3.31 \cdot 10^{-1}g'^{2}$ \\
	$\mathrm{Im}\:V_{1i}^{\dagger}V_{j3}C_{\underset{ij31}{qd}}^{'(1)}$&---&$8.56 \cdot 10^{-2}g'^{2}$ \\
	$\mathrm{Im}\:V_{1i}^{\dagger}V_{j3}C_{\underset{ij31}{qd}}^{'(8)}$&---&$7.37 \cdot 10^{-3}g'^{2}$ \\	
	$\mathrm{Im}\:V_{2i}^{\dagger}V_{j1}C_{\underset{ij12}{qd}}^{'(1)}$&---&$3.35 \cdot 10^{3}g'^{2}$ \\
	$\mathrm{Im}\:V_{2i}^{\dagger}V_{j1}C_{\underset{ij12}{qd}}^{'(8)}$&---&$2.52 \cdot 10^{3}g'^{2}$ \\
	$\mathrm{Im}\:V_{2i}^{\dagger}V_{j2}C_{\underset{ij22}{qd}}^{'(1)}$&---&$1.68 \cdot 10^{2}g'^{2}$ \\
	$\mathrm{Im}\:V_{2i}^{\dagger}V_{j2}C_{\underset{ij22}{qd}}^{'(8)}$&---&$1.26 \cdot 10^{2}g'^{2}$ \\
	$\mathrm{Im}\:V_{2i}^{\dagger}V_{j3}C_{\underset{ij32}{qd}}^{'(1)}$&---&$3.75 \cdot 10^{0}g'^{2}$ \\
	$\mathrm{Im}\:V_{2i}^{\dagger}V_{j3}C_{\underset{ij32}{qd}}^{'(8)}$&---&$2.81 \cdot 10^{0}g'^{2}$ \\
	$\mathrm{Im}\:C_{\underset{1331}{ud}}^{'(1)}$&$9.30\cdot 10^{0}g'^{2}$&$7.17\cdot 10^{0}g'^{2}$\\
	$\mathrm{Im}\:C_{\underset{1321}{ud}}^{'(8)}$&$6.98\cdot 10^{0}g'^{2}$&$5.38\cdot 10^{0}g'^{2}$\\[10pt]
	\hline\vspace{-10pt}
\end{tabular}
\end{minipage}
\end{center}
\caption{Upper bounds on the Wilson coefficients of the 4-fermion operators assuming $\Lambda=5$ TeV and no further assumptions. Notice that each entry of the table corresponds to one the mass-basis Wilson coefficients that enter the expression of the neutron EDM. For all of them, however, the corresponding $C^\prime$ Wilson coefficients in the up-quark gauge basis are indicated, together with the CKM transformations needed to the change of basis. Wherever the phase of the CKM matrix enters in the bound, the bound is given on the real instead of the imaginary part. If the summation over the CKM elements gives a symmetric contribution for the operator $O_{qd}^{(1,8)}$, they have to be ignored because they are CP even and cannot give rise to an EDM. Note also, that the 'RGE + finite' column for $V_{1i}^{\dagger}C_{\underset{i111}{quqd}}^{'(1,8)}$ includes the tree level contribution.}
\label{tab:nEDMBounds4FWC}
\end{table}
\renewcommand{\arraystretch}{1}

\renewcommand{\arraystretch}{1.3}
\begin{table}[H]
\begin{center}
\begin{minipage}{.45\linewidth}
\begin{tabular}{ccc}
	\hline
	Operator & Tree & Tree+Loop\\
	\hline
	$\mathrm{Im}\:C_{\underset{11}{uG}}$&$3.93 \cdot 10^{1}$&$8.55 \cdot 10^{1}$\\
	$\mathrm{Im}\:C_{\underset{11}{uB}}$&$3.11 \cdot 10^{1}$&$1.98 \cdot 10^{1}$\\
	$\mathrm{Im}\:C_{\underset{11}{uW}}$&$3.11 \cdot 10^{1}$&$2.30 \cdot 10^{1}$\\
	$\mathrm{Im}\:C_{\underset{11}{dG}}$&$8.19 \cdot 10^{1}$&$1.65 \cdot 10^{2}$\\
	$\mathrm{Im}\:C_{\underset{11}{dB}}$&$8.96 \cdot 10^{1}$&$4.97 \cdot 10^{1}$\\
	$\mathrm{Im}\:C_{\underset{11}{dW}}$&$8.96 \cdot 10^{1}$&$5.94 \cdot 10^{1}$\\
	$\mathrm{Im}\:C_{\underset{22}{dB}}$&$2.35 \cdot 10^{1}$&$1.47 \cdot 10^{1}$\\
	$\mathrm{Im}\:C_{\underset{22}{dW}}$&$2.35 \cdot 10^{1}$&$1.54 \cdot 10^{1}$\\[5pt]
	\hline
\end{tabular}
\end{minipage}
\hspace{10pt}
\begin{minipage}{.45\linewidth}
\begin{tabular}{ccc}
	\hline
	Operator & RGE only & RGE + finite \\
	\hline
	$\mathrm{Im}\:C_{\underset{22}{dG}}$&$3.26 \cdot 10^{1}$&$1.99 \cdot 10^{1}$\\
	$\mathrm{Im}\:C_{\underset{22}{uG}}$&---&$3.89 \cdot 10^{1}$\\
	$\mathrm{Im}\:C_{\underset{33}{dG}}$&---&$3.89 \cdot 10^{1}$\\
	$\mathrm{Im}\:C_{\underset{33}{uG}}$&---&$3.89 \cdot 10^{1}$\\[5pt]
	\hline
\end{tabular}
\end{minipage}
\end{center}
\caption{Lower bounds on the UV scale $\Lambda$ in TeV assuming natural scaling of the dipole Wilson coefficients and no further assumptions. On the left-hand side the coefficients are presented which can enter at tree level. The column 'Tree+Loop' presents bounds including the tree level contribution, the RG running and all finite terms. On the right-hand side one can find all elements which cannot enter at tree level. The left column shows only RG running, while the right column shows both RG running and finite terms.}
\label{tab:nEDMBoundsDipLam}
\end{table}
\renewcommand{\arraystretch}{1.0}
\renewcommand{\arraystretch}{1.3}
\begin{table}[H]
\begin{center}
\begin{minipage}{.45\linewidth}
\begin{tabular}{ccc}
	\hline
	Operator & RGE only & RGE + finite\\
	\hline
	$C_{H\wtilde{G}}$&$6.73 \cdot 10^{1}$&$7.16 \cdot 10^{1}$\\
	$C_{H\wtilde{B}}$&$3.29 \cdot 10^{0}$&$3.94 \cdot 10^{0}$\\
	$C_{H\wtilde{W}}$&$9.97 \cdot 10^{0}$&$1.06 \cdot 10^{1}$\\
	$C_{HW\wtilde{B}}$&$1.33 \cdot 10^{1}$&$1.38 \cdot 10^{1}$\\
	$C_{\wtilde{W}}$&---&$2.69 \cdot 10^{0}$\\
	$C_{\wtilde{G}}$&$1.09 \cdot 10^{3}$&$1.01 \cdot 10^{3}$\\
	\hline
\end{tabular}
\end{minipage}
\hspace{20pt}
\begin{minipage}{.45\linewidth}
\begin{tabular}{ccc}
	\hline
	Operator & RGE only & RGE + finite\\
	\hline
	$\mathrm{Im}\:C_{\underset{11}{Hud}}$&$3.67 \cdot 10^{1}$&$3.53 \cdot 10^{1}$\\
	$\mathrm{Im}\:C_{\underset{31}{Hud}}$&---&$4.93 \cdot 10^{1}$\\
	$\mathrm{Re}\:C_{\underset{31}{Hud}}$&---&$8.42 \cdot 10^{1}$\\
	$\mathrm{Im}\:C_{\underset{11}{uH}}$&---& $1.37 \cdot 10^{-4}$\\
	$\mathrm{Im}\:C_{\underset{11}{dH}}$&---&$1.37 \cdot 10^{-4}$\\
	\hline
\end{tabular}
\end{minipage}
\end{center}
\caption{Lower bounds on the UV scale $\Lambda$ in TeV assuming natural scaling for the Wilson coefficients of the bosonic operators and no further assumptions. The 'RGE + finite' column for $C_{\underset{11}{Hud}}$ also includes the tree level contribution.}
\label{tab:nEDMBoundsBosonLam}
\end{table}
\renewcommand{\arraystretch}{1}
\renewcommand{\arraystretch}{1.3}
\begin{table}[H]
\begin{center}
\begin{minipage}{.45\linewidth}
\begin{tabular}{ccc}
	\hline
	Operator & RGE only & RGE + finite\\
	\hline
	$\mathrm{Im}\:C_{\underset{1111}{lequ}}^{'(3)}$&$4.02 \cdot 10^{-5}$&--- \\
	$\mathrm{Im}\:C_{\underset{2211}{lequ}}^{'(3)}$&$1.62 \cdot 10^{-3}$&--- \\
	$\mathrm{Im}\:C_{\underset{3311}{lequ}}^{'(3)}$&$1.49 \cdot 10^{-1}$&--- \\
	$\mathrm{Im}\:V_{1i}^{\dagger}C_{\underset{i111}{quqd}}^{'(1)}$&$5.85 \cdot 10^{-4}$&$6.14 \cdot 10^{-4}$ \\
	$\mathrm{Im}\:V_{1i}^{\dagger}C_{\underset{i111}{quqd}}^{'(8)}$&$6.42 \cdot 10^{-4}$&$6.55 \cdot 10^{-4}$ \\
	$\mathrm{Im}\:V_{1i}^{\dagger}C_{\underset{i221}{quqd}}^{'(1)}$&$1.32 \cdot 10^{-1}$&--- \\
	$\mathrm{Im}\:V_{1i}^{\dagger}C_{\underset{i221}{quqd}}^{'(8)}$&$8.84 \cdot 10^{-2}$&--- \\
	$\mathrm{Im}\:V_{1i}^{\dagger}C_{\underset{i331}{quqd}}^{'(1)}$&$1.88 \cdot 10^{1}$&--- \\
	$\mathrm{Im}\:V_{1i}^{\dagger}C_{\underset{i331}{quqd}}^{'(8)}$&$1.27 \cdot 10^{1}$&--- \\	
	$\mathrm{Im}\:V_{2i}^{\dagger}C_{\underset{i112}{quqd}}^{'(1)}$&$1.22 \cdot 10^{-3}$&--- \\
	$\mathrm{Im}\:V_{2i}^{\dagger}C_{\underset{i112}{quqd}}^{'(8)}$&$5.96 \cdot 10^{-4}$&--- \\
	$\mathrm{Im}\:V_{2i}^{\dagger}C_{\underset{i222}{quqd}}^{'(1)}$&$1.64 \cdot 10^{-2}$&--- \\
	$\mathrm{Im}\:V_{2i}^{\dagger}C_{\underset{i222}{quqd}}^{'(8)}$&$1.97 \cdot 10^{-2}$&--- \\
	$\mathrm{Im}\:V_{2i}^{\dagger}C_{\underset{i332}{quqd}}^{'(1)}$&$2.47 \cdot 10^{0}$&--- \\
	$\mathrm{Im}\:V_{2i}^{\dagger}C_{\underset{i332}{quqd}}^{'(8)}$&$2.94 \cdot 10^{0}$&--- \\
	$\mathrm{Im}\:V_{3i}^{\dagger}C_{\underset{i113}{quqd}}^{'(1)}$&$1.58 \cdot 10^{-1}$&--- \\
	$\mathrm{Im}\:V_{3i}^{\dagger}C_{\underset{i113}{quqd}}^{'(8)}$&$3.69 \cdot 10^{-2}$&--- \\[10pt]
	\hline
\end{tabular}
\end{minipage}
\hspace{35pt}
\begin{minipage}{.45\linewidth}
\begin{tabular}{ccc}
	\hline
	Operator & RGE only & RGE + finite\\
	\hline
$\mathrm{Im}\:C_{\underset{1221}{qu}}^{'(1)}$&---&$2.08 \cdot 10^{1}$ \\
	$\mathrm{Im}\:C_{\underset{1221}{qu}}^{'(8)}$&---&$2.31 \cdot 10^{1}$ \\
	$\mathrm{Im}\:C_{\underset{1331}{qu}}^{'(1)}$&---&$2.50 \cdot 10^{2}$ \\
	$\mathrm{Im}\:C_{\underset{1331}{qu}}^{'(8)}$&---&$2.59 \cdot 10^{2}$ \\	
	$\mathrm{Im}\:V_{1i}^{\dagger}V_{j1}C_{\underset{ij11}{qd}}^{'(1)}$&---&$1.81 \cdot 10^{0}$ \\
	$\mathrm{Im}\:V_{1i}^{\dagger}V_{j1}C_{\underset{ij11}{qd}}^{'(8)}$&---&$1.95 \cdot 10^{0}$ \\
	$\mathrm{Im}\:V_{1i}^{\dagger}V_{j2}C_{\underset{ij21}{qd}}^{'(1)}$&---&$8.06 \cdot 10^{0}$ \\
	$\mathrm{Im}\:V_{1i}^{\dagger}V_{j2}C_{\underset{ij21}{qd}}^{'(8)}$&---&$8.69 \cdot 10^{0}$ \\
	$\mathrm{Im}\:V_{1i}^{\dagger}V_{j3}C_{\underset{ij31}{qd}}^{'(1)}$&---&$5.40 \cdot 10^{1}$ \\
	$\mathrm{Im}\:V_{1i}^{\dagger}V_{j3}C_{\underset{ij31}{qd}}^{'(8)}$&---&$5.82 \cdot 10^{1}$ \\	
	$\mathrm{Im}\:V_{2i}^{\dagger}V_{j1}C_{\underset{ij12}{qd}}^{'(1)}$&---&$8.63 \cdot 10^{-2}$ \\
	$\mathrm{Im}\:V_{2i}^{\dagger}V_{j1}C_{\underset{ij12}{qd}}^{'(8)}$&---&$9.97 \cdot 10^{-2}$ \\
	$\mathrm{Im}\:V_{2i}^{\dagger}V_{j2}C_{\underset{ij22}{qd}}^{'(1)}$&---&$3.85 \cdot 10^{-1}$ \\
	$\mathrm{Im}\:V_{2i}^{\dagger}V_{j2}C_{\underset{ij22}{qd}}^{'(8)}$&---&$4.45 \cdot 10^{-1}$ \\
	$\mathrm{Im}\:V_{2i}^{\dagger}V_{j3}C_{\underset{ij32}{qd}}^{'(1)}$&---&$2.58 \cdot 10^{0}$ \\
	$\mathrm{Im}\:V_{2i}^{\dagger}V_{j3}C_{\underset{ij32}{qd}}^{'(8)}$&---&$2.98\cdot 10^{0}$ \\
	$\mathrm{Im}\:C_{\underset{1331}{ud}}^{'(1)}$&$1.26\cdot 10^{0}$&$1.61\cdot 10^{0}$\\
	$\mathrm{Im}\:C_{\underset{1321}{ud}}^{'(8)}$&$1.52\cdot 10^{0}$&$1.90\cdot 10^{0}$\\[10pt]
	\hline
\end{tabular}
\end{minipage}
\end{center}
\caption{Lower bounds on the UV scale $\Lambda$ in TeV assuming natural scaling of the Wilson coefficients of the 4-fermion operators. Notice that each entry of the table corresponds to one the mass-basis Wilson coefficients that enter the expression of the neutron EDM. For all of them, however, the corresponding $C^\prime$ Wilson coefficients in the up-quark gauge basis are indicated, together with the CKM transformations needed to the change of basis. Wherever the phase of the CKM matrix enters in the bound, the bound is given from the real instead of the imaginary part of the Wilson coefficient. If the summation over the CKM elements gives a symmetric contribution for the operator $O_{qd}^{(1,8)}$, they have to be ignored because they are CP even and cannot give rise to an EDM. Note also, that the 'RGE + finite' column for $V_{1i}^{\dagger}C_{\underset{i111}{quqd}}^{'(1,8)}$ includes the tree level contribution.}
\label{tab:nEDMBounds4FLam}
\end{table}
\renewcommand{\arraystretch}{1}

\subsubsection{Bounds with Flavor Symmetries}
\label{sec:nEDMwFlavor}

We present here the bounds obtained when a $U(3)^5$ or $U(2)^5$ flavor symmetry is imposed in the SMEFT. The results for the purely bosonic operators are not presented, since they are left unchanged with respect the flavor generic scenario discussed previously.
\renewcommand{\arraystretch}{1.3}
\begin{table}[H]
\begin{center}
\begin{minipage}{.45\linewidth}
\begin{tabular}{ccc}
	\hline
	Operator & Tree & Tree+Loop\\
	\hline
	$\mathrm{Im}\:F_{uG}$&$2.99 \cdot 10^{-2}$&$4.93 \cdot 10^{-3}$\\
	$\mathrm{Im}\:F_{uB}$&$9.25 \cdot 10^{-3}$&$1.83 \cdot 10^{-2}$\\
	$\mathrm{Im}\:F_{uW}$&$1.69 \cdot 10^{-2}$&$2.73 \cdot 10^{-2}$\\
	$\mathrm{Im}\:F_{dG}$&$7.10 \cdot 10^{-3}$&$1.89 \cdot 10^{-3}$\\
	$\mathrm{Im}\:F_{dB}$&$1.23 \cdot 10^{-3}$&$2.55 \cdot 10^{-3}$\\
	$\mathrm{Im}\:F_{dW}$&$2.24 \cdot 10^{-3}$&$3.88 \cdot 10^{-3}$\\
	$\mathrm{Im}\:F_{quqd}^{(1)}$&$5.90 \cdot 10^{7}$&$3.40 \cdot 10^{3}$\\
	$\mathrm{Im}\:F_{quqd}^{(8)}$&$5.90 \cdot 10^{7}$&$2.93 \cdot 10^{3}$\\
	$\mathrm{Im}\:F_{Hud}$&---&---\\
	\hline
\end{tabular}
\end{minipage}
\hspace{10pt}
\begin{minipage}{.45\linewidth}
\begin{tabular}{ccc}
	\hline
	Operator & RGE only & RGE + finite \\
	\hline
	$\mathrm{Im}\:F_{lequ}^{(3)}$&$6.19 \cdot 10^{2}$&---\\
	$\mathrm{Im}\:F_{qu}^{(1,8)}$&---&---\\
	$\mathrm{Im}\:F_{qd}^{(1,8)}$&---&---\\
	$\mathrm{Im}\:F_{ud}^{(1,8)}$&---&---\\
	$\mathrm{Im}\:F_{uH}$& --- &$1.33 \cdot 10^{9}$\\
	$\mathrm{Im}\:F_{dH}$&---&$1.33 \cdot 10^{9}$\\
	\hline
\end{tabular}
\end{minipage}
\vspace{-5pt}
\end{center}
\caption{Upper bounds on the Wilson coefficients assuming $\Lambda=5$ TeV and a $U(3)^5$ flavor symmetry, keeping terms up to $\mathcal{O}(y_{u,d,e}^2)$. The dipoles can again enter at tree level while the 4-fermion operators all only contribute via RG running. The operators $O_{qu}^{(1,8)},O_{qd}^{(1,8)},O_{ud}^{(1,8)}$ and $O_{Hud}$ are forbidden at the considered order in the spurions. The bounds on all bosonic operators are obviously the same as above.}
\label{tab:nEDMBoundsU3WC}
\end{table}
\renewcommand{\arraystretch}{1.0}

\renewcommand{\arraystretch}{1.3}
\begin{table}[H]
\begin{center}
\begin{minipage}{.45\linewidth}
\begin{tabular}{ccc}
	\hline
	Operator & Tree & Tree+Loop\\
	\hline
	$\mathrm{Im}\:F_{uG}$&$2.89 \cdot 10^{1}$&$7.45 \cdot 10^{1}$\\
	$\mathrm{Im}\:F_{uB}$&$5.20 \cdot 10^{1}$&$3.16 \cdot 10^{1}$\\
	$\mathrm{Im}\:F_{uW}$&$3.85 \cdot 10^{1}$&$2.82 \cdot 10^{1}$\\
	$\mathrm{Im}\:F_{dG}$&$5.93 \cdot 10^{1}$&$1.24 \cdot 10^{2}$\\
	$\mathrm{Im}\:F_{dB}$&$1.43 \cdot 10^{2}$&$7.51 \cdot 10^{1}$\\
	$\mathrm{Im}\:F_{dW}$&$1.06 \cdot 10^{2}$&$7.00 \cdot 10^{1}$\\
	$\mathrm{Im}\:F_{quqd}^{(1)}$&$6.51 \cdot 10^{-4}$&$4.76 \cdot 10^{-2}$\\
	$\mathrm{Im}\:F_{quqd}^{(8)}$&$6.51 \cdot 10^{-4}$&$5.21 \cdot 10^{-2}$\\
	$\mathrm{Im}\:F_{Hud}$&---&---\\
	\hline
\end{tabular}
\end{minipage}
\hspace{10pt}
\begin{minipage}{.45\linewidth}
\begin{tabular}{ccc}
	\hline
	Operator & RGE only & RGE + finite \\
	\hline
	$\mathrm{Im}\:F_{lequ}^{(3)}$&$1.50 \cdot 10^{-1}$&---\\
	$\mathrm{Im}\:F_{qu}^{(1,8)}$&---&---\\
	$\mathrm{Im}\:F_{qd}^{(1,8)}$&---&---\\
	$\mathrm{Im}\:F_{ud}^{(1,8)}$&---&---\\
	$\mathrm{Im}\:F_{uH}$&---&$1.37 \cdot 10^{-4}$\\
	$\mathrm{Im}\:F_{dH}$&---&$1.37 \cdot 10^{-4}$\\
	\hline
\end{tabular}
\end{minipage}
\end{center}
\caption{Lower bounds on the UV scale $\Lambda$ in TeV, assuming $F_i=1$ for all the coefficients of the operators, under a $U(3)^5$ flavor symmetry, keeping terms up to $\mathcal{O}(y_{u,d,e}^2)$.}
\label{tab:nEDMBoundsU3Lam}
\end{table}
\renewcommand{\arraystretch}{1.0}

\renewcommand{\arraystretch}{1.3}
\begin{table}[H]
\begin{center}
\begin{minipage}{.45\linewidth}
\begin{tabular}{ccc}
	\hline
	Operator & Tree & Tree+Loop\\
	\hline
	$\mathrm{Im}\:C_{uG}(\alpha_1)$&---&$3.05 \cdot 10^{-2}$\\
	$\mathrm{Im}\:C_{uG}(\rho_1)$&$2.97 \cdot 10^{-2}$&$5.83 \cdot 10^{-3}$\\
	$\mathrm{Im}\:C_{uB}(\rho_1)$&$9.19 \cdot 10^{-3}$&$1.82 \cdot 10^{-2}$\\
	$\mathrm{Im}\:C_{uW}(\rho_1)$&$1.68 \cdot 10^{-2}$&$2.71 \cdot 10^{-2}$\\
	$\mathrm{Im}\:C_{dG}(\alpha_1)$&---&$7.38 \cdot 10^{-4}$\\
	$\mathrm{Im}\:C_{dG}(\rho_1)$&$1.83 \cdot 10^{-4}$&$5.22 \cdot 10^{-5}$\\
	$\mathrm{Im}\:C_{dG}(\beta_1)$&---&$4.23 \cdot 10^{-1}$\\
	$\mathrm{Re}\:C_{dG}(\beta_1)$&---&$3.09 \cdot 10^{1}$\\
	$\mathrm{Im}\:C_{dB}(\rho_1)$&$3.17 \cdot 10^{-5}$&$6.59 \cdot 10^{-5}$\\
	$\mathrm{Im}\:C_{dW}(\rho_1)$&$5.78 \cdot 10^{-5}$&$1.00 \cdot 10^{-4}$\\
	$\mathrm{Im}\:C_{quqd}^{(1)}(\rho_1)$&---&$1.23 \cdot 10^{1}$\\
	$\mathrm{Im}\:C_{quqd}^{(1)}(\rho_2)$&---&$2.68 \cdot 10^{-3}$\\
	$\mathrm{Im}\:C_{quqd}^{(8)}(\rho_1)$&---&$1.35 \cdot 10^{2}$\\
	$\mathrm{Im}\:C_{quqd}^{(8)}(\rho_2)$&---&$5.66 \cdot 10^{-3}$\\
	\hline
\end{tabular}
\end{minipage}
\hspace{10pt}
\begin{minipage}{.45\linewidth}
\begin{tabular}{ccc}
	\hline
	Operator & RGE only & RGE + finite \\
	\hline
	$\mathrm{Im}\:C_{lequ}^{(3)}(\rho_1)$&$6.30 \cdot 10^{0}$&---\\	
	$\mathrm{Im}\:C_{qu}^{(1)}(\rho_1)$&---&$3.85 \cdot 10^{0}$\\
	$\mathrm{Im}\:C_{qu}^{(8)}(\rho_1)$&---&$4.12 \cdot 10^{0}$\\
	$\mathrm{Im}\:C_{qd}^{(1)}(\rho_1)$&---&$1.03 \cdot 10^{0}$\\
	$\mathrm{Im}\:C_{qd}^{(1)}(\beta_1)$&---&$2.35 \cdot 10^{3}$\\
	$\mathrm{Re}\:C_{qd}^{(1)}(c_1)$&---&$6.78 \cdot 10^{3}$\\
	$\mathrm{Im}\:C_{qd}^{(8)}(\rho_1)$&---&$9.80 \cdot 10^{-1}$\\
	$\mathrm{Im}\:C_{qd}^{(8)}(\beta_1)$&---&$8.73 \cdot 10^{3}$\\
	$\mathrm{Re}\:C_{qd}^{(8)}(c_1)$&---&$5.30 \cdot 10^{3}$\\
	$\mathrm{Im}\:C_{uH}(\rho_1)$&---&$1.32 \cdot 10^{9}$\\
	$\mathrm{Im}\:C_{dH}(\rho_1)$&---&$3.53 \cdot 10^{7}$\\
	\hline
\end{tabular}
\end{minipage}
\vspace{-5pt}
\end{center}
\caption{Upper bounds on the Wilson coefficients assuming $\Lambda=5$ TeV and a $U(2)^5$ flavor symmetry, keeping terms up to $\mathcal{O}(\Delta,V^2)$. We use the notation of Ref. \cite{SMEFTFlavorSym2} for the Wilson coefficients (see also App. \ref{app:flavorsym}). The operators $O_{Hud}$ and $O_{ud}^{(1,8)}$ don't contribute at the considered order.}
\label{tab:nEDMBoundsU2WC}
\end{table}
\renewcommand{\arraystretch}{1.0}

\renewcommand{\arraystretch}{1.3}
\begin{table}[H]
\begin{center}
\begin{minipage}{.45\linewidth}
\begin{tabular}{ccc}
	\hline
	Operator & Tree & Tree+Loop\\
	\hline
	$\mathrm{Im}\:C_{uG}(\alpha_1)$&---&$2.86 \cdot 10^{1}$\\
	$\mathrm{Im}\:C_{uG}(\rho_1)$&$2.90 \cdot 10^{1}$&$6.89 \cdot 10^{1}$\\
	$\mathrm{Im}\:C_{uB}(\rho_1)$&$5.22 \cdot 10^{1}$&$3.17 \cdot 10^{1}$\\
	$\mathrm{Im}\:C_{uW}(\rho_1)$&$3.86 \cdot 10^{1}$&$2.83 \cdot 10^{1}$\\
	$\mathrm{Im}\:C_{dG}(\alpha_1)$&---&$1.84 \cdot 10^{2}$\\
	$\mathrm{Im}\:C_{dG}(\rho_1)$&$3.69 \cdot 10^{2}$&$7.85 \cdot 10^{2}$\\
	$\mathrm{Im}\:C_{dG}(\beta_1)$&---&$7.69 \cdot 10^{0}$\\
	$\mathrm{Re}\:C_{dG}(\beta_1)$&---&$8.99 \cdot 10^{-1}$\\
	$\mathrm{Im}\:C_{dB}(\rho_1)$&$8.88 \cdot 10^{2}$&$3.55 \cdot 10^{2}$\\
	$\mathrm{Im}\:C_{dW}(\rho_1)$&$6.58 \cdot 10^{2}$&$3.88 \cdot 10^{2}$\\
	$\mathrm{Im}\:C_{quqd}^{(1)}(\rho_1)$&---&$1.28 \cdot 10^{0}$\\
	$\mathrm{Im}\:C_{quqd}^{(1)}(\rho_2)$&---&$1.36 \cdot 10^{2}$\\
	$\mathrm{Im}\:C_{quqd}^{(8)}(\rho_1)$&---&$3.39 \cdot 10^{-1}$\\
	$\mathrm{Im}\:C_{quqd}^{(8)}(\rho_2)$&---&$9.06 \cdot 10^{1}$\\
	\hline
\end{tabular}
\end{minipage}
\hspace{10pt}
\begin{minipage}{.45\linewidth}
\begin{tabular}{ccc}
	\hline
	Operator & RGE only & RGE + finite \\
	\hline
	$\mathrm{Im}\:C_{lequ}^{(3)}(\rho_1)$&$1.86 \cdot 10^{0}$&---\\
	$\mathrm{Im}\:C_{qu}^{(1)}(\rho_1)$&---&$2.58 \cdot 10^{0}$\\
	$\mathrm{Im}\:C_{qu}^{(8)}(\rho_1)$&---&$2.42 \cdot 10^{0}$\\
	$\mathrm{Im}\:C_{qd}^{(1)}(\rho_1)$&---&$4.92 \cdot 10^{0}$\\
	$\mathrm{Im}\:C_{qd}^{(1)}(\beta_1)$&---&$1.03 \cdot 10^{-1}$\\
	$\mathrm{Re}\:C_{qd}^{(1)}(c_1)$&---&$6.07 \cdot 10^{-2}$\\
	$\mathrm{Im}\:C_{qd}^{(8)}(\rho_1)$&---&$5.05 \cdot 10^{0}$\\
	$\mathrm{Im}\:C_{qd}^{(8)}(\beta_1)$&---&$5.35 \cdot 10^{-2}$\\
	$\mathrm{Re}\:C_{qd}^{(8)}(c_1)$&---&$6.87 \cdot 10^{-2}$\\
	$\mathrm{Im}\:C_{uH}(\rho_1)$&---&$1.37 \cdot 10^{-4}$\\
	$\mathrm{Im}\:C_{dH}(\rho_1)$&---&$8.41 \cdot 10^{-4}$\\
	\hline
\end{tabular}
\end{minipage}
\end{center}
\caption{Lower bounds on the UV scale $\Lambda$ in TeV assuming $C_X(Y)=1$ for all the coefficients of the operators, under a $U(2)^5$ flavor symmetry, keeping terms up to $\mathcal{O}(\Delta,V^2)$.}
\label{tab:nEDMBoundsU2Lam}
\end{table}



\bibliography{bib/loopEDMs}

\providecommand{\href}[2]{#2}\begingroup\raggedright\begin{thebibliography}{100}

\bibitem{Andreev2018}
V.~Andreev, \emph{Improved limit on the electric dipole moment of the
  electron}, \href{https://doi.org/10.1038/s41586-018-0599-8}{\emph{Nature}
  {\bfseries 562} (oct, 2018) 355--360}.

\bibitem{Bennett:2008dy}
{\scshape Muon (g-2)} collaboration, G.~W. Bennett et~al., \emph{{An Improved
  Limit on the Muon Electric Dipole Moment}},
  \href{https://doi.org/10.1103/PhysRevD.80.052008}{\emph{Phys. Rev. D}
  {\bfseries 80} (2009) 052008},
  [\href{https://arxiv.org/abs/0811.1207}{{\ttfamily 0811.1207}}].

\bibitem{Grozin:2008nw}
A.~G. Grozin, I.~B. Khriplovich and A.~S. Rudenko, \emph{{Electric dipole
  moments, from e to tau}},
  \href{https://doi.org/10.1134/S1063778809070138}{\emph{Phys. Atom. Nucl.}
  {\bfseries 72} (2009) 1203--1205},
  [\href{https://arxiv.org/abs/0811.1641}{{\ttfamily 0811.1641}}].

\bibitem{PhysRevLett.124.081803}
C.~Abel, S.~Afach, N.~J. Ayres, C.~A. Baker, G.~Ban, G.~Bison et~al.,
  \emph{Measurement of the permanent electric dipole moment of the neutron},
  \href{https://doi.org/10.1103/PhysRevLett.124.081803}{\emph{Phys. Rev. Lett.}
  {\bfseries 124} (Feb, 2020) 081803}.

\bibitem{Buttazzo:2020eyl}
D.~Buttazzo and P.~Paradisi, \emph{{Probing the muon g-2 anomaly at a Muon
  Collider}},  \href{https://arxiv.org/abs/2012.02769}{{\ttfamily 2012.02769}}.

\bibitem{Doyle:2016}
J.~Doyle, ``{{\it Search for the Electric Dipole Moment of the Electron with
  Thorium Monoxide -- The ACME Experiment}}.''
  {\href{https://online.kitp.ucsb.edu/online/nuclear_c16/doyle/}{Talk at the
  KITP, September 2016}}.

\bibitem{Ayres:2021hoq}
{\scshape n2EDM} collaboration, N.~J. Ayres et~al., \emph{{The design of the
  n2EDM experiment}},  \href{https://arxiv.org/abs/2101.08730}{{\ttfamily
  2101.08730}}.

\bibitem{Pospelov:2013sca}
M.~Pospelov and A.~Ritz, \emph{{CKM benchmarks for electron electric dipole
  moment experiments}},
  \href{https://doi.org/10.1103/PhysRevD.89.056006}{\emph{Phys. Rev. D}
  {\bfseries 89} (2014) 056006},
  [\href{https://arxiv.org/abs/1311.5537}{{\ttfamily 1311.5537}}].

\bibitem{Pospelov:1991zt}
M.~E. Pospelov and I.~B. Khriplovich, \emph{{Electric dipole moment of the W
  boson and the electron in the Kobayashi-Maskawa model}}, {\emph{Sov. J. Nucl.
  Phys.} {\bfseries 53} (1991) 638--640}.

\bibitem{Booth:1993af}
M.~J. Booth, \emph{{The Electric dipole moment of the W and electron in the
  Standard Model}},  \href{https://arxiv.org/abs/hep-ph/9301293}{{\ttfamily
  hep-ph/9301293}}.

\bibitem{KHRIPLOVICH1982490}
I.~Khriplovich and A.~Zhitnitsky, \emph{What is the value of the neutron
  electric dipole moment in the kobayashi-maskawa model?},
  \href{https://doi.org/https://doi.org/10.1016/0370-2693(82)91121-2}{\emph{Physics
  Letters B} {\bfseries 109} (1982) 490--492}.

\bibitem{Czarnecki:1997bu}
A.~Czarnecki and B.~Krause, \emph{{Neutron electric dipole moment in the
  standard model: Valence quark contributions}},
  \href{https://doi.org/10.1103/PhysRevLett.78.4339}{\emph{Phys. Rev. Lett.}
  {\bfseries 78} (1997) 4339--4342},
  [\href{https://arxiv.org/abs/hep-ph/9704355}{{\ttfamily hep-ph/9704355}}].

\bibitem{Yamaguchi:2020eub}
Y.~Yamaguchi and N.~Yamanaka, \emph{{Large long-distance contributions to the
  electric dipole moments of charged leptons in the standard model}},
  \href{https://doi.org/10.1103/PhysRevLett.125.241802}{\emph{Phys. Rev. Lett.}
  {\bfseries 125} (2020) 241802},
  [\href{https://arxiv.org/abs/2003.08195}{{\ttfamily 2003.08195}}].

\bibitem{Jarlkog85}
C.~Jarlskog, \emph{Commutator of the quark mass matrices in the standard
  electroweak model and a measure of maximal $\mathrm{CP}$ nonconservation},
  \href{https://doi.org/10.1103/PhysRevLett.55.1039}{\emph{Phys. Rev. Lett.}
  {\bfseries 55} (Sep, 1985) 1039--1042}.

\bibitem{Smith2017}
C.~Smith and S.~Touati, \emph{Edm with and beyond flavor invariants},
  \href{https://arxiv.org/abs/1707.06805}{{\ttfamily 1707.06805}}.

\bibitem{Bona:2007vi}
{\scshape UTfit} collaboration, M.~Bona et~al., \emph{{Model-independent
  constraints on $\Delta F=2$ operators and the scale of new physics}},
  \href{https://doi.org/10.1088/1126-6708/2008/03/049}{\emph{JHEP} {\bfseries
  03} (2008) 049}, [\href{https://arxiv.org/abs/0707.0636}{{\ttfamily
  0707.0636}}].

\bibitem{UTFIT:2016}
{\scshape UTfit} collaboration, \emph{{ Latest results from UTfit }},  2016.

\bibitem{Giudice:2005rz}
G.~F. Giudice and A.~Romanino, \emph{{Electric dipole moments in split
  supersymmetry}},
  \href{https://doi.org/10.1016/j.physletb.2006.01.027}{\emph{Phys. Lett. B}
  {\bfseries 634} (2006) 307--314},
  [\href{https://arxiv.org/abs/hep-ph/0510197}{{\ttfamily hep-ph/0510197}}].

\bibitem{Nakai:2016atk}
Y.~Nakai and M.~Reece, \emph{{Electric Dipole Moments in Natural
  Supersymmetry}}, \href{https://doi.org/10.1007/JHEP08(2017)031}{\emph{JHEP}
  {\bfseries 08} (2017) 031},
  [\href{https://arxiv.org/abs/1612.08090}{{\ttfamily 1612.08090}}].

\bibitem{Cesarotti:2018huy}
C.~Cesarotti, Q.~Lu, Y.~Nakai, A.~Parikh and M.~Reece, \emph{{Interpreting the
  Electron EDM Constraint}},
  \href{https://doi.org/10.1007/JHEP05(2019)059}{\emph{JHEP} {\bfseries 05}
  (2019) 059}, [\href{https://arxiv.org/abs/1810.07736}{{\ttfamily
  1810.07736}}].

\bibitem{Aloni:2021wzk}
D.~Aloni, P.~Asadi, Y.~Nakai, M.~Reece and M.~Suzuki, \emph{{Spontaneous CP
  Violation and Horizontal Symmetry in the MSSM: Toward Lepton Flavor
  Naturalness}},  \href{https://arxiv.org/abs/2104.02679}{{\ttfamily
  2104.02679}}.

\bibitem{KerenZur:2012fr}
B.~Keren-Zur, P.~Lodone, M.~Nardecchia, D.~Pappadopulo, R.~Rattazzi and
  L.~Vecchi, \emph{{On Partial Compositeness and the CP asymmetry in charm
  decays}}, \href{https://doi.org/10.1016/j.nuclphysb.2012.10.012}{\emph{Nucl.
  Phys. B} {\bfseries 867} (2013) 394--428},
  [\href{https://arxiv.org/abs/1205.5803}{{\ttfamily 1205.5803}}].

\bibitem{Konig:2014iqa}
M.~K\"onig, M.~Neubert and D.~M. Straub, \emph{{Dipole operator constraints on
  composite Higgs models}},
  \href{https://doi.org/10.1140/epjc/s10052-014-2945-9}{\emph{Eur. Phys. J. C}
  {\bfseries 74} (2014) 2945},
  [\href{https://arxiv.org/abs/1403.2756}{{\ttfamily 1403.2756}}].

\bibitem{Panico:2015jxa}
G.~Panico and A.~Wulzer, \emph{{The Composite Nambu-Goldstone Higgs}},
  vol.~913.
\newblock Springer, 2016,
  \href{https://doi.org/10.1007/978-3-319-22617-0}{10.1007/978-3-319-22617-0}.

\bibitem{Dorsner:2016wpm}
I.~Dor\v{s}ner, S.~Fajfer, A.~Greljo, J.~F. Kamenik and N.~Ko\v{s}nik,
  \emph{{Physics of leptoquarks in precision experiments and at particle
  colliders}}, \href{https://doi.org/10.1016/j.physrep.2016.06.001}{\emph{Phys.
  Rept.} {\bfseries 641} (2016) 1--68},
  [\href{https://arxiv.org/abs/1603.04993}{{\ttfamily 1603.04993}}].

\bibitem{Fuyuto:2018scm}
K.~Fuyuto, M.~Ramsey-Musolf and T.~Shen, \emph{{Electric Dipole Moments from
  CP-Violating Scalar Leptoquark Interactions}},
  \href{https://doi.org/10.1016/j.physletb.2018.11.016}{\emph{Phys. Lett. B}
  {\bfseries 788} (2019) 52--57},
  [\href{https://arxiv.org/abs/1804.01137}{{\ttfamily 1804.01137}}].

\bibitem{Dekens:2018bci}
W.~Dekens, J.~de~Vries, M.~Jung and K.~K. Vos, \emph{{The phenomenology of
  electric dipole moments in models of scalar leptoquarks}},
  \href{https://doi.org/10.1007/JHEP01(2019)069}{\emph{JHEP} {\bfseries 01}
  (2019) 069}, [\href{https://arxiv.org/abs/1809.09114}{{\ttfamily
  1809.09114}}].

\bibitem{Altmannshofer:2020ywf}
W.~Altmannshofer, S.~Gori, H.~H. Patel, S.~Profumo and D.~Tuckler,
  \emph{{Electric dipole moments in a leptoquark scenario for the $B$-physics
  anomalies}}, \href{https://doi.org/10.1007/JHEP05(2020)069}{\emph{JHEP}
  {\bfseries 05} (2020) 069},
  [\href{https://arxiv.org/abs/2002.01400}{{\ttfamily 2002.01400}}].

\bibitem{Altmannshofer:2020shb}
W.~Altmannshofer, S.~Gori, N.~Hamer and H.~H. Patel, \emph{{Electron EDM in the
  complex two-Higgs doublet model}},
  \href{https://doi.org/10.1103/PhysRevD.102.115042}{\emph{Phys. Rev. D}
  {\bfseries 102} (2020) 115042},
  [\href{https://arxiv.org/abs/2009.01258}{{\ttfamily 2009.01258}}].

\bibitem{Hou:2021zqq}
W.-S. Hou, G.~Kumar and S.~Teunissen, \emph{{Charged Lepton EDM with Extra
  Yukawa Couplings}},  \href{https://arxiv.org/abs/2109.08936}{{\ttfamily
  2109.08936}}.

\bibitem{Logan:2020mdz}
H.~E. Logan, S.~Moretti, D.~Rojas-Ciofalo and M.~Song, \emph{{CP violation from
  charged Higgs bosons in the three Higgs doublet model}},
  \href{https://doi.org/10.1007/JHEP07(2021)158}{\emph{JHEP} {\bfseries 07}
  (2021) 158}, [\href{https://arxiv.org/abs/2012.08846}{{\ttfamily
  2012.08846}}].

\bibitem{Cheung:2020ugr}
K.~Cheung, A.~Jueid, Y.-N. Mao and S.~Moretti, \emph{{Two-Higgs-doublet model
  with soft $CP$ violation confronting electric dipole moments and colliders}},
  \href{https://doi.org/10.1103/PhysRevD.102.075029}{\emph{Phys. Rev. D}
  {\bfseries 102} (2020) 075029},
  [\href{https://arxiv.org/abs/2003.04178}{{\ttfamily 2003.04178}}].

\bibitem{Chun:2019oix}
E.~J. Chun, J.~Kim and T.~Mondal, \emph{{Electron EDM and Muon anomalous
  magnetic moment in Two-Higgs-Doublet Models}},
  \href{https://doi.org/10.1007/JHEP12(2019)068}{\emph{JHEP} {\bfseries 12}
  (2019) 068}, [\href{https://arxiv.org/abs/1906.00612}{{\ttfamily
  1906.00612}}].

\bibitem{Davoudiasl:2019lcg}
H.~Davoudiasl, I.~M. Lewis and M.~Sullivan, \emph{{Higgs Troika for Baryon
  Asymmetry}}, \href{https://doi.org/10.1103/PhysRevD.101.055010}{\emph{Phys.
  Rev. D} {\bfseries 101} (2020) 055010},
  [\href{https://arxiv.org/abs/1909.02044}{{\ttfamily 1909.02044}}].

\bibitem{Davoudiasl:2021syn}
H.~Davoudiasl, I.~M. Lewis and M.~Sullivan, \emph{{Multi-TeV signals of
  baryogenesis in a Higgs troika model}},
  \href{https://doi.org/10.1103/PhysRevD.104.015024}{\emph{Phys. Rev. D}
  {\bfseries 104} (2021) 015024},
  [\href{https://arxiv.org/abs/2103.12089}{{\ttfamily 2103.12089}}].

\bibitem{Abada:2018zra}
A.~Abada and T.~Toma, \emph{{Electric dipole moments in the minimal scotogenic
  model}}, \href{https://doi.org/10.1007/JHEP04(2018)030}{\emph{JHEP}
  {\bfseries 04} (2018) 030},
  [\href{https://arxiv.org/abs/1802.00007}{{\ttfamily 1802.00007}}].

\bibitem{FileviezPerez:2020oke}
P.~Fileviez~Perez and A.~D. Plascencia, \emph{{Electric dipole moments, new
  forces and dark matter}},
  \href{https://doi.org/10.1007/JHEP03(2021)185}{\emph{JHEP} {\bfseries 03}
  (2021) 185}, [\href{https://arxiv.org/abs/2008.09116}{{\ttfamily
  2008.09116}}].

\bibitem{Panico2018}
G.~Panico, A.~Pomarol and M.~Riembau, \emph{Eft approach to the electron
  electric dipole moment at the two-loop level},
  \href{https://arxiv.org/abs/1810.09413}{{\ttfamily 1810.09413}}.

\bibitem{Aebischer:2021uvt}
J.~Aebischer, W.~Dekens, E.~E. Jenkins, A.~V. Manohar, D.~Sengupta and
  P.~Stoffer, \emph{{Effective field theory interpretation of lepton magnetic
  and electric dipole moments}},
  \href{https://doi.org/10.1007/JHEP07(2021)107}{\emph{JHEP} {\bfseries 07}
  (2021) 107}, [\href{https://arxiv.org/abs/2102.08954}{{\ttfamily
  2102.08954}}].

\bibitem{Haisch2019}
U.~Haisch and A.~Hala, \emph{Bounds on cp-violating higgs-gluon interactions:
  the case of vanishing light-quark yukawa couplings},
  \href{https://arxiv.org/abs/1909.09373}{{\ttfamily 1909.09373}}.

\bibitem{Haisch:2021hcg}
U.~Haisch and G.~Koole, \emph{{Beautiful and charming chromodipole moments}},
  \href{https://doi.org/10.1007/JHEP09(2021)133}{\emph{JHEP} {\bfseries 09}
  (2021) 133}, [\href{https://arxiv.org/abs/2106.01289}{{\ttfamily
  2106.01289}}].

\bibitem{Kamenik:2011dk}
J.~F. Kamenik, M.~Papucci and A.~Weiler, \emph{{Constraining the dipole moments
  of the top quark}},
  \href{https://doi.org/10.1103/PhysRevD.85.071501}{\emph{Phys. Rev. D}
  {\bfseries 85} (2012) 071501},
  [\href{https://arxiv.org/abs/1107.3143}{{\ttfamily 1107.3143}}].

\bibitem{Brod:2013cka}
J.~Brod, U.~Haisch and J.~Zupan, \emph{{Constraints on CP-violating Higgs
  couplings to the third generation}},
  \href{https://doi.org/10.1007/JHEP11(2013)180}{\emph{JHEP} {\bfseries 11}
  (2013) 180}, [\href{https://arxiv.org/abs/1310.1385}{{\ttfamily 1310.1385}}].

\bibitem{Brod:2018pli}
J.~Brod and E.~Stamou, \emph{{Electric dipole moment constraints on
  CP-violating heavy-quark Yukawas at next-to-leading order}},
  \href{https://doi.org/10.1007/JHEP07(2021)080}{\emph{JHEP} {\bfseries 07}
  (2021) 080}, [\href{https://arxiv.org/abs/1810.12303}{{\ttfamily
  1810.12303}}].

\bibitem{Fuchs:2020uoc}
E.~Fuchs, M.~Losada, Y.~Nir and Y.~Viernik, \emph{{$CP$ violation from $\tau$,
  $t$ and $b$ dimension-6 Yukawa couplings - interplay of baryogenesis, EDM and
  Higgs physics}}, \href{https://doi.org/10.1007/JHEP05(2020)056}{\emph{JHEP}
  {\bfseries 05} (2020) 056},
  [\href{https://arxiv.org/abs/2003.00099}{{\ttfamily 2003.00099}}].

\bibitem{Fuyuto:2017xup}
K.~Fuyuto and M.~Ramsey-Musolf, \emph{{Top Down Electroweak Dipole Operators}},
  \href{https://doi.org/10.1016/j.physletb.2018.04.022}{\emph{Phys. Lett. B}
  {\bfseries 781} (2018) 492--498},
  [\href{https://arxiv.org/abs/1706.08548}{{\ttfamily 1706.08548}}].

\bibitem{Cirigliano:2016njn}
V.~Cirigliano, W.~Dekens, J.~de~Vries and E.~Mereghetti, \emph{{Is there room
  for CP violation in the top-Higgs sector?}},
  \href{https://doi.org/10.1103/PhysRevD.94.016002}{\emph{Phys. Rev. D}
  {\bfseries 94} (2016) 016002},
  [\href{https://arxiv.org/abs/1603.03049}{{\ttfamily 1603.03049}}].

\bibitem{Cirigliano:2019vfc}
V.~Cirigliano, A.~Crivellin, W.~Dekens, J.~de~Vries, M.~Hoferichter and
  E.~Mereghetti, \emph{{CP Violation in Higgs-Gauge Interactions: From Tabletop
  Experiments to the LHC}},
  \href{https://doi.org/10.1103/PhysRevLett.123.051801}{\emph{Phys. Rev. Lett.}
  {\bfseries 123} (2019) 051801},
  [\href{https://arxiv.org/abs/1903.03625}{{\ttfamily 1903.03625}}].

\bibitem{Cirigliano:2016nyn}
V.~Cirigliano, W.~Dekens, J.~de~Vries and E.~Mereghetti, \emph{{Constraining
  the top-Higgs sector of the Standard Model Effective Field Theory}},
  \href{https://doi.org/10.1103/PhysRevD.94.034031}{\emph{Phys. Rev. D}
  {\bfseries 94} (2016) 034031},
  [\href{https://arxiv.org/abs/1605.04311}{{\ttfamily 1605.04311}}].

\bibitem{Altmannshofer:2015qra}
W.~Altmannshofer, J.~Brod and M.~Schmaltz, \emph{{Experimental constraints on
  the coupling of the Higgs boson to electrons}},
  \href{https://doi.org/10.1007/JHEP05(2015)125}{\emph{JHEP} {\bfseries 05}
  (2015) 125}, [\href{https://arxiv.org/abs/1503.04830}{{\ttfamily
  1503.04830}}].

\bibitem{Chien:2015xha}
Y.~T. Chien, V.~Cirigliano, W.~Dekens, J.~de~Vries and E.~Mereghetti,
  \emph{{Direct and indirect constraints on CP-violating Higgs-quark and
  Higgs-gluon interactions}},
  \href{https://doi.org/10.1007/JHEP02(2016)011}{\emph{JHEP} {\bfseries 02}
  (2016) 011}, [\href{https://arxiv.org/abs/1510.00725}{{\ttfamily
  1510.00725}}].

\bibitem{Bonnefoy:2021tbt}
Q.~Bonnefoy, E.~Gendy, C.~Grojean and J.~T. Ruderman, \emph{{Beyond Jarlskog:
  699 invariants for CP violation in SMEFT}},
  \href{https://arxiv.org/abs/2112.03889}{{\ttfamily 2112.03889}}.

\bibitem{Cohen2020b}
T.~Cohen, N.~Craig, X.~Lu and D.~Sutherland, \emph{Is smeft enough?},
  \href{https://arxiv.org/abs/2008.08597}{{\ttfamily 2008.08597}}.

\bibitem{Grzadkowski2010}
B.~Grzadkowski, M.~Iskrzynski, M.~Misiak and J.~Rosiek, \emph{Dimension-six
  terms in the standard model lagrangian},
  \href{https://arxiv.org/abs/1008.4884}{{\ttfamily 1008.4884}}.

\bibitem{Jiang:2018pbd}
M.~Jiang, N.~Craig, Y.-Y. Li and D.~Sutherland, \emph{{Complete one-loop
  matching for a singlet scalar in the Standard Model EFT}},
  \href{https://doi.org/10.1007/JHEP02(2019)031}{\emph{JHEP} {\bfseries 02}
  (2019) 031}, [\href{https://arxiv.org/abs/1811.08878}{{\ttfamily
  1811.08878}}].

\bibitem{Gherardi:2020det}
V.~Gherardi, D.~Marzocca and E.~Venturini, \emph{{Matching scalar leptoquarks
  to the SMEFT at one loop}},
  \href{https://doi.org/10.1007/JHEP07(2020)225}{\emph{JHEP} {\bfseries 07}
  (2020) 225}, [\href{https://arxiv.org/abs/2003.12525}{{\ttfamily
  2003.12525}}].

\bibitem{Jenkins2017}
E.~E. Jenkins, A.~V. Manohar and P.~Stoffer, \emph{Low-energy effective field
  theory below the electroweak scale: Operators and matching},
  \href{https://doi.org/10.1007/JHEP03(2018)016}{\emph{JHEP 1803 (2018) 016}
  (Sept., 2017) },
  [\href{https://arxiv.org/abs/http://arxiv.org/abs/1709.04486v2}{{\ttfamily
  http://arxiv.org/abs/1709.04486v2}}].

\bibitem{Dekens2019}
W.~Dekens and P.~Stoffer, \emph{Low-energy effective field theory below the
  electroweak scale: matching at one loop},
  \href{https://doi.org/10.1007/JHEP10(2019)197}{\emph{JHEP 1910 (2019) 197}
  (Aug., 2019) }, [\href{https://arxiv.org/abs/1908.05295}{{\ttfamily
  1908.05295}}].

\bibitem{Jenkins2013a}
E.~E. Jenkins, A.~V. Manohar and M.~Trott, \emph{Renormalization group
  evolution of the standard model dimension six operators i: Formalism and
  lambda dependence},
  \href{https://arxiv.org/abs/http://arxiv.org/abs/1308.2627v4}{{\ttfamily
  http://arxiv.org/abs/1308.2627v4}}.

\bibitem{Jenkins2013}
E.~E. Jenkins, A.~V. Manohar and M.~Trott, \emph{Renormalization group
  evolution of the standard model dimension six operators ii: Yukawa
  dependence},
  \href{https://arxiv.org/abs/http://arxiv.org/abs/1310.4838v3}{{\ttfamily
  http://arxiv.org/abs/1310.4838v3}}.

\bibitem{Alonso2013}
R.~Alonso, E.~E. Jenkins, A.~V. Manohar and M.~Trott, \emph{Renormalization
  group evolution of the standard model dimension six operators iii: Gauge
  coupling dependence and phenomenology},
  \href{https://arxiv.org/abs/http://arxiv.org/abs/1312.2014v4}{{\ttfamily
  http://arxiv.org/abs/1312.2014v4}}.

\bibitem{Jenkins2017a}
E.~E. Jenkins, A.~V. Manohar and P.~Stoffer, \emph{Low-energy effective field
  theory below the electroweak scale: Anomalous dimensions},
  \href{https://arxiv.org/abs/http://arxiv.org/abs/1711.05270v2}{{\ttfamily
  http://arxiv.org/abs/1711.05270v2}}.

\bibitem{Pospelov:2005pr}
M.~Pospelov and A.~Ritz, \emph{{Electric dipole moments as probes of new
  physics}}, \href{https://doi.org/10.1016/j.aop.2005.04.002}{\emph{Annals
  Phys.} {\bfseries 318} (2005) 119--169},
  [\href{https://arxiv.org/abs/hep-ph/0504231}{{\ttfamily hep-ph/0504231}}].

\bibitem{Gupta2018}
R.~Gupta, B.~Yoon, T.~Bhattacharya, V.~Cirigliano, Y.-C. Jang and H.-W. Lin,
  \emph{Flavor diagonal tensor charges of the nucleon from 2+1+1 flavor lattice
  qcd}, \href{https://doi.org/10.1103/PhysRevD.98.091501}{\emph{Phys. Rev. D
  98, 091501 (2018)} (Aug., 2018) },
  [\href{https://arxiv.org/abs/1808.07597}{{\ttfamily 1808.07597}}].

\bibitem{Engel2013}
J.~Engel, M.~J. Ramsey-Musolf and U.~van Kolck, \emph{Electric dipole moments
  of nucleons, nuclei, and atoms: The standard model and beyond},
  \href{https://arxiv.org/abs/1303.2371}{{\ttfamily 1303.2371}}.

\bibitem{Hisano2012}
J.~Hisano, J.~Y. Lee, N.~Nagata and Y.~Shimizu, \emph{Reevaluation of neutron
  electric dipole moment with qcd sum rules},
  \href{https://doi.org/10.1103/PhysRevD.85.114044}{\emph{Phys.Rev.D85:114044,2012}
  (Apr., 2012) }, [\href{https://arxiv.org/abs/1204.2653}{{\ttfamily
  1204.2653}}].

\bibitem{deVries:2012ab}
J.~de~Vries, E.~Mereghetti, R.~G.~E. Timmermans and U.~van Kolck, \emph{{The
  Effective Chiral Lagrangian From Dimension-Six Parity and Time-Reversal
  Violation}}, \href{https://doi.org/10.1016/j.aop.2013.05.022}{\emph{Annals
  Phys.} {\bfseries 338} (2013) 50--96},
  [\href{https://arxiv.org/abs/1212.0990}{{\ttfamily 1212.0990}}].

\bibitem{Yamanaka:2018uud}
{\scshape JLQCD} collaboration, N.~Yamanaka, S.~Hashimoto, T.~Kaneko and
  H.~Ohki, \emph{{Nucleon charges with dynamical overlap fermions}},
  \href{https://doi.org/10.1103/PhysRevD.98.054516}{\emph{Phys. Rev. D}
  {\bfseries 98} (2018) 054516},
  [\href{https://arxiv.org/abs/1805.10507}{{\ttfamily 1805.10507}}].

\bibitem{Yamanaka:2020kjo}
N.~Yamanaka and E.~Hiyama, \emph{{Weinberg operator contribution to the nucleon
  electric dipole moment in the quark model}},
  \href{https://doi.org/10.1103/PhysRevD.103.035023}{\emph{Phys. Rev. D}
  {\bfseries 103} (2021) 035023},
  [\href{https://arxiv.org/abs/2011.02531}{{\ttfamily 2011.02531}}].

\bibitem{Demir:2002gg}
D.~A. Demir, M.~Pospelov and A.~Ritz, \emph{{Hadronic EDMs, the Weinberg
  operator, and light gluinos}},
  \href{https://doi.org/10.1103/PhysRevD.67.015007}{\emph{Phys. Rev. D}
  {\bfseries 67} (2003) 015007},
  [\href{https://arxiv.org/abs/hep-ph/0208257}{{\ttfamily hep-ph/0208257}}].

\bibitem{Haisch:2019bml}
U.~Haisch and A.~Hala, \emph{{Sum rules for CP-violating operators of Weinberg
  type}}, \href{https://doi.org/10.1007/JHEP11(2019)154}{\emph{JHEP} {\bfseries
  11} (2019) 154}, [\href{https://arxiv.org/abs/1909.08955}{{\ttfamily
  1909.08955}}].

\bibitem{Weinberg1989}
S.~Weinberg, \emph{Larger higgs-boson-exchange terms in the neutron electric
  dipole moment},
  \href{https://doi.org/10.1103/physrevlett.63.2333}{\emph{Physical Review
  Letters} {\bfseries 63} (nov, 1989) 2333--2336}.

\bibitem{PhysRevD.16.1791}
R.~D. Peccei and H.~R. Quinn, \emph{Constraints imposed by $\mathrm{CP}$
  conservation in the presence of pseudoparticles},
  \href{https://doi.org/10.1103/PhysRevD.16.1791}{\emph{Phys. Rev. D}
  {\bfseries 16} (Sep, 1977) 1791--1797}.

\bibitem{Hook:2018dlk}
A.~Hook, \emph{{TASI Lectures on the Strong CP Problem and Axions}},
  {\emph{PoS} {\bfseries TASI2018} (2019) 004},
  [\href{https://arxiv.org/abs/1812.02669}{{\ttfamily 1812.02669}}].

\bibitem{Pospelov1999}
M.~Pospelov and A.~Ritz, \emph{Hadron electric dipole moments from cp-odd
  operators of dimension five via qcd sum rules: The vector meson},
  \href{https://doi.org/10.1016/S0370-2693(99)01343-X}{\emph{Phys.Lett. B471
  (2000) 388-395} (Oct., 1999) },
  [\href{https://arxiv.org/abs/hep-ph/9910273}{{\ttfamily hep-ph/9910273}}].

\bibitem{Pospelov2000}
M.~Pospelov and A.~Ritz, \emph{Neutron edm from electric and chromoelectric
  dipole moments of quarks},
  \href{https://doi.org/10.1103/PhysRevD.63.073015}{\emph{Phys.Rev. D63 (2001)
  073015} (Oct., 2000) },
  [\href{https://arxiv.org/abs/hep-ph/0010037}{{\ttfamily hep-ph/0010037}}].

\bibitem{Cheung2015}
C.~Cheung and C.-H. Shen, \emph{Non-renormalization theorems without
  supersymmetry},
  \href{https://doi.org/10.1103/PhysRevLett.115.071601}{\emph{Phys. Rev. Lett.
  115, 071601 (2015)} (May, 2015) },
  [\href{https://arxiv.org/abs/1505.01844}{{\ttfamily 1505.01844}}].

\bibitem{Azatov2016}
A.~Azatov, R.~Contino, C.~S. Machado and F.~Riva, \emph{Helicity selection
  rules and non-interference for bsm amplitudes},
  \href{https://doi.org/10.1103/PhysRevD.95.065014}{\emph{Phys. Rev. D 95,
  065014 (2017)} (July, 2016) },
  [\href{https://arxiv.org/abs/1607.05236}{{\ttfamily 1607.05236}}].

\bibitem{Craig2019}
N.~Craig, M.~Jiang, Y.-Y. Li and D.~Sutherland, \emph{Loops and trees in
  generic efts},
  \href{https://arxiv.org/abs/http://arxiv.org/abs/2001.00017v1}{{\ttfamily
  http://arxiv.org/abs/2001.00017v1}}.

\bibitem{Jiang2020}
M.~Jiang, J.~Shu, M.-L. Xiao and Y.-H. Zheng, \emph{New selection rules from
  angular momentum conservation},
  \href{https://doi.org/10.1103/PhysRevLett.126.011601}{\emph{Phys. Rev. Lett.
  126, 011601 (2021)} (Jan., 2020) },
  [\href{https://arxiv.org/abs/2001.04481}{{\ttfamily 2001.04481}}].

\bibitem{Anastasiou2006}
C.~Anastasiou, R.~Britto, B.~Feng, Z.~Kunszt and P.~Mastrolia, \emph{{Unitarity
  cuts and Reduction to master integrals in d dimensions for one-loop
  amplitudes}},
  \href{https://doi.org/10.1088/1126-6708/2007/03/111}{\emph{JHEP} {\bfseries
  03} (2007) 111}, [\href{https://arxiv.org/abs/hep-ph/0612277}{{\ttfamily
  hep-ph/0612277}}].

\bibitem{Badger2008}
S.~D. Badger, \emph{{Direct Extraction Of One Loop Rational Terms}},
  \href{https://doi.org/10.1088/1126-6708/2009/01/049}{\emph{JHEP} {\bfseries
  01} (2009) 049}, [\href{https://arxiv.org/abs/0806.4600}{{\ttfamily
  0806.4600}}].

\bibitem{ArkaniHamed2017}
N.~Arkani-Hamed, T.-C. Huang and Y.-t. Huang, \emph{{Scattering Amplitudes For
  All Masses and Spins}},  \href{https://arxiv.org/abs/1709.04891}{{\ttfamily
  1709.04891}}.

\bibitem{Boudjema1991}
F.~Boudjema, K.~Hagiwara, C.~Hamzaoui and K.~Numata, \emph{Anomalous moments of
  quarks and leptons from {nonstandardWW}$\gamma$couplings},
  \href{https://doi.org/10.1103/physrevd.43.2223}{\emph{Physical Review D}
  {\bfseries 43} (apr, 1991) 2223--2232}.

\bibitem{Gripaios2013}
B.~Gripaios and D.~Sutherland, \emph{On lhc searches for cp-violating,
  dimension-6 electroweak gauge boson operators},
  \href{https://doi.org/10.1103/PhysRevD.89.076004}{\emph{Phys. Rev. D 89,
  076004 (2014)} (Sept., 2013) },
  [\href{https://arxiv.org/abs/1309.7822}{{\ttfamily 1309.7822}}].

\bibitem{Baratella2020}
P.~Baratella, C.~Fernandez and A.~Pomarol, \emph{Renormalization of
  higher-dimensional operators from on-shell amplitudes},
  \href{https://arxiv.org/abs/2005.07129}{{\ttfamily 2005.07129}}.

\bibitem{Barr1990}
S.~M. Barr, E.~M. Freire and A.~Zee, \emph{Mechanism for large neutrino
  magnetic moments},
  \href{https://doi.org/10.1103/physrevlett.65.2626}{\emph{Physical Review
  Letters} {\bfseries 65} (nov, 1990) 2626--2629}.

\bibitem{SEKHARCHIVUKULA198799}
B.~{Sekhar Chivukula} and H.~Georgi, \emph{Composite-technicolor standard
  model},
  \href{https://doi.org/https://doi.org/10.1016/0370-2693(87)90713-1}{\emph{Physics
  Letters B} {\bfseries 188} (1987) 99--104}.

\bibitem{DAmbrosio2002}
G.~D'Ambrosio, G.~F. Giudice, G.~Isidori and A.~Strumia, \emph{Minimal flavour
  violation: an effective field theory approach},
  \href{https://doi.org/10.1016/S0550-3213(02)00836-2}{\emph{Nucl.Phys.B645:155-187,2002}
  (July, 2002) }, [\href{https://arxiv.org/abs/hep-ph/0207036}{{\ttfamily
  hep-ph/0207036}}].

\bibitem{Isidori2012}
G.~Isidori and D.~M. Straub, \emph{Minimal flavour violation and beyond},
  \href{https://arxiv.org/abs/1202.0464}{{\ttfamily 1202.0464}}.

\bibitem{SMEFTFlavorSym}
D.~A. Faroughy, G.~Isidori, F.~Wilsch and K.~Yamamoto, \emph{{Flavour
  symmetries in the SMEFT}},
  \href{https://doi.org/10.1007/JHEP08(2020)166}{\emph{JHEP} {\bfseries 08}
  (2020) 166}, [\href{https://arxiv.org/abs/2005.05366}{{\ttfamily
  2005.05366}}].

\bibitem{SMEFTFlavorSym2}
J.~Fuentes-Mart\'\i{}n, G.~Isidori, J.~Pag\`es and K.~Yamamoto, \emph{{With or
  without U(2)? Probing non-standard flavor and helicity structures in
  semileptonic B decays}},
  \href{https://doi.org/10.1016/j.physletb.2019.135080}{\emph{Phys. Lett. B}
  {\bfseries 800} (2020) 135080},
  [\href{https://arxiv.org/abs/1909.02519}{{\ttfamily 1909.02519}}].

\bibitem{FlavorU2}
R.~Barbieri, D.~Buttazzo, F.~Sala and D.~M. Straub, \emph{{Flavour physics from
  an approximate $U(2)^3$ symmetry}},
  \href{https://doi.org/10.1007/JHEP07(2012)181}{\emph{JHEP} {\bfseries 07}
  (2012) 181}, [\href{https://arxiv.org/abs/1203.4218}{{\ttfamily 1203.4218}}].

\bibitem{Hartmann:2015oia}
C.~Hartmann and M.~Trott, \emph{{On one-loop corrections in the standard model
  effective field theory; the $\Gamma(h \rightarrow \gamma \, \gamma)$ case}},
  \href{https://doi.org/10.1007/JHEP07(2015)151}{\emph{JHEP} {\bfseries 07}
  (2015) 151}, [\href{https://arxiv.org/abs/1505.02646}{{\ttfamily
  1505.02646}}].

\bibitem{Hooft1972}
G.~t~Hooft and M.~Veltman, \emph{Regularization and renormalization of gauge
  fields}, \href{https://doi.org/10.1016/0550-3213(72)90279-9}{\emph{Nuclear
  Physics B} {\bfseries 44} (jul, 1972) 189--213}.

\bibitem{Breitenlohner1977}
P.~Breitenlohner and D.~Maison, \emph{Dimensional renormalization and the
  action principle},
  \href{https://doi.org/10.1007/bf01609069}{\emph{Communications in
  Mathematical Physics} {\bfseries 52} (feb, 1977) 11--38}.

\bibitem{Bonneau1980}
G.~Bonneau, \emph{Trace and axial anomalies in dimensional renormalization
  through zimmermann-like identities},
  \href{https://doi.org/10.1016/0550-3213(80)90382-x}{\emph{Nuclear Physics B}
  {\bfseries 171} (1980) 477--508}.

\bibitem{Abbott1981}
L.~Abbott, \emph{The background field method beyond one loop},
  \href{https://doi.org/10.1016/0550-3213(81)90371-0}{\emph{Nuclear Physics B}
  {\bfseries 185} (jul, 1981) 189--203}.

\bibitem{Abbott1983}
L.~Abbott, M.~Grisaru and R.~Schaefer, \emph{The background field method and
  the s-matrix},
  \href{https://doi.org/10.1016/0550-3213(83)90337-1}{\emph{Nuclear Physics B}
  {\bfseries 229} (dec, 1983) 372--380}.

\bibitem{Denner1994}
A.~Denner, S.~Dittmaier and G.~Weiglein, \emph{Application of the
  background-field method to the electroweak standard model},
  \href{https://doi.org/10.1016/0550-3213(95)00037-S}{\emph{Nucl.Phys. B440
  (1995) 95-128} (Oct., 1994) },
  [\href{https://arxiv.org/abs/hep-ph/9410338}{{\ttfamily hep-ph/9410338}}].

\bibitem{Denner1996}
A.~Denner, S.~Dittmaier and G.~Weiglein, \emph{The background-field formulation
  of the electroweak standard model}, {\emph{Acta
  Phys.Polon.B27:3645-3660,1996} (Sept., 1996) },
  [\href{https://arxiv.org/abs/hep-ph/9609422}{{\ttfamily hep-ph/9609422}}].

\bibitem{Helset2018}
A.~Helset, M.~Paraskevas and M.~Trott, \emph{Gauge fixing the standard model
  effective field theory},
  \href{https://doi.org/10.1103/PhysRevLett.120.251801}{\emph{Phys. Rev. Lett.
  120, 251801 (2018)} (Mar., 2018) },
  [\href{https://arxiv.org/abs/1803.08001}{{\ttfamily 1803.08001}}].

\bibitem{Corbett2020}
T.~Corbett, \emph{The feynman rules for the smeft in the background field
  gauge},  \href{https://arxiv.org/abs/2010.15852}{{\ttfamily 2010.15852}}.

\bibitem{Corbett2020a}
T.~Corbett and M.~Trott, \emph{One loop verification of smeft ward identities},
   \href{https://arxiv.org/abs/2010.08451}{{\ttfamily 2010.08451}}.

\bibitem{Patel2015}
H.~H. Patel, \emph{Package-x: A mathematica package for the analytic
  calculation of one-loop integrals},
  \href{https://arxiv.org/abs/http://arxiv.org/abs/1503.01469v2}{{\ttfamily
  http://arxiv.org/abs/1503.01469v2}}.

\bibitem{Alloul2013}
A.~Alloul, N.~D. Christensen, C.~Degrande, C.~Duhr and B.~Fuks, \emph{Feynrules
  2.0 - a complete toolbox for tree-level phenomenology},
  \href{https://doi.org/10.1016/j.cpc.2014.04.012}{\emph{Comput.Phys.Commun.
  185 (2014) 2250-2300} (Oct., 2013) },
  [\href{https://arxiv.org/abs/1310.1921}{{\ttfamily 1310.1921}}].

\bibitem{Hahn2000}
T.~Hahn, \emph{Generating feynman diagrams and amplitudes with feynarts 3},
  \href{https://doi.org/10.1016/S0010-4655(01)00290-9}{\emph{Comput.Phys.Commun.
  140 (2001) 418-431} (Dec., 2000) },
  [\href{https://arxiv.org/abs/hep-ph/0012260}{{\ttfamily hep-ph/0012260}}].

\bibitem{Hahn1998}
T.~Hahn and M.~Perez-Victoria, \emph{Automatized one-loop calculations in 4 and
  d dimensions},
  \href{https://arxiv.org/abs/http://arxiv.org/abs/hep-ph/9807565v1}{{\ttfamily
  http://arxiv.org/abs/hep-ph/9807565v1}}.

\bibitem{Jegerlehner2000}
F.~Jegerlehner, \emph{Facts of life with gamma(5)},
  \href{https://doi.org/10.1007/s100520100573}{\emph{Eur.Phys.J.C18:673-679,2001}
  (May, 2000) }, [\href{https://arxiv.org/abs/hep-th/0005255}{{\ttfamily
  hep-th/0005255}}].

\bibitem{BECCHI1974344}
C.~Becchi, A.~Rouet and R.~Stora, \emph{The abelian higgs kibble model,
  unitarity of the s-operator},
  \href{https://doi.org/https://doi.org/10.1016/0370-2693(74)90058-6}{\emph{Physics
  Letters B} {\bfseries 52} (1974) 344--346}.

\bibitem{Becchi1975}
C.~Becchi, A.~Rouet and R.~Stora, \emph{Renormalization of the abelian
  higgs-kibble model},
  \href{https://doi.org/10.1007/BF01614158}{\emph{Communications in
  Mathematical Physics} {\bfseries 42} (1975) 127--162}.

\bibitem{BECCHI1976287}
C.~Becchi, A.~Rouet and R.~Stora, \emph{Renormalization of gauge theories},
  \href{https://doi.org/https://doi.org/10.1016/0003-4916(76)90156-1}{\emph{Annals
  of Physics} {\bfseries 98} (1976) 287--321}.

\bibitem{Bhattacharya2015}
T.~Bhattacharya, V.~Cirigliano, R.~Gupta, E.~Mereghetti and B.~Yoon,
  \emph{Dimension-5 cp-odd operators: Qcd mixing and renormalization},
  \href{https://arxiv.org/abs/1502.07325}{{\ttfamily 1502.07325}}.

\end{thebibliography}\endgroup
\bibliographystyle{JHEP}
\nocite{*}
\end{document}